 \documentclass[final,3p,times]{elsarticle}

\usepackage{amssymb}
\usepackage{babel,tabularx,ragged2e,booktabs}

 \usepackage{lineno}

\usepackage[utf8]{inputenc}
\usepackage{textcomp}
\usepackage[mathscr]{euscript}
\usepackage{graphicx}
\usepackage{amsmath}
\usepackage[version=4]{mhchem}
\usepackage{siunitx}
\usepackage{longtable,tabularx}
\usepackage{color}
\usepackage{cancel}
\usepackage{gensymb}
\usepackage{color}
\usepackage{tcolorbox}
\usepackage{float}
\usepackage{multirow}
\usepackage{caption}
\usepackage{subcaption}
\usepackage{graphicx}
\usepackage{amsmath}
\usepackage{multicol}
\usepackage{wrapfig}
\usepackage{amsmath,amssymb}
\usepackage{nicematrix}
\usepackage{paracol}
\usepackage{ragged2e}
\usepackage{algpseudocode}
\usepackage{algorithm2e}
\RestyleAlgo{ruled}

\makeatletter
\newsavebox\myboxA
\newsavebox\myboxB
\newlength\mylenA

\newcommand*\xoverline[2][0.75]{%
    \sbox{\myboxA}{$\m@th#2$}%
    \setbox\myboxB\null
    \ht\myboxB=\ht\myboxA%
    \dp\myboxB=\dp\myboxA%
    \wd\myboxB=#1\wd\myboxA
    \sbox\myboxB{$\m@th\overline{\copy\myboxB}$}
    \setlength\mylenA{\the\wd\myboxA}
    \addtolength\mylenA{-\the\wd\myboxB}%
    \ifdim\wd\myboxB<\wd\myboxA%
       \rlap{\hskip 0.5\mylenA\usebox\myboxB}{\usebox\myboxA}%
    \else
        \hskip -0.5\mylenA\rlap{\usebox\myboxA}{\hskip 0.5\mylenA\usebox\myboxB}%
    \fi}
\makeatother

\usepackage[colorlinks]{hyperref}

\colorlet{colorRev1}{blue!80!black} 
\colorlet{colorRev2}{red!80!black} 

\journal{Journal of Sound and Vibration}

\begin{document}

\begin{frontmatter}



\title{Modeling and analysis of a flexible spinning Euler-Bernoulli beam with centrifugal stiffening and softening: A Linear Fractional Representation approach with application to spinning spacecraft}


\author[ISAE]{R. Rodrigues\corref{cor1}}
\ead{ricardo.rodrigues@isae-supaero.fr}
\cortext[cor1]{Corresponding author; Ph.D. student.}

\author[ISAE]{D. Alazard\fnref{label2}}
\ead{daniel.alazard@isae-supaero.fr}
\fntext[label2]{Professor}

\author[ISAE]{F. Sanfedino\fnref{label3}}
\ead{francesco.sanfedino@isae-supaero.fr}
\fntext[label3]{Associate Professor}

\author[ISAE]{T. Mauriello\fnref{label4}}
\ead{tommaso.mauriello@student.isae-supaero.fr}
\fntext[label4]{MSc Student}

\author[ISAE]{P. Iannelli\fnref{label5}}
\ead{paolo.iannelli2@isae-supaero.fr}
\fntext[label5]{Postdoctoral Researcher}


\address[ISAE]{Institut Supérieur de l’Aéronautique et de l’Espace (ISAE-SUPAERO), Université de Toulouse, 10 Avenue Edouard Belin, BP-54032, 31055, Toulouse
Cedex 4, France}


\begin{abstract} 

The derivation of a linear fractional representation (LFR) model for a flexible, spinning and uniform Euler-Bernoulli beam is accomplished using the {Lagrange} technique, fully capturing the centrifugal force generated by the spinning motion and accounting for its dependence on the angular velocity. This six degrees of freedom (DOF) model accounts for the behavior of deflection in the moving body frame, encompassing the bending, traction and torsion dynamics. The model is also designed to be compliant with the Two-Input-Two-Output Port (TITOP) approach, which offers the possibility to model complex multibody mechanical systems, while keeping the uncertain nature of the plant and condensing all the possible mechanical configurations in a single LFR. To evaluate the effectiveness of the model, various scenarios are considered and their results are tabulated. These scenarios include uniform beams with fixed root boundary conditions for different values of tip mass, root offset and angular velocity. The results from the analysis of the uniform cantilever beam are compared with solutions found in the literature and obtained from a commercial finite element software. Ultimately, this paper presents a multibody model for a spinning spacecraft mission scenario. A comprehensive analysis of the system dynamics is conducted, providing insights into the behavior of the spacecraft under spinning conditions.

\end{abstract}


\end{frontmatter}

\nolinenumbers

\section*{Nomenclature}

{

\noindent\begin{longtable*}{@{}l @{\quad \quad} l@{}}
{AMDM} & {Adomian Modified Decomposition Method} \\
{AMM} & {Assumed Modes Method} \\ 
{AOCS} & {Attitude and Orbital Control System} \\
{CMS} & {Component Mode Synthesis} \\
{DCM} & {Direct Cosine Matrix} \\
{DOF} & {Degrees Of Freedom} \\
{FEM} & {Finite Element Method} \\
{GNC} & {Guidance, Navigation and Control} \\
{LFR} & {Linear Fractional Representation} \\
{LPV} & {Linear Parameter-Varying} \\
{SDTlib} & {Satellite Dynamics Toolbox library} \\
{THOR} & {Turbulence Heating ObserveR} \\
{TITOP} & {Two-Input Two-Output Ports} \\
{TMM} & {Transfer Matrix Method} \\
{3D} & {Three-Dimensional} \\
$\displaystyle \mathbf{a}_{P}^{\mathcal{B}} \equiv \left.\frac{d {\mathbf{v}}^{\mathcal{B}}_{P}}{d t}\right|_{\mathcal{R}_i}$ & Inertial linear acceleration vector of the body ${\mathcal{B}}$ at the point $P$, expressed in \si{\meter\per\square\second}.\\

$\displaystyle \dot{\mathbf{v}}_{P}^{\mathcal{B}} \equiv \left.\frac{d {\mathbf{v}}^{\mathcal{B}}_{P}}{d t}\right|_{\mathcal{R}_b}$ & Linear acceleration vector of the body ${\mathcal{B}}$ computed at the point $P$ with respect to the \\ 
& body frame $\mathcal{R}_b$, expressed in \si{\meter\per\square\second}. \\
$\boldsymbol{\dot{\omega}}^{\mathcal{B}}_P$ & Inertial angular acceleration vector of the body $\mathcal{B}$ at the point $P$, expressed in \si{\radian\per\square\second}.\\

$\displaystyle \mathbf{v}_{P}^{\mathcal{B}} \equiv \left.\frac{d {\mathbf{x}}_{P}}{d t}\right|_{\mathcal{R}_i}$ & Inertial velocity vector of the body $\mathcal{B}$ at the point $P$, expressed in \si{\meter\per\second}. \\
$\boldsymbol{{\omega}}^{\mathcal{B}}_P$ & Inertial angular velocity vector of the body $\mathcal{B}$ at the point $P$, expressed in \si{\radian\per\second}. \\

$\mathbf{BP}$ & Distance vector defined between the points $B$ and $P$, expressed in \si{\meter}. \\
$\mathbf{x}^{\mathcal{B}}_P \equiv \mathbf{OP}$ & Distance vector defined between the point $O$, the origin of the inertial frame, \\
& and the point ${P}$ of the body $\mathcal{B}$, expressed in \si{\meter}. \\
$\mathbf{BP}$ & Distance vector defined between the points $B$ and $P$, expressed in \si{\meter}. \\
$\boldsymbol{{\Theta}}^{\mathcal{B}}_{P}$ & \textsc{Euler} angles vector of the body ${\mathcal{B}}$ computed at the point $P$ with respect to the inertial \\
& frame, expressed in \si{\radian} and using the 'ZYX' sequence of rotations. \\
$\mathbf{F}_{\mathrm{ext} / \mathcal{B}, G}$  & External forces vector applied to the body $\mathcal{B}$ at the point $G$, expressed in \si{\newton}.\\
$\mathbf{T}_{\mathrm{ext} / \mathcal{B}, G}$ & External torques vector applied to the body 
$\mathcal{B}$ at the point $G$, expressed in \si{\newton\meter}.\\
$\mathbf{F}_{\mathcal{B/A},G}$ & Forces vector applied by the body 
$\mathcal{B}$ to the body $\mathcal{A}$ at the point $G$, expressed in \si{\newton}.\\
$\mathbf{T}_{\mathcal{B/A},G}$ & Torques vector applied by the body 
$\mathcal{B}$ to the body $\mathcal{A}$ at the point $G$, expressed in 
\si{\newton\meter}.\\
$\mathbf{W}_{\mathcal{B/A},G}$ & Wrench vector applied by the body $\mathcal{B}$ to the body $\mathcal{A}$ at the point $G$:\\
& $\mathbf{W}_{\mathcal{B/A},G}=\left[\begin{array}{cc}\mathbf{F}^{\mathrm{T}}_{\mathcal{B/A},G}\;\;\mathbf{T}^{\mathrm{T}}_{\mathcal{B/A},G}\end{array}\right]^{\mathrm{T}}$, expressed in $\left[\si{\newton}\;\; \si{\newton\meter}\right]$.\\

$\left[\mathbf{X}\right]_{\mathcal{R}_{\bullet}}$ & $\mathbf{X}$ (model, vector 
or tensor) expressed in the frame $\mathcal{R}_{\bullet}$.\\
$\mathbf{X}$ & $\mathbf{X}$ (model, vector 
or tensor) expressed in the body frame, unless stated otherwise.\\

${{\mathbf{m}}^{\mathcal{B}}_{P}}$ & Motion vector of the body $\mathcal{B}$ at the point $P$:  ${{\mathbf{m}}^{\mathcal{B}}_{P}}=\left[\begin{array}{cccccc} {\dot{\mathbf{v}}}^{\mathcal{B}^{{\mathrm{T}}}}_{P} & {\boldsymbol{\dot{\omega}}}_P^{\mathcal{B}^{{\mathrm{T}}}} & \mathbf{v}_{P}^{{\mathcal{B}}^{\mathrm{T}}} & \boldsymbol{{\omega}}_P^{{\mathcal{B}}^{\mathrm{T}}} &
\mathbf{x}_P^{{\mathrm{T}}} & \boldsymbol{{\Theta}}_P^{{\mathcal{B}}^{\mathrm{T}}}\end{array}\right]^{\mathrm{T}}$. \\
$\mathbf P_{\mathcal R_a/\mathcal R_b}$ & DCM from the frame $\mathcal R_a$ to the frame $\mathcal R_b$ ($\left[\mathbf{v}\right]_{\mathcal{R}_{b}}=\mathbf P_{\mathcal R_a/\mathcal R_b}\left[\mathbf{v}\right]_{\mathcal{R}_{a}}$ for any vector $\mathbf v$).\\
$\mathbf P_{\mathcal R_b/\mathcal R_i}=\mathrm{DCM}(\boldsymbol{{\Theta}}^{\mathcal{B}})$ & Function that computes the DCM from the \textsc{Euler} angles vector of a rigid body $\mathcal B$.\\
$\textbf{I}_{n}$ &  Identity matrix $n \times n$.\\
$\mathbf{0}_{n\times m}$  & Zero matrix $n \times m$.\\
$\mathbf{v}\{i\}$ & Component $i$ of vector $\mathbf{v}$.\\
$(^*\mathbf{v})$ & Skew symmetric matrix associated with vector
$\mathbf{v}$: $\left[(^*\mathbf{v})\right]_{\mathcal{R}_{\bullet}}=\left[\begin{array}{ccc}0 & -\mathbf{v}\{3\} & \mathbf{v}\{2\} \\ \mathbf{v}\{3\} & 0 & -\mathbf{v}\{1\} \\ -\mathbf{v}\{2\} & \mathbf{v}\{1\} & 0 \end{array}\right]_{\mathcal{R}_{\bullet}}$.\\
$\boldsymbol{\tau}_{PB}$ &  Kinematic model between the points $P$ and $B$: $\left[\boldsymbol{\tau}_{PB}\right]_{\mathcal{R}_{\bullet}}=\left[\begin{array}{cc} \textbf{I}_{3}& (^*\mathbf{PB}) \\  \mathbf{0}_{3\times 3}& \textbf{I}_{3} \end{array}\right]_{\mathcal{R}_{\bullet}}$.\\

$\mathbf{u} \wedge \mathbf{v}$ & Cross product of vectors $\mathbf{u}$ and $\mathbf{v}\left(\mathbf{u} \wedge \mathbf{v}=\left({ }^* \mathbf{u}\right) \mathbf{v}\right)$. \\
$\|\mathbf{v}\|_2$ & Euclidean norm of the vector $\mathbf{v}$. \\
$\displaystyle \hat{\mathbf{v}}=\frac{\mathbf{v}}{\|\mathbf{v}\|_2}$ & Normalized vector $\hat{\mathbf{v}}$ of a non-zero vector ${\mathbf{v}}$ (or unit vector in the direction of ${\mathbf{v}}$). \\
$\boldsymbol{\Phi}^{\prime}(x)$ & Derivative of $\boldsymbol{\Phi}(x)$ with respect to the spatial derivative $x$. \\
$\dot{\mathbf{x}}(t)$ & First time derivative of $\mathbf{x}$ with respect to the body frame. \\

$\mathbf{X}_{(\mathbf{I},\mathbf{J})}$ & Subsystem of $\mathbf{X}$ from the inputs indexed in the vector $\mathbf{J}$ to the outputs indexed \\ 
& in the vector $\mathbf{I}$ (if $(\mathbf{I},\mathbf{J})=(:,7:10)$, it means that one is considering the subsystem \\ 
& between the inputs from $7$ until $10$ to all the outputs). \\
$\mbox{diag}(\mathbf A,\mathbf B,\cdots)$ & Block-diagonal augmented matrix.\\
$\overline{\mathbf{x}}$ & Equilibrium value of $\mathbf{x}$. \\
$\delta{\mathbf{x}}$ & Small variation of $\mathbf{x}$ around a defined equilibrium. \\
$\textbf{J}_{G}^{\mathcal{B}}$ & Inertia tensor of $\mathcal{B}$, computed at the point $G$, written in $\mathcal{R}_{b}$ and expressed in \si{\kilogram\square\meter}.\\
$m^{\mathcal{B}}$ & Mass of $\mathcal{B}$ expressed in \si{\kilogram}.\\
$\mathrm{s}$ & \textsc{Laplace}'s variable.

\end{longtable*}}

\section{Introduction}

\subsection{Background and motivation}

The design of helicopter rotor systems, wind turbines and various other devices has motivated numerous studies into the flexural vibration modes of rotating beams \cite{wright}. The use of structures that can be modeled as a rotating beam is also very common in spacecraft applications, which usually employ long, flexible and rotating space booms \cite{PUIG201012} or high gain antennas. Some examples are the Cluster mission \cite{cluster} and the Turbulence Heating ObserveR (THOR) satellite mission proposal \cite{thor}. On this basis, it is not surprising that there is extensive literature concerning the dynamics and vibrations of such structures \cite{Shabana1997}. Traditionally, these components were designed using rigid materials. However, the current trend is to employ more flexible and lightweight structures. This trend is particularly significant in space missions.

Numerous previous studies have focused on the analysis of non-rotating non-uniform blades \cite{AUCIELLO1997522, ZHOU2000203, ELISHAKOFF2000529, KHAN20111963}. In particular, Auciello {\it et al.} \cite{AUCIELLO1997522} and Zhou {\it et al.} \cite{ZHOU2000203} employed the Rayleigh-Ritz method to investigate the vibration behavior of non-uniform Euler-Bernoulli beams with polynomial cross-sections, incorporating discrete elements to account for moment of inertia variations and different boundary conditions. Recent studies have also utilized the homotopy perturbation transform method \cite{KHAN20111963, KHAN20132702, Khan2012, KhanUsman+2014+19+25, KHAN2014229} to address the same problem.

Moreover, several research groups have conducted studies using the Adomian decomposition method and the Adomian Modified Decomposition Method (AMDM) for uniform beams, according to either the Euler-Bernoulli or the Timoshenko formulations \cite{Mao2013, HSU2009451, lai2008free, LAI20083204}. Mao \cite{Mao2013} utilized the AMDM for rotating uniform beams, incorporating a centrifugal stiffening term, while Hsu {\it et al.} \cite{HSU2009451} applied the AMDM to Timoshenko beams.  Khan {\it et al.} \cite{KhanAlHayani+2013+355+361} demonstrated the potential for enhanced accuracy in AMDM solutions by incorporating Green functions. Recent studies have also explored the analysis of rotating cantilever beams and related nonlinear oscillations in structural engineering \cite{VzquezLeal2013, AKBARZADE2012480, KhanAkbarzade+2012+435+440}. Additionally, Adair {\it et al.} \cite{ADAIR20163230} employed the AMDM to analyze the free transverse vibration of rotating non-uniform Euler-Bernoulli beams, considering various boundary conditions, rotation speeds and beam lengths. Previous studies \cite{hodges, wright, NAGULESWARAN1994613} have generally utilized the Euler-Bernoulli beam theory and employed various approximate solution techniques to determine the dynamic characteristics of such rotating beams. Yoo et al. \cite{YOO1998807} specifically investigated the influence of the centrifugal force and utilized a modal formulation to determine the natural frequencies. Furthermore, other methods have also been used to compute the natural frequencies of rotating beams in \cite{wang, BANERJEE20061034, OZDEMIR2006413, Hashemi, Kumar}.

From the perspective of the Attitude and Orbit Control System (AOCS) and Guidance, Navigation, and Control (GNC), the challenges associated with spinning spacecraft missions arise primarily due to the impact of centrifugal stiffening on the dynamics of the satellite's flexible rotating booms. Therefore, the success of such missions relies heavily on the ability to develop an accurate system model. In this context, the comprehensive modeling of complex multibody structures becomes crucial as it enables the prediction of worst-case scenarios early on and offers the capacity to push the control system to its performance limits. Furthermore, having a model that is valid for any possible configuration simplifies the synthesis of the controller, eliminating the need for frequent switching between control modes. The transition between different control phases poses a critical aspect in control design, often involving intermediate tranquilization time windows. Consequently, the objective of this paper is not only to achieve an accurate model that characterizes the flexible modes of a rotating beam but also to construct a model which can be integrated in a multibody flexible system that facilitates robust control design and analysis.

In multibody dynamics, link flexibility is usually approximated using finite element methods (FEM) or assumed mode methods (AMM) \cite{Theodore1995} and the complete system model is derived using the Euler-Lagrange formalism. For space applications with a flexible robotic arm on a chaser spacecraft, the AMM approach is commonly used \cite{Theodore1995, DELUCA}, considering the first clamped-mass eigenfunctions. However, when dealing with scenarios like capturing massive space debris, the boundary conditions at the flexible link tips may resemble clamped-clamped rather than clamped-mass conditions. Therefore, it is very important to have valid models for each link or substructure under arbitrary boundary conditions prior to assembling the entire structure. The transfer matrix method (TMM) has been extensively researched in substructuring approaches, establishing a relationship between state vectors at different points of a flexible body \cite{LECKIE1960137}. It has proven effective for modeling structures with serially connected bodies or open-chain-like structures. The TMM has been combined with the finite element (FE) method, leading to the FE-TM method \cite{Dokainish}, enabling reduced models, handling complex-shaped bodies and reducing computational efforts \cite{Mucino1981, rong}. Extensions have been developed for control design, including non-collocated feedback \cite{TAN199047, Krauss}, and for hybrid multibody systems with both rigid and flexible components \cite{Rui2007}. However, the TMM has limitations. Inverting submatrices can be challenging when dealing with non-square or non-invertible matrices, depending on the boundary conditions \cite{TAN199047}. Moreover, the TMM cannot be directly applied to tree-like structures, such as a spacecraft with a main rigid body and multiple flexible appendages. In these cases, establishing a relationship between generalized forces and displacements at the root of the structure is crucial for designing the AOCS. Efficient approaches have been developed to address this scenario, such as the effective mass/inertia model for each appendage or the utilization of the impedance matrix \cite{Sylla2008}. These methods, commonly employed in space engineering, analyze dynamic coupling between substructures using component mode synthesis (CMS) \cite{Guy2014}. However, the effective mass/inertia approach has limitations as it loses information about the appendage beyond the attachment point, such as the deflection at the free tip. Therefore, it cannot accurately model arbitrary chain-like structures with flexible bodies.

The main contribution of this paper is to propose a methodology for modeling the six degrees of freedom (DOF) linear dynamic behavior of a flexible rotating Euler-Bernoulli beam within a multibody system, which includes the analysis of vibration in rotating uniform cantilever beams with a central hub and tip mass. The proposed model is built using the Two-Input-Two-Output Port (TITOP) approach \cite{alazard}, which considers forces and accelerations at the connection points as inputs and outputs. Unlike traditional methods, this approach is not reliant on specific boundary conditions at the connection points of the link. By integrating direct and inverse dynamic models in a concise state-space representation, the TITOP model ensures invertible input-output channels. This allows for seamless integration into the overall multibody system model, treating it as a block-diagram model. The boundary conditions at the connection points are treated as external feedback loops between forces and accelerations. This approach also offers the possibility to model complex multibody mechanical systems, while keeping the uncertain nature of the plant and condensing all the possible mechanical configurations in a single Linear Fractional Representation (LFR). The TITOP model effectively bridges the gap between the transfer matrix method and the effective mass-inertia method. This approach has been previously introduced in \cite{alazard} and has found practical applications in space engineering, as demonstrated in \cite{MURALI, Perez2015, Perez2016, SANFEDINO2018128, RODRIGUES2022107865, rodrigues}. The beam model is derived using the Lagrange technique combined with a finite element model which considers fifth-order polynomial shape functions \cite{CHEBBI}. The obtained LFR model is characterized by its parameterization with respect to the angular velocity, resulting in a Linear Parameter-Varying (LPV) system that accommodates the centrifugal stiffening effect.

The models constructed using the TITOP approach are well-suited for robust control synthesis, as well as robust performance assessment. Additionally, all the models developed with the TITOP approach have been integrated into the most recent release of the Satellite Dynamics Toolbox library (SDTlib) \cite{userguide}.

The paper also includes a case study involving a spinning spacecraft, where the spacecraft's design is inspired by the THOR and Cluster missions. The main equipment platform is located around a central rigid body. Two flexible booms, resembling long rods, are connected to the rigid hub of the spacecraft. These booms become active when the spacecraft starts spinning. Their purpose can be to measure the varying electrical and magnetic fields surrounding the spacecraft (Cluster mission) or to carry plasma measurement instruments at their tips (THOR project). In this paper, these measurement instruments will be treated as tip masses.

\subsection{Contributions and paper organization}

The paper introduces the following key contributions:

\begin{itemize}

\item the development of a six DOF analytical two-port model of a spinning, flexible and uniform Euler-Bernoulli beam, fully capturing the bending, traction and torsion dynamics in a single LFR. Furthermore, this LPV model is parameterized according to the angular velocity of the beam, which facilitates robust control design.

\item the validation of the TITOP six DOF beam model, by comparing the natural frequencies with a commercial finite element software. This analysis pretends to fill a gap in the literature since the authors did not find any results regarding the natural frequencies of a flexible, spinning, cantilever and uniform six DOF beam for the different conditions presented in this paper.

\item the illustration in tables of the frequency ratios associated with the beam's different dynamics.

\item the multibody modeling and analysis of a spinning satellite mission scenario composed of one main central body, two flexible booms and two tip masses.

\end{itemize}

This paper is organized into five parts. In the first part (section \ref{modeling}), an introduction to the TITOP approach is provided and the scope of the proposed model is discussed, together with some generalities. In the second part (section \ref{rigibody}), the spinning rigid body dynamics are derived. In the third part of the paper (section \ref{beam}), the analytical two-port model of a spinning, flexible and uniform beam taking into account a hub radius and a tip mass is computed. Finally, the fourth part (section \ref{results}) showcases a model validation and thorough analysis of all the obtained results, while the fifth part (section \ref{casestudy}) details the rigorous modeling procedure of a spinning spacecraft mission scenario.

\section{Scope of the proposed model and generalities}
\label{modeling}

\begin{figure}[!ht]
	\centering
	\includegraphics[width=0.8\textwidth]{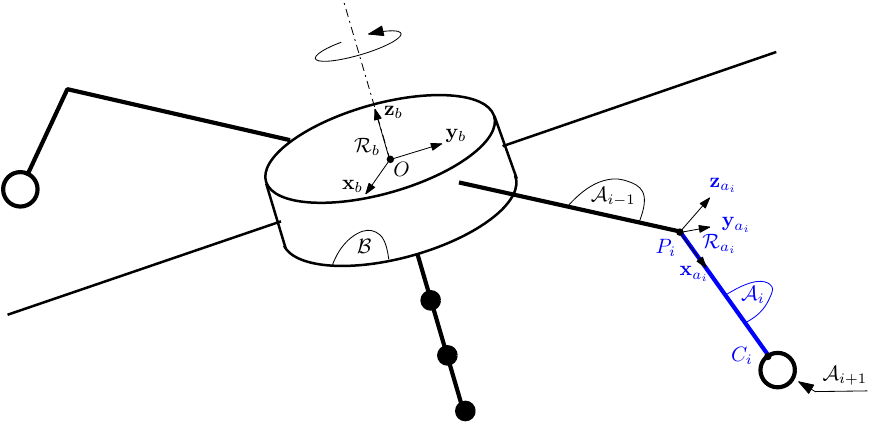}
	\caption{A spinning satellite with various rigid and/or flexible appendages.}
	\label{generic} 
\end{figure}

One of the objectives of this paper is to develop a generic multibody system modeling tool dedicated to the dynamics of spinning spacecraft composed of several rigid and flexible appendages. A general sketch of such a satellite, inspired by previous space missions like Cluster and THOR, is depicted in Fig. \ref{generic}. This spacecraft comprises the main body $\mathcal{B}$ and various links $\mathcal{A}_i$. Each body $\mathcal{A}_i$ is connected to a parent substructure $\mathcal{A}_{i-1}$ at the point $P_i$ and to a child substructure $\mathcal{A}_{i+1}$ at the point $C_i$. Without loss of generality, it is assumed that the spin axis of the main body $\mathcal{B}$ is along the $\mathbf{z}_b$-axis of its body frame $\mathcal{R}_b$.

\subsection{Assumptions}

The following assumptions are taken into account throughout this work:

\begin{itemize}
 \item the proposed model considers any rigid appendages or substructures but flexible appendages or substructures are restricted to Euler-Bernoulli beams.
\item only the linear behavior of the spacecraft is addressed; i.e., only small motions and deformations around an equilibrium are considered.
\item the considered equilibrium is stable. This implies that the spacecraft is symmetric around its spin axis $\mathbf{z}_b$ and that this axis does not correspond to the intermediate principal moment of inertia of the satellite, in order to avoid the \textsc{Dzhanibekov} effect \cite{Murakami2016}.
\item the static deformations of flexible beams due to centrifugal loads are small and neglected in the computation of the equilibrium conditions. Note that these static deformations can be evaluated to verify this assumption, as proposed in section \ref{beam}.
\item zero-gravity environment.
\end{itemize}

Under these assumptions, it is possible to compute for each body $\mathcal{A}_i$ the equilibrium conditions which are characterized by:

\begin{itemize}

\item $\overline{\mathbf{v}}_{P_i}^{\mathcal{A}_i}$ and $\overline{\boldsymbol{\omega}}_{P_i}^{\mathcal{A}_i}$ : the linear and angular velocity vectors of the body $\mathcal{A}_i$ at the point $P_i$.
\item $\overline{\mathbf{W}}_{\mathcal{A}_{i+1} / \mathcal{A}_i, C_i}=\left[\overline{\mathbf{F}}_{\mathcal{A}_{i+1} / \mathcal{A}_i, C_i}^{\mathrm{T}}, \overline{\mathbf{T}}_{\mathcal{A}_{i+1} / \mathcal{A}_i, C_i}^{\mathrm{T}}\right]^{\mathrm{T}}$: the wrench applied by the child substructure $\mathcal{A}_{i+1}$ to the current body $\mathcal{A}_i$ at the connection point $C_i$ due to centrifugal loads.
\end{itemize}

\subsection{TITOP approach with motion vectors}

The TITOP approach that has been previously presented in \cite{alazard, MURALI, Perez2015, Perez2016, SANFEDINO2018128, RODRIGUES2022107865, rodrigues} links the 6 components of the interaction wrenches to the 6 components of the inertial acceleration dual vectors at the points $P_i$ and $C_i$, assuming small motions around null equilibrium conditions. To linearize the model of a spinning satellite around non-null kinematic equilibrium conditions $(\overline{\mathbf{v}}_{P_i}^{\mathcal{A}_i}$ and $\overline{\boldsymbol{\omega}}_{P_i}^{\mathcal{A}_i})$, the TITOP model developed hereafter considers the 18-components motion vectors $\mathbf{m}_{P_i}^{\mathcal{A}_i}$ and $\mathbf{m}_{C_i}^{\mathcal{A}_i}$, as defined in the nomenclature, as well as their corresponding equilibrium values and variations. For example, it follows that:

\begin{equation}
\mathbf{m}_{P_i}^{\mathcal{A}_i}=\underbrace{\left[\mathbf{0}_{1 \times 6}, \mathbf{0}_{1 \times 6}, \overline{\mathbf{v}}_{P_i}^{\mathcal{A}_i^\mathrm{T}}, \overline{\boldsymbol{\omega}}_{P_i}^{\mathcal{A}_i^\mathrm{T}}, \overline{\mathbf{x}}_{P_i}^{\mathcal{A}_i^\mathrm{T}}, \overline{\boldsymbol{\Theta}}_{P_i}^{\mathcal{A}_i^ \mathrm{T}}\right]^{\mathrm{T}}}_{\overline{\mathbf{m}}_{P_i}^{\mathcal{A}_i}}+\underbrace{\left[\delta \dot{\mathbf{v}}_{P_i}^{\mathcal{A}_i^\mathrm{T}}, \delta \dot{\boldsymbol{\omega}}_{P_i}^{\mathcal{A}_i ^\mathrm{T}}, \delta \mathbf{v}_{P_i}^{\mathcal{A}_i^ \mathrm{T}}, \delta \boldsymbol{\omega}_{P_i}^{\mathcal{A}_i ^\mathrm{T}}, \delta \mathbf{x}_{P_i}^{\mathcal{A}_i ^\mathrm{T}}, \delta \boldsymbol{\Theta}_{P_i}^{\mathcal{A}_i^ \mathrm{T}}\right]^{\mathrm{T}}}_{\delta \mathbf{m}_{P_i}^{\mathcal{A}_i}}
\label{eq:matPbar}
\end{equation}

\begin{figure}[!ht]
	\centering
	\includegraphics[width=1\textwidth]{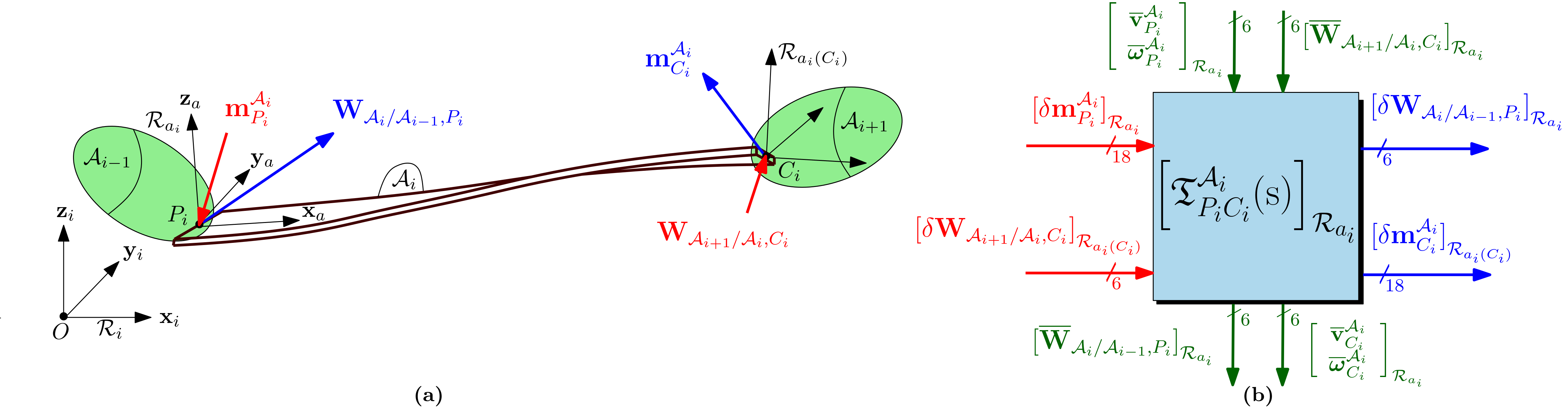}
	\caption{(a) Three-Dimensional (3D) representation of a flexible spinning beam $\mathcal{A}_i$. (b) TITOP model ${\left[\mathfrak{T}_{P_i C_i}^{\mathcal{A}_i}(\mathrm{s})\right]}_{\mathcal{R}_{a_i}}$ block-diagram.}
	\label{test} 
\end{figure}

Similarly, since the wrench $\overline{\mathbf{W}}_{\mathcal{A}_{i+1} / \mathcal{A}_i, C_i}$ applied at the point $C_i$ at the equilibrium is no longer null, the linear TITOP model must now consider the deformed body frame $\mathcal{R}_{a_i\left(C_i\right)}$ attached to the body $\mathcal{A}_i$ at the point $C_i$, in addition to the "rigid" body frame $\mathcal{R}_{a_i}$ attached to the body $\mathcal{A}_i$ at the point $P_i$, as depicted in Fig. \ref{test}a.

Therefore, the double-port or TITOP model $\left[\mathfrak{T}_{P_{i} C_{i}}^{\mathcal{A}_{i}}(\mathrm s)\right]_{\mathcal{R}_{a_i}}$ of the body $\mathcal{A}_i$ is a linear dynamic model between 24 inputs:
\begin{itemize}
  
 \item the six components of the wrench variation $\left[\delta \mathbf{W}_{\mathcal{A}_{i+1} / \mathcal{A}_i, C}\right]_{\mathcal{R}_{a_i\left(C_i\right)}}$ applied by the child substructure $\mathcal{A}_{i+1}$ to the body $\mathcal{A}_i$ at the point $C_i$, expressed in the deformed body frame $\mathcal{R}_{a_i\left(C_i\right)}$.
 \item  the eighteen components of the motion vector variation $\left[\delta \mathbf{m}_{P_i}^{\mathcal{A}_i}\right]_{\mathcal{R}_{a_i}}$ defined at the point $P_i$ and expressed in the rigid body frame $\mathcal{R}_{a_i}$.

\end{itemize}
and 24 outputs:

\begin{itemize}
\item the eighteen components of the motion vector variation $\left[\delta \mathbf{m}_{C_i}^{\mathcal{A}_i}\right]_{\mathcal{R}_{a_i\left(C_i\right)}}$ defined at the point $C_i$ and expressed in the deformed body frame $\mathcal{R}_{a_i\left(C_i\right)}$.
\item the six components of the wrench variation $\left[\delta \mathbf{W}_{\mathcal{A}_i / \mathcal{A}_{i-1}, P_i}\right]_{\mathcal{R}_{a_i}}$ that is applied by the body $\mathcal{A}_i$ to the parent substructure $\mathcal{A}_{i-1}$ at the point $P_i$, expressed in the rigid body frame $\mathcal{R}_{a_i}$.

\end{itemize}

and can be represented by the block-diagram depicted in Fig. \ref{test}b. The TITOP model is composed of the direct dynamic model (transfer from motion vector variation to wrench variation) at the point $P_i$ and the inverse dynamic model (transfer from wrench variation to motion vector variation) at the point $C_i$. This model depends on the kinematic equilibrium conditions $\left[\overline{\mathbf{v}}_{P_i}^{\mathcal{A}_i^ \mathrm{T}}, \overline{\boldsymbol{\omega}}_{P_i}^{\mathcal{A}_i ^\mathrm{T}}\right]_{\mathcal{R}_{a_i}}^{\mathrm{T}}$ at the point $P_i$ and the equilibrium wrench at the point $C_i\left[\overline{\mathbf{W}}_{\mathcal{A}_{i+1} / \mathcal{A}_i, C_i}\right]_{\mathcal{R}_{a_i\left(C_i\right)}}=$ $\left[\overline{\mathbf{W}}_{\mathcal{A}_{i+1} / \mathcal{A}_i, C_i}\right]_{\mathcal{R}_{a_i}}$. Indeed, at the equilibrium and under the adopted assumptions, the frames $\mathcal{R}_{a_i}$ and $\mathcal{R}_{a_i\left(C_i\right)}$ are aligned. It should also be noted that the equilibrium conditions can be propagated forward (from parent to child) to provide $\left[\overline{\mathbf{v}}_{C_i}^{\mathcal{A}_i ^\mathrm{T}}, \overline{\boldsymbol{\omega}}_{C_i}^{\mathcal{A}_i ^\mathrm{T}}\right]_{\mathcal{R}_{a_i}}^{\mathrm{T}}$ and backwards (from child to parent) to provide $\left[\overline{\mathbf{W}}_{\mathcal{A}_i / \mathcal{A}_{i-1}, P_i}\right]_{\mathcal{R}_{a_i}}$.

Moreover, Eq. \eqref{eq:matPbar} also introduced the geometric equilibrium conditions $\left[\overline{\mathbf{x}}_{P_i}^{\mathcal{A}_i{ }^{\mathrm{T}}}, \overline{\mathbf{\Theta}}_{P_i}^{\mathcal{A}_i ^\mathrm{T}}\right]_{\mathcal{R}_{a_i}}^{\mathrm{T}}$. However, the dynamics of the various TITOP models developed hereafter do not depend on these geometric conditions. For this reason, they do not appear on the block-diagram depicted in Fig. \ref{test}b.

Ultimately, the subscript $i$ will be ommitted hereafter and the substructures (parent or child) connected to the body in question will be denoted as $\bullet$ for brevity. Similarly, the equilibrium signals (represented in green in Fig. \ref{test}b) will not be displayed in the block-diagrams presented henceforth.

\subsection{\textsc{Lagrange} equations in terms of quasi-coordinates}

The TITOP models derived in the following sections are based on the \textsc{Lagrange} equations in terms of quasi-coordinates \cite{Meirovitch1991HybridSE}, which are rewritten here in the non-dissipative case:

\begin{equation}
\displaystyle\frac{d}{d t} \frac{\partial \mathcal{L}}{\partial\left[\begin{array}{c}
\delta \mathbf{v}_P^{\mathcal{A}} \\
\delta \boldsymbol{\omega}_P^{\mathcal{A}}
\end{array}\right]_{\mathcal{R}_a}}+\left[\begin{array}{cc}
({ }^* \boldsymbol{\omega}_P^{\mathcal{A}}) & \mathbf{0}_{3 \times 3} \\
({ }^* \mathbf{v}_P^{\mathcal{A}}) & ({ }^* \boldsymbol{\omega}_P^{\mathcal{A}})
\end{array}\right]_{\mathcal{R}_a} \frac{\partial \mathcal{L}}{\partial\left[\begin{array}{c}
\delta \mathbf{v}_P^{\mathcal{A}} \\
\delta \boldsymbol{\omega}_P^{\mathcal{A}}
\end{array}\right]_{\mathcal{R}_a}}-\left[\begin{array}{cc}
\mathbf{0}_{3 \times 3} & \mathbf{0}_{3 \times 3} \\
\mathbf{0}_{3 \times 3} & \boldsymbol{\Gamma}^{-\mathrm{T}}\left(\boldsymbol{\Theta}_P^{\mathcal{A}}\right)
\end{array}\right] \frac{\partial \mathcal{L}}{\partial\left[\begin{array}{c}
{\left[\delta \mathbf{x}_P^{\mathcal{A}}\right]_{\mathcal{R}_a}} \\
\delta \mathbf{\Theta}_P^{\mathcal{A}}
\end{array}\right]} =\left[\begin{array}{c}
\mathbf{F}_{\mathrm{ext} / \mathcal{A}, P} \\
\mathbf{T}_{\mathrm{ext} / \mathcal{A}, P}
\end{array}\right]_{\mathcal{R}_a}
\label{eq:lagrange1}
\end{equation}

\begin{equation}
    \displaystyle\frac{d}{d t} \frac{\partial \mathcal{L}}{\partial \delta\dot{\mathbf{q}}_f}-\frac{\partial \mathcal{L}}{\partial \delta{\mathbf{q}}_f}=\widehat{\mathbf{U}}
    \label{eq:lagrange2}
\end{equation}

In Eq. \eqref{eq:lagrange1}, $\mathcal{L}=\mathcal{T}-\mathcal{V}$ is the Lagrangian, where $\mathcal{T}$ and $\mathcal{V}$ are the kinetic and potential energies, respectively. Furthermore, $\boldsymbol{\Gamma}\left(\boldsymbol{\Theta}_P^{\mathcal{A}}\right)$ represents the mapping between the angular velocity vector and the time derivative of the \textsc{Euler} angles vector \cite{Schaub2018}, with $\left[\boldsymbol{\omega}_P^{\mathcal{A}}\right]_{\mathcal{R}_a}=\boldsymbol{\Gamma}\left(\boldsymbol{\Theta}_P^{\mathcal{A}}\right) \dot{\boldsymbol{\Theta}}_P^{\mathcal{A}}$. Eq. \eqref{eq:lagrange1} is related to the 6 rigid DOF (3 translations and 3 rotations) of the body, while Eq. \eqref{eq:lagrange2} is related to the internal elastic deformations. The right-hand sides of Eqs. \eqref{eq:lagrange1} and \eqref{eq:lagrange2} are composed of the resultant wrench $\left[\mathbf{W}_{\mathrm{ext} / \mathcal{A}, P}\right]_{\mathcal{R}_a}=\left[\mathbf{F}_{\mathrm{ext} / \mathcal{A}, P}^{\mathrm{T}}, \mathbf{T}_{\mathrm{ext} / \mathcal{A}, P}^{\mathrm{T}}\right]_{\mathcal{R}_a}^{\mathrm{T}}$ applied to the body $\mathcal{A}$ at the point $P$ and the nonconservative generalized forces $\widehat{\mathbf{U}}$ associated with the elastic motions. Finally, $\mathbf{q}_f$ are distributed coordinates describing elastic motions relative to the rigid body motions.

\section{Analytical two-port model of a spinning rigid body}
\label{rigibody}

\subsection{Case of a rigid appendage}

For a rigid body $\mathcal{A}$ that is free to move in space, the frames $\mathcal{R}_a$ and $\mathcal{R}_{a(C)}$ are always aligned. Besides that, gravity is not taken into account, meaning that:

\begin{equation}
\mathcal{L}=\mathcal{T}=\frac{1}{2}\left[\begin{array}{c}
\mathbf{v}_P^{\mathcal{A}} \\
\boldsymbol{\omega}_P^{\mathcal{A}}
\end{array}\right]^{\mathrm{T}} \mathbf{D}_P^{\mathcal{A}}\left[\begin{array}{c}
\mathbf{v}_P^{\mathcal{A}} \\
\boldsymbol{\omega}_P^{\mathcal{A}}
\end{array}\right]
\label{lagrigid}
\end{equation}

It should be noted that all the quantities in Eq. \eqref{lagrigid} are projected in the body frame $\mathcal{R}_a$ (omitted for brevity). $\mathbf{D}_P^{\mathcal{A}}$ is the $6 \times 6$ direct rigid dynamic model of the body $\mathcal{A}$ at the point $P$, which can be computed from the dynamic model $\mathbf{D}_A^{\mathcal{A}}$ written at the body's center of mass $A$ and the kinematic model $\boldsymbol{\tau}_{A P}$ (see nomenclature):

\begin{equation}
\mathbf{D}_P^{\mathcal{A}}=\boldsymbol{\tau}_{A P}^{\mathrm{T}} \underbrace{\left[\begin{array}{cc}
m^{\mathcal{A}} \mathbf{I}_3 & \mathbf{0}_{3 \times 3} \\
\mathbf{0}_{3 \times 3} & \mathbf{J}_A^{\mathcal{A}}
\end{array}\right]}_{\mathbf{D}_A^{\mathcal{A}}} \boldsymbol{\tau}_{A P}=\left[\begin{array}{cc}
m^{\mathcal{A}} \mathbf{I}_3 & m^{\mathcal{A}}({ }^* \mathbf{A} \mathbf{P}) \\
-m^{\mathcal{A}}({ }^* \mathbf{A} \mathbf{P}) & \underbrace{\mathbf{J}_A^{\mathcal{A}}-m^{\mathcal{A}}({ }^* \mathbf{A} \mathbf{P})^2}_{\mathbf{J}_P^{\mathcal{A}}}
\end{array}\right]
\end{equation}

Considering that the body $\mathcal{A}$ is only subjected to the interaction wrenches at the points $P$ and $C$, the direct application of Eq. \eqref{eq:lagrange1} leads to the following non-linear model:

\begin{equation}
\mathbf{D}_P^{\mathcal{A}}\left[\begin{array}{c}
\dot{\mathbf{v}}_P^{\mathcal{A}} \\
\dot{\boldsymbol{\omega}}_P^{\mathcal{A}}
\end{array}\right]+\left[\begin{array}{cc}
({ }^* \boldsymbol{\omega}_P^{\mathcal{A}}) & \mathbf{0}_{3 \times 3} \\
({ }^* \mathbf{v}_P^{\mathcal{A}}) & ({ }^* \boldsymbol{\omega}_P^{\mathcal{A}})
\end{array}\right] \mathbf{D}_P^{\mathcal{A}}\left[\begin{array}{c}
\mathbf{v}_P^{\mathcal{A}} \\
\boldsymbol{\omega}_P^{\mathcal{A}}
\end{array}\right]=-\mathbf{W}_{\mathcal{A} / \bullet, P}+\boldsymbol{\tau}_{C P}^{\mathrm{T}} \mathbf{W}_{\bullet / \mathcal{A} ., C}
\end{equation}

Given the equilibrium conditions $\overline{\mathbf{v}}_P^{\mathcal{A}}, \overline{\boldsymbol{\omega}}_P^{\mathcal{A}}$ and $\overline{\mathbf{W}}_{\bullet / \mathcal{A}, C}$, one can compute the equilibrium wrench at the point $\mathrm{P}$:

\begin{equation}
\overline{\mathbf{W}}_{\mathcal{A} / \bullet, P}={\boldsymbol{\tau}}_{C P}^{\mathrm{T}} \overline{\mathbf{W}}_{\bullet / \mathcal{A},, C}-\left[\begin{array}{cc}
({ }^* \overline{\boldsymbol{\omega}}_P^{\mathcal{A}}) & \mathbf{0}_{3 \times 3} \\
({ }^* \overline{\mathbf{v}}_P^{\mathcal{A}}) & ({ }^* \overline{\boldsymbol{\omega}}_P^{\mathcal{A}})
\end{array}\right] \mathbf{D}_P^{\mathcal{A}}\left[\begin{array}{c}
\overline{\mathbf{v}}_P^{\mathcal{A}} \\
\overline{\boldsymbol{\omega}}_P^{\mathcal{A}}
\end{array}\right]
\end{equation}

After performing some algebraic operations based on the properties of the skew symmetric matrix, the first-order linearized model becomes:

\begin{equation}
\begin{aligned}
& \mathbf{D}_{{P}}^{\mathcal{A}}{\left[\begin{array}{c}
\delta\dot{\mathbf{v}}^{\mathcal{A}}_{P} \\
\delta\dot{\boldsymbol{\omega}}^{\mathcal{A}}_P
\end{array}\right]}+\underbrace{\left[\begin{array}{cc}
m^{\mathcal A} ({ }^{*} \overline{\boldsymbol{\omega}}_P^{\mathcal{A}})&  m^{\mathcal A} \left( 2({ }^{*} \overline{\boldsymbol{\omega}}_P^{\mathcal{A}})(^*\mathbf{AP})-(^*\mathbf{AP})({ }^{*} \overline{\boldsymbol{\omega}}_P^{\mathcal{A}})-  	({ }^{*} \overline{\mathbf{v}}^{\mathcal{A}}_{{P}})  \right)  \\
- m^{\mathcal A}(^*\mathbf{AP})({ }^{*} \overline{\boldsymbol{\omega}}^{\mathcal{A}}_{{P}}) &  m^{\mathcal A}(^*\mathbf{AP})({ }^{*} \overline{\mathbf{v}}^{\mathcal{A}}_{{P}})+({ }^{*} \overline{\boldsymbol{\omega}}_P^{\mathcal{A}})\textbf J^{\mathcal{A}}_P -({ }^{*} (\textbf J^{\mathcal{A}}_P\overline{\boldsymbol{\omega}}_P^{\mathcal{A}}))
\end{array}\right]}_{\mathbf{X}_{{P}}^{\mathcal{A}}(\overline{\mathbf{v}}^{\mathcal{A}}_{{P}},\overline{\boldsymbol{\omega}}^{\mathcal{A}}_{{P}})}{\left[\begin{array}{c}
\delta\mathbf{v}^{\mathcal{A}}_{{P}} \\
\delta\boldsymbol{\omega}_P^{\mathcal{A}}
\end{array}\right]}=\\
& -\delta \mathbf{W}_{\mathcal{A} / \bullet, P}+\boldsymbol{\tau}_{C P}^{\mathrm{T}} \delta \mathbf{W}_{\bullet / \mathcal{A} ., C}
\end{aligned}
\end{equation}

In the context of a rigid body, the propagation of the variation of the motion vector from the point $P$ to the point $C$ simply reads:

\begin{equation}
\delta\mathbf{m}_C^{\mathcal{A}}=\underbrace{\operatorname{diag}\left(\boldsymbol{\tau}_{C P}, \boldsymbol{\tau}_{C P}, \mathbf{I}_6\right)}_{\boldsymbol{v}_{C P}} \delta \mathbf{m}_P^{\mathcal{A}}
\end{equation}

\begin{figure}[!ht]
	\centering
	\includegraphics[width=0.8\textwidth]{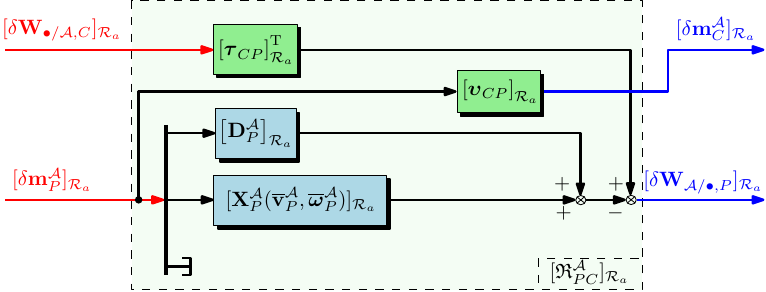}
	\caption{Block-diagram representation of $[\mathfrak{R}_{PC}^{\mathcal{A}}]_{\mathcal{R}_a}$: the linear $24\times 24$ model
		of a rigid body $\mathcal{A}$ computed at the points $P$ and $C$.}
	\label{modelLINEARtwopointone} 
\end{figure}

The block-diagram representation of the TITOP model $\mathfrak{R}_{P C}^{\mathcal{A}}$ of the rigid body $\mathcal{A}$ computed at the points $P$ and $C$ is summarized in Fig. \ref{modelLINEARtwopointone}. This linear model is static and does not depend on the equilibrium wrench $\overline{\mathbf{W}}_{\bullet / \mathcal{A}, C}$, which will no longer be the case for the TITOP model of a spinning flexible beam, as the one presented in section \ref{beam}. Furthermore, one can also define the one-port model of the rigid body $\mathcal{A}$ at the point $P$ in case there is no child substructure connected to it, with $\mathfrak{R}_{P}^{\mathcal{A}}=\mathfrak{R}_{PC(19:24,7:24)}^{\mathcal{A}}$.

\subsection{Case of the rigid main body}

For the particular case of the main body $\mathcal{B}$ represented in Fig. \ref{generic}, one has to consider the inverse dynamic model (transfer from wrench variation to the motion vector variation) written at the body's center of mass $B$ and including the 6 rigid modes of the whole system. Thus, the model of the main body $\mathcal{B}$ subjected to an external wrench and connected to several appendages $\mathcal{A}_i$ at the points $C_i$ reads:

\begin{equation}\label{eq:Z1}
\mathbf{D}_{{B}}^{\mathcal{B}}{\left[\begin{array}{c}
\delta\dot{\mathbf{v}}^{\mathcal{B}}_{B} \\
\delta\dot{\boldsymbol{\omega}}^{\mathcal{B}}_B
\end{array}\right]}+\mathbf{X}_{{B}}^{\mathcal{B}}(\overline{\mathbf{v}}^{\mathcal{B}}_{{B}},\overline{\boldsymbol{\omega}}^{\mathcal{B}}_{{B}}){\left[\begin{array}{c}
\delta\mathbf{v}^{\mathcal{B}}_{{B}} \\
\delta\boldsymbol{\omega}_B^{\mathcal{B}}
\end{array}\right]}
= \delta\mathbf W_{\mathrm{ext}/\mathcal B,B}+ \Sigma_i \boldsymbol{\tau}_{C_iB}^{\mathrm T} \delta\mathbf W_{\mathcal A_i/\mathcal  B,C_i}
\end{equation}

This model must then be completed with the following kinematic equation:
\begin{equation}
\left[\begin{array}{c}
\dot{\mathbf{x}}_B^{\mathcal{B}} \\
\dot{\mathbf{\Theta}}_B^{\mathcal{B}}
\end{array}\right]=\left[\begin{array}{c}
\left.\frac{d \mathbf{x}_B^{\mathcal{B}}}{d t}\right|_{\mathcal{R}_b} \\
\dot{\mathbf{\Theta}}_B^{\mathcal{B}}
\end{array}\right]=\left[\begin{array}{cc}
\mathbf{I}_3 & \left({ }^* \mathbf{x}_B^{\mathcal{B}}\right) \\
\mathbf{0}_{3 \times 3} & \boldsymbol{\Gamma}^{-1}\left(\mathbf{\Theta}_B^{\mathcal{B}}\right)
\end{array}\right]\left[\begin{array}{c}
\mathbf{v}_B^{\mathcal{B}} \\
\boldsymbol{\omega}_B^{\mathcal{B}}
\end{array}\right]
\end{equation}

Without loss of generality, this equation is linearized around an equilibrium that is defined by: $\overline{\mathbf{x}}_B^{\mathcal{B}}=\mathbf{0}_{3 \times 1}$, $\overline{\boldsymbol{\omega}}_B^{\mathcal{B}}=$ $\left[0,0, \Omega\right]^{\mathrm{T}}$ (the $\mathbf{z}_b$-axis is the spin axis) and $\overline{\boldsymbol{\Theta}}_B^{\mathcal{B}}=\left[0,0, \Omega t\right]^{\mathrm{T}}$, where the variable $t$ represents the time. Then, considering the chosen 'ZYX' sequence of rotations for the \textsc{Euler} angles vector $\boldsymbol{\Theta}_B^{\mathcal{B}}=\left[\delta \phi_B^{\mathcal{B}}, \delta \theta_B^{\mathcal{B}}, \psi_B^{\mathcal{B}}\right]^{\mathrm{T}}$, one can verify that $\mathbf{\Gamma}\left(\boldsymbol{\Theta}_B^{\mathcal{B}}\right)$ does not depend on $\psi_B^{\mathcal{B}}$, the spin angle, and write:

\begin{equation}
\boldsymbol{\Gamma}\left(\boldsymbol{\Theta}_B^{\mathcal{B}}\right)=\left[\def\arraystretch{1.2}\begin{array}{ccc}
1 & 0 & -\sin \delta \theta_B^{\mathcal{B}} \\
0 & \cos \delta \phi_B^{\mathcal{B}} & \sin \delta \phi_B^{\mathcal{B}} \cos \delta \theta_B^{\mathcal{B}} \\
0 & -\sin \delta \phi_B^{\mathcal{B}} & \cos \delta \phi_B^{\mathcal{B}} \cos \delta \theta_B^{\mathcal{B}}
\end{array}\right] \approx\left[\def\arraystretch{1.2}\begin{array}{ccc}
1 & 0 & -\delta \theta_B^{\mathcal{B}} \\
0 & 1 & \delta \phi_B^{\mathcal{B}} \\
0 & -\delta \phi_B^{\mathcal{B}} & 1
\end{array}\right] 
\end{equation}

Consequently, the first-order linear kinematic model reads:

\begin{equation}\label{eq:Z2}
\left[\begin{array}{c}
{\delta\dot{\mathbf{x}}}^{\mathcal B}_B \\
{\delta\dot{\boldsymbol{\Theta}}}_B^{\mathcal{B}}
\end{array}\right]=\left[\begin{array}{c}
\delta\mathbf v_B^{\mathcal B} \\
\delta\boldsymbol \omega_B^{\mathcal B}  
\end{array}\right]-\underbrace{\left[\begin{array}{cc}
({ }^{*} \overline{\boldsymbol{\omega}}^{\mathcal{B}}_{{B}} )  & \mathbf{0}_{3 \times 3}\\
\mathbf{0}_{3 \times 3} & ({ }^{*} \overline{\boldsymbol{\omega}}^{\mathcal{B}}_{{B}} )
\end{array}\right]}_{\mathbf H(\overline{\boldsymbol{\omega}}^{\mathcal{B}}_{{B}})}\left[\begin{array}{c}
\delta\mathbf{x}^{\mathcal B}_B \\
\delta\boldsymbol{\Theta}_B^{\mathcal{B}}
\end{array}\right]
\end{equation}

From Eqs. \ref{eq:Z1} and \ref{eq:Z2}, the block-diagram representation of the twelfth-order model $\mathfrak{X}_B^{\mathcal{B}}(\mathrm{s})$ of the main body $\mathcal{B}$ written at its center of mass $B$ can be obtained, as depicted in Fig. \ref{modelLINEARtwopointthree_v2}.

Finally, in order to assemble linear TITOP models between each other and/or to connect them to the main body, the DCM $\mathbf{P}_{\mathcal{R}_a / \mathcal{R}_b}$ between the frame $\mathcal{R}_a$ attached to the body $\mathcal{A}$ and the frame $\mathcal{R}_b$ attached to the body $\mathcal{B}$ must be taken into account in the propagation of the wrench and motion vector variations. Let $P$ be the point where $\mathcal{A}$ is connected to $\mathcal{B}$. Then, it follows that:

\begin{equation}
\left[\delta \mathbf{W}_{\mathcal{A} / \mathcal{B}, P}\right]_{\mathcal{R}_b}=\underbrace{\operatorname{diag}\left(\mathbf{P}_{\mathcal{R}_a / \mathcal{R}_b}, \mathbf{P}_{\mathcal{R}_a / \mathcal{R}_b}\right)}_{\mathbf{P}_{\mathcal{R}_a / \mathcal{R}_b}^{\times 2}}\left[\delta \mathbf{W}_{\mathcal{A} / \mathcal{B}, P}\right]_{\mathcal{R}_a} \quad \text{and} \quad \left[\delta \mathbf{m}_P^{\mathcal{B}}\right]_{\mathcal{R}_b}=\underbrace{\operatorname{diag}\left(\mathbf{P}_{\mathcal{R}_a / \mathcal{R}_b}^{\times 2}, \mathbf{P}_{\mathcal{R}_a / \mathcal{R}_b}^{\times 2}, \mathbf{P}_{\mathcal{R}_a / \mathcal{R}_b}^{\times 2}\right)}_{\mathbf{P}_{\mathcal{R}_a / \mathcal{R}_b}^{\times 6}}\left[\delta \mathbf{m}_P^{\mathcal{A}}\right]_{\mathcal{R}_a}
\end{equation}

Indeed, the DCM can also be applied to the variation of the \textsc{Euler} angles vector, since $\operatorname{DCM}(\delta \boldsymbol{\Theta}) \approx \mathbf{I}_3+\left({ }^* \delta \boldsymbol{\Theta}\right)$. Then, the variation $\delta \mathbf{\Theta}_P^{\mathcal{B}}$ on the attitude of the body $\mathcal{B}$ or the variation $\delta \mathbf{\Theta}_P^{\mathcal{A}}$ on the attitude of the body $\mathcal{A}$ must have the same effect on the whole DCM between $\mathcal{R}_a$ and $\mathcal{R}_i$:

\begin{figure}[!ht]
	\centering
	\includegraphics[width=1\textwidth]{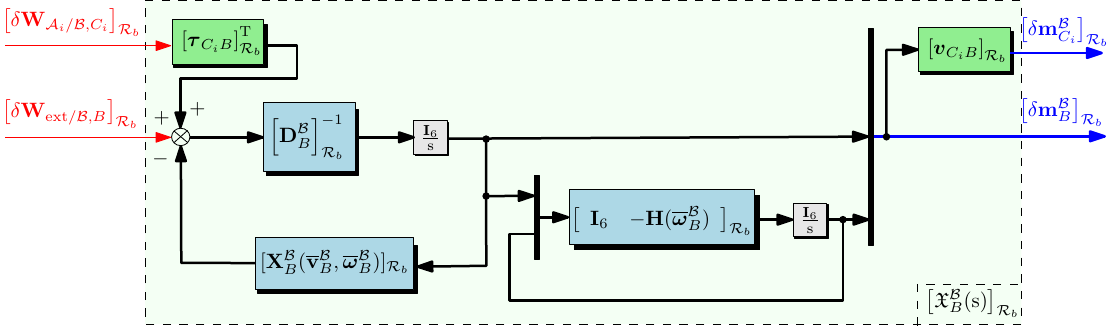}
	\caption{Detailed block-diagram representation of ${\left[\mathfrak{X}_{B}^{\mathcal{B}}(\mathrm s)\right]_{\mathcal{R}_{b}}}$.}
	\label{modelLINEARtwopointthree_v2} 
\end{figure}

\begin{equation}
\begin{aligned}
\left(\mathbf{I}_3+ (^*\delta\boldsymbol\Theta^{\mathcal B}_P)\right)\mathbf P_{\mathcal R_a/\mathcal R_b}=\mathbf P_{\mathcal R_a/\mathcal R_b}	\left(\mathbf{I}_3+ (^*\delta\boldsymbol\Theta^{\mathcal A}_P)\right) \; &  \Leftrightarrow\; (^*\delta\boldsymbol\Theta^{\mathcal B}_P)=\mathbf P_{\mathcal R_a/\mathcal R_b}  (^*\delta\boldsymbol\Theta^{\mathcal A}_P) \mathbf P_{\mathcal R_a/\mathcal R_b}^{\mathrm T}= (^*(\mathbf P_{\mathcal R_a/\mathcal R_b} \delta\boldsymbol\Theta^{\mathcal A}_P)) \\
& \Leftrightarrow \;\delta\boldsymbol\Theta^{\mathcal B}_P=\mathbf P_{\mathcal R_a/\mathcal R_b} \delta\boldsymbol\Theta^{\mathcal A}_P
\end{aligned}
\end{equation}

All the elementary blocks and gains defined in this section ($\mathfrak{X}_{B}^{\mathcal{B}}(\mathrm s)$, $\mathfrak{R}_{P}^{\mathcal{A}}$, $\mathfrak{R}_{PC}^{\mathcal{A}}$, $\mathbf P^{\times 6}_{\mathcal R_a/\mathcal R_b}$) can be used to build the model of the whole rigid system. Since all these blocks are analytically defined, it is possible to take into account uncertainties on the mechanical parameters of the system and also on the spin rate of the spacecraft by using the LFR formalism, as proposed in \cite{Perez2015, RODRIGUES2022107865, rodrigues, userguide}. In the following section, the TITOP modeling approach of spinning bodies  is extended to flexible beams. Then, the way to use these elementary blocks is  presented in section \ref{casestudy} (see Fig. \ref{studycasefullmodel} as an example).

\section{Analytical two-port model of a spinning, flexible and uniform Euler-Bernoulli beam}
\label{beam}

The objective of this section is to compute the TITOP model $\mathfrak{T}_{PC}^{\mathcal{A}}(\mathrm{s})$ of a spinning, flexible and uniform beam $\mathcal{A}$, as the one depicted in Fig. \ref{test}. This model will be first developed in the body frame $\mathcal R_a$.  As a result, in this section, all the quantities are consistently projected in the frame ${\mathcal{R}_a}$, unless stated otherwise. For this reason, this frame is omitted in the notation for the sake of simplicity. Additionally, the deformed body frame $\mathcal R_{a(C)}$ is introduced in section \ref{sect:assembly}. This beam possesses various parameters, such as mass density $\rho^{\mathcal{A}}$ (\si{\kilogram\per\cubic\meter}), cross-sectional area $S^{\mathcal{A}}$ (\si{\square\meter}), length $l^{\mathcal{A}}$ (\si{\meter}), Young's modulus $E^{\mathcal{A}}$ (\si{\newton\per\square\meter}), Poisson's ratio $\nu^{\mathcal{A}}$, shear modulus $G^{\mathcal{A}}$ (\si{\newton\per\square\meter}), with $G^{\mathcal{A}}=\frac{E^{\mathcal{A}}}{2(1+v^{\mathcal{A}})}$, second moment of area $J^{\mathcal{A}}_{y}$ with respect to the $\mathbf{y}_a$-axis, second moment of area $J^{\mathcal{A}}_{z}$ with respect to the $\mathbf{z}_a$-axis and second polar moment of area $J^{\mathcal{A}}_{px}$ with respect to the $\mathbf{x}_a$-axis.


\subsection{Parametrization of the internal elastic deformations}

\begin{figure}[!ht]
\centering
 \includegraphics[width=1\textwidth]{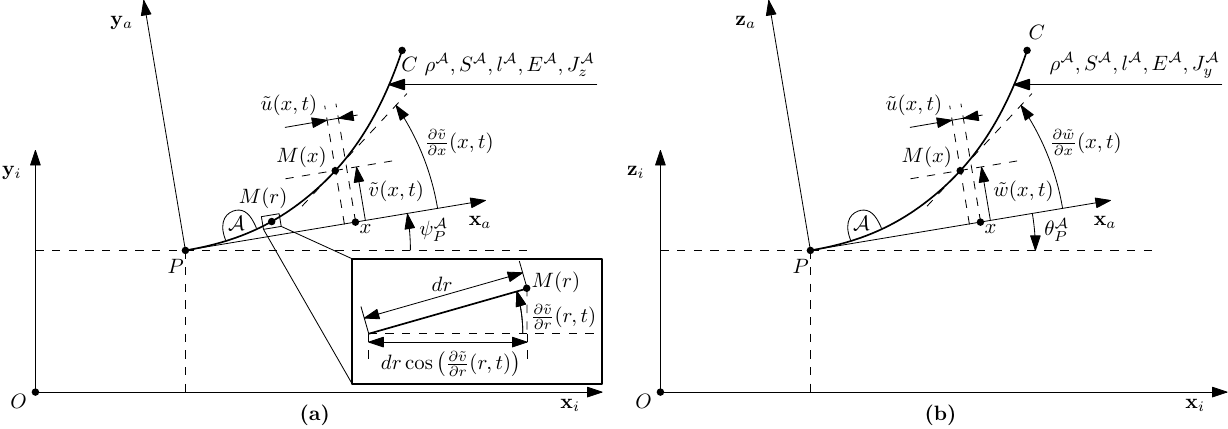}
\caption{(a) Parameterization of the beam deflection at a given time $t$ (Note: the displacement of a specific point $M(x)$ on the deformed flexible beam ${u}(x,t)$ along the $\mathbf{x}_a$-axis is displayed for both cases): (a) bending in the plane $\left(P, \mathbf{x}_a, \mathbf{y}_a\right)$. (b) bending in the plane $\left(P, \mathbf{x}_a, \mathbf{z}_a\right)$.}
\label{beam_v3} 
\end{figure}

The \textsc{Lagrange} technique combined with a finite element model which considers fifth-order polynomial shape functions \cite{CHEBBI} is used for obtaining the dynamic model of a flexible beam. In this approach, a moving body frame ${\mathcal{R}_a}=(P; {\mathbf{x}_a}, {\mathbf{y}_a}, \mathbf{z}_a)$ is attached to the beam at the point $P$, as displayed in Fig. \ref{test}. Furthermore, Figs. \ref{beam_v3}a and \ref{beam_v3}b display the parameterization of the beam deflection at a given time $t$ for two different cases: bending in the plane $\left(P, \mathbf{x}_a, \mathbf{y}_a\right)$ and bending within the plane defined by $\left(P, \mathbf{x}_a, \mathbf{z}_a\right)$. 

The deflections $\tilde{v}(x, t)$ and $\tilde{w}(x, t)$ in the moving frame ${\mathcal{R}}_a$ are split up into $N$ mode shapes $\varphi^{\bullet}_i(x)$ (for $i=1, \ldots, N$), where $\boldsymbol{\Phi}_{\bullet}(x)=\left[\varphi^{\bullet}_1(x), \cdots, \varphi^{\bullet}_N(x)\right]^{\mathrm{T}}$ represents the mode shapes vectors. These vectors are linked with $N$ time-varying variables $q^{\bullet}_i(t)$, with ${\mathbf{q}_{y}}(t)=\left[q^{y}_1(t), \cdots, q^{y}_N(t)\right]^{\mathrm{T}}$ and ${\mathbf{q}_{z}}(t)=\left[q^{z}_1(t), \cdots, q^{z}_N(t)\right]^{\mathrm{T}}$, as follows:

\begin{equation}
\tilde{v}(x, t)=\sum_{i=1}^{N} \varphi_{i}^{y}(x) q^{y}_{i}(t)=\boldsymbol{\Phi}_y^{\mathrm{T}}(x)\mathbf{{q}}_{y}(t) 
\quad \text{and} \quad \tilde{w}(x, t)=\sum_{i=1}^{N} \varphi_{i}^{z}(x) q^{z}_{i}(t)=\boldsymbol{\Phi}_z^{\mathrm{T}}(x)\mathbf{{q}}_{z}(t) 
\label{vtilde}
\end{equation}

The mode shapes vectors which are used to model the beam dynamics in this paper are displayed in \ref{modal}, where $N=4$. Moreover, the inertial velocity $\mathbf{v}^{\mathcal{A}}_{M}$ projected in the body frame and defined at a general point $M(x)$ can be computed as follows:

\begin{equation}
\mathbf{v}^{\mathcal{A}}_{M}=\left.\frac{d \mathbf{O M}}{d t}\right|_{\mathcal{R}_i}=\left.\frac{d \mathbf{O P}}{d t}\right|_{\mathcal{R}_i}+\left.\frac{d \mathbf{P M}}{d t}\right|_{\mathcal{R}_a}+\boldsymbol{{\omega}}_P^{\mathcal{A}} \wedge \mathbf{P M}\text{, with} \quad \mathbf{OP} \equiv \mathbf{x}^{\mathcal{A}}_P \quad \text{and} \quad \mathbf{OM} \equiv \mathbf{x}^{\mathcal{A}}_M  
\label{vm}
\end{equation}
%
Furthermore, the distance vector $\mathbf{PM}$ displayed in Eq. \eqref{vm} can be computed as:
\begin{equation}
\mathbf{P M}(x,t)=\left[\begin{array}{c}
x+\tilde{u}(x,t) + \Delta u(x, t)\\
\tilde{v}(x, t) \\
{\tilde{w}}(x, t) 
\end{array}\right]
\label{positions1}
\end{equation}
In Eq. \eqref{positions1}, the variable $\tilde{u}(x,t)$ represents the displacement of a specific point $M(x)$ on the deformed flexible beam along the $\mathbf{x}_a$-axis and $\Delta u(x, t)$ denotes the deformation of the beam along the $\mathbf{x}_a$-axis due to traction-compression. Simply put, $\tilde{u}(x,t)$ measures the deviation of the point $M(x)$ from its position when the beam is in its undeformed state. From Figs. \ref{beam_v3}a and \ref{beam_v3}b, it can be observable that $\tilde{u}(x,t)$ is equal to:
\begin{equation}
\begin{aligned}
\displaystyle \tilde{u}(x, t) & =-\int_{0}^{x}\left[1-\cos\left({{\frac{\partial \tilde{v}}{\partial r}(r, t)}}\right)\right] d r-\int_{0}^{x}\left[1-\cos\left({{\frac{\partial \tilde{w}}{\partial r}(r, t)}}\right)\right] d r \\
& =-\int_{0}^{x}\left[1-\cos \left({\boldsymbol{\Phi}}_y^{\prime^\mathrm{T}}(r) {\mathbf{q}_{y}}(t)\right)\right] dr-\int_{0}^{x}\left[1-\cos \left({\boldsymbol{\Phi}}_z^{\prime^\mathrm{T}}(r) {\mathbf{q}_{z}}(t)\right)\right] dr
\end{aligned}
\label{traction1}
\end{equation}
Eq. \eqref{traction1} can be rewritten by considering a second-order Taylor expansion, as follows:
\begin{equation}
\displaystyle \tilde{u}(x, t) \approx -\int_{0}^{x} \frac{\left({\boldsymbol{\Phi}}_y^{\prime^\mathrm{T}}(r) {\mathbf{q}_{y}}(t)\right)^{2}}{2} d r-\int_{0}^{x} \frac{\left({\boldsymbol{\Phi}}_z^{\prime^\mathrm{T}}(r) {\mathbf{q}_{z}}(t)\right)^{2}}{2} d r=-\frac{1}{2} \mathbf{q}^{\mathrm{T}}(t) {\mathbf{E}^{yz}(x)} {\mathbf{q}}(t) 
\label{traction}
\end{equation}
In Eq. \eqref{traction}, ${\mathbf{q}}(t)=\left[ {\mathbf{q}}^{\mathrm{T}}_{y}(t),{\mathbf{q}}^{\mathrm{T}}_{z}(t)\right]^{\mathrm{T}}$ and ${\mathbf{E}}^{yz}(x)=\operatorname{diag}\left({\mathbf{E}}_y(x), {\mathbf{E}}_z(x)\right)$, with ${\mathbf{E}_{\bullet}(x)}=\int_{0}^{x} {\boldsymbol{\Phi}}_{\bullet}^{\prime}\left(r\right) {{{\boldsymbol{\Phi}}}_{\bullet}^{\prime^\mathrm{T}}}\left(r\right)  d r$. In addition, only one flexible mode is considered to model the traction-compression dynamics. Let $\Delta u(t)$ be the overall deformation of the beam due to the axial load. Then, it follows that: 
\begin{equation}
  \Delta u(x, t) =\tau(x) \Delta u(t) \mbox{, with}  \quad \tau(x)= \frac{x}{l^{\mathcal{A}}}
\end{equation}

Similarly, only one flexible mode is considered to represent the torsion dynamics due to transverse shear around the $\mathbf{x}_a$-axis. Let us now consider $\Delta\phi(x, t)$ to be the axial angular deformation at any point of abscissa $x$ $\left(x \in[0, l^{\mathcal{A}}]\right)$, which is caused by the axial torques $\mathbf{W}_{\mathcal{A}/\bullet,{P}}\{4\}$ and $\mathbf{W}_{\bullet/\mathcal{A},{C}}\{4\}$, applied by the beam $\mathcal{A}$ at the point $P$ and to the beam $\mathcal{A}$ at the point $C$, respectively. Then, it is assumed that:

\begin{equation}
\Delta\phi(x, t)=\frac{x}{l^{\mathcal{A}}}\left(\phi^{\mathcal{A}}_C(t)-\phi_P^{\mathcal{A}}(t)\right)=\frac{x}{l^{\mathcal{A}}} \Delta\phi(t)=\sigma(x)\Delta\phi(t)
\text{, with} \quad \sigma(x) = \frac{x}{l^{\mathcal{A}}}
\label{phiprop}
\end{equation}

Let us also consider $\mathbf{q}_f(t)=\left[ {\mathbf{q}}_y^{\mathrm{T}}(t) , {\mathbf{q}}_z^{\mathrm{T}}(t)  , {\Delta} u(t) , \Delta\phi(t)\right]^{\mathrm{T}}$ to be the vector (with $2N+2$ components) of all the internal elastic DOF of the body $\mathcal{A}$ to be used in the \textsc{Lagrange} derivation (Eq. \eqref{eq:lagrange2}). Then, Eq. \eqref{positions1} can be rewritten in the following way:
\begin{equation}
\mathbf{P M}(x,t)=\underbrace{\left[\begin{array}{c}
x\\
0 \\
0
\end{array}\right]}_{\mathbf{P}_0(x)}+\underbrace{\left[\begin{array}{cccc}
\mathbf{0}_{1\times N} & \mathbf{0}_{1\times N} & \tau(x) & 0\\
\boldsymbol{\Phi}_y^{\mathrm{T}}(x) & \mathbf{0}_{1\times N} & 0 & 0\\
\mathbf{0}_{1\times N} & \boldsymbol{\Phi}_z^{\mathrm{T}}(x) & 0 & 0
\end{array}\right]}_{\mathbf{M}^{\mathrm{T}}_1(x)}\underbrace{\left[\begin{array}{c}
\mathbf{q}_y(t)\\
\mathbf{q}_z(t)\\
\Delta u(t)\\
\Delta \phi(t)
\end{array}\right]}_{\mathbf{q}_f(t)}+\underbrace{\left[\begin{array}{cccc}
-\frac{1}{2} \mathbf{q}_f^{\mathrm{T}}(t)\left[\begin{array}{cc}\mathbf{E}^{yz}(x) & \mathbf{0}_{2N\times 2} \\ \mathbf{0}_{2\times 2N} & \mathbf{0}_{2\times 2}\end{array}\right] {\mathbf{q}_f}(t) \\
0\\
0
\end{array}\right]}_{\mathbf{P}_2(x,\mathbf{q}_f)}
\label{PMnew}
\end{equation}
Following the notation defined in Eq. \eqref{PMnew}, the inertial velocity $\mathbf{v}^{\mathcal{A}}_{M}$ expressed in the frame $\mathcal{R}_a$ is equal to:
\begin{equation}
\mathbf{v}^{\mathcal{A}}_{M}=\mathbf{v}^{\mathcal{A}}_{P}+\mathbf{M}^{\mathrm{T}}_1(x)\dot{\mathbf{q}}_f(t)+\dot{\mathbf{P}}_2(x,\mathbf{q}_f)+({ }^{*} \boldsymbol{\omega}_P^{\mathcal{A}})\left(\mathbf{P}_0(x)+\mathbf{M}^{\mathrm{T}}_1(x)\mathbf{q}_f(t)+\mathbf{P}_2(x,\mathbf{q}_f)\right)
\end{equation}

\subsection{Lagrangian computation and derivation}

Let us define the overall vectors of quasi-coordinates $\mathbf Q_{p}(t)$ and quasi-velocities $\mathbf Q_{v}(t)$, as well as their equilibrium values and variations:
\begin{eqnarray}\label{qp}
\mathbf{Q}_{p}(t) ={\left[\begin{array}{ccc}\mathbf{x}^{{\mathcal{A}}^\mathrm{T}}_P(t)   & \boldsymbol{{\Theta}}_P^{{\mathcal{A}}^{\mathrm{T}}}(t) & \mathbf{q}_f^{\mathrm{T}}(t)\end{array}\right]}^{\mathrm{T}} = \overline{\mathbf{Q}}_{p}+\delta\mathbf{Q}_{p}(t) \mbox{, with} \quad \overline{\mathbf{Q}}_{p}=\left[\begin{array}{ccc}\overline{\mathbf{x}}^{{\mathcal{A}}^\mathrm{T}}_P   & \overline{\boldsymbol{\Theta}}_P^{{\mathcal{A}}^{\mathrm{T}}} & \mathbf 0_{1\times (2N+2)} \end{array}\right]^{\mathrm{T}} \\
\mathbf{Q}_{v}(t)= {\left[\begin{array}{ccc} \mathbf{v}_{P}^{{\mathcal{A}}^{\mathrm{T}}}(t) & \boldsymbol{{\omega}}_P^{{\mathcal{A}}^{\mathrm{T}}}(t) & \dot{\mathbf{q}}_f^{\mathrm{T}}(t) \end{array}\right]}^{\mathrm{T}}= \overline{\mathbf{Q}}_{v}+\delta\mathbf{Q}_{v}(t) \mbox{, with} \quad\overline{\mathbf{Q}}_{v}=\left[\begin{array}{ccc}\overline{\mathbf{v}}^{{\mathcal{A}}^\mathrm{T}}_P   & \overline{\boldsymbol{\omega}}_P^{{\mathcal{A}}^{\mathrm{T}}} & \mathbf 0_{1\times (2N+2)} \end{array}\right]^{\mathrm{T}} 
\label{qv}
\end{eqnarray}
Indeed, according to the general assumptions, it is considered that $\overline{\mathbf q}_f=\mathbf 0_{ (2N+2)\times 1}$. The kinetic energy $\mathcal{T}$ computation of the beam element $\mathcal{A}$ can be outlined as follows:
\begin{equation}
\mathcal{T}=\frac{1}{2} \int_{0}^{l^{\mathcal{A}}} \rho^{\mathcal{A}} S^{\mathcal{A}}
{\mathbf{v}_{M}^{\mathcal{A}}}^{\mathrm T} \mathbf{v}_{M}^{\mathcal{A}}  d x + \underbrace{\frac{1}{2} \int_0^{l^{\mathcal{A}}} \rho^{\mathcal{A}} J^{\mathcal{A}}_{p x} \left(\boldsymbol{{\omega}}_P^{\mathcal{A}}\{1\}+\sigma(x)\dot{{\Delta \phi}}(t)\right)^2 d x}_{\text{torsion contribution}}
\label{kinetic}
\end{equation}
By considering a second-order \textsc{Taylor} expansion of $\mathcal{T}$ with respect to $\delta\mathbf{Q}_{p}$ and $\delta\mathbf{Q}_{v}$, Eq. \eqref{kinetic} can be expressed as:
\begin{equation}
\mathcal{T}=\frac{1}{2} \delta{\mathbf{Q}}_{v}^{\mathrm{T}}(t) \mathbf{M}_{\mathcal{T}}\delta{\mathbf{Q}}_{v}(t)+\frac{1}{2} \delta{\mathbf{Q}}_{p}^{\mathrm{T}}(t) \mathbf{K}_{\mathcal{T}}{{\delta{\mathbf{Q}}_{p}}}(t)+\frac{1}{2} \delta{\mathbf{Q}}_{p}^{\mathrm{T}}(t) \mathbf{G}_{\mathcal{T}}\delta{\mathbf{Q}}_{v}(t)+\mathbf{C}_{\dot{{Q}}}^{\mathrm{T}}\delta{\mathbf{Q}}_{v}(t)+\mathbf{C}_{{{Q}}}^{\mathrm{T}}\delta{{{\mathbf{Q}}_{p}}}(t)+\frac{1}{2} \overline{\mathbf{Q}}_{v}^{\mathrm{T}}\mathbf{M}_{\mathcal{T}}\overline{\mathbf{Q}}_{v}
\label{kinetic2}
\end{equation}

The kinetic energy of the beam does not depend on $\overline{\mathbf{Q}}_{p}$. However, the matrices $\mathbf{K}_{\mathcal{T}}$ (the centrifugal stiffness matrix), $\mathbf{G}_{\mathcal{T}}$, $\mathbf{C}_{\dot{Q}}$ and $\mathbf{C}_{{Q}}$ are naturally dependent on $\overline{\mathbf{Q}}_{v}$. In addition, the calculation of the elastic potential energy, denoted as $\mathcal{V}$, consists of four distinct components: $\mathcal{V}_b^{y}$, $\mathcal{V}_b^{z}$, $\mathcal{V}_t^x$ and $\mathcal{V}_r^x$. The terms $\mathcal{V}_b^{y}$ and $\mathcal{V}_b^{z}$ correspond to the elastic potential energy resulting from pure bending in the planes $\left(P, \mathbf{x}_a, \mathbf{y}_a\right)$ and $\left(P, \mathbf{x}_a, \mathbf{z}_a\right)$, respectively, while $\mathcal{V}_t^x$ and $\mathcal{V}_r^x$ represent the contributions from the traction and torsion dynamics:

\begin{equation}
\begin{aligned}
& \def\arraystretch{1.2}\begin{array}{ll}
\begin{cases}
\mathcal{V}^{y}_b=\frac{1}{2}\mathbf{q}_{y}^{\mathrm{T}}(t) \underbrace{\int_{0}^{l^{\mathcal{A}}} E^{\mathcal{A}}J^{\mathcal{A}}_z{\boldsymbol{\Phi}_y^{\prime\prime}}(x){\boldsymbol{\Phi}}_y^{{\prime\prime}^{\mathrm{T}}}(x) d x}_{\mathbf{K}^{y}_b}{{\mathbf{q}_{y}}}(t) \\
\mathcal{V}^{z}_b=\frac{1}{2}\mathbf{q}_{z}^{\mathrm{T}}(t) \underbrace{\int_{0}^{l^{\mathcal{A}}} E^{\mathcal{A}}J^{\mathcal{A}}_y{\boldsymbol{\Phi}_z^{\prime\prime}}(x){\boldsymbol{\Phi}}_z^{{\prime\prime}^{\mathrm{T}}}(x) d x}_{\mathbf{K}^{z}_b}{{\mathbf{q}_{z}}}(t) \\
\end{cases} &
\begin{cases}
\mathcal{V}^{x}_t=\frac{1}{2}{\Delta u}(t) \underbrace{\int_{0}^{l^{\mathcal{A}}}E^{\mathcal{A}}S^{\mathcal{A}}{{\tau^\prime}(x){\tau^\prime}(x) d x}}_{\mathbf{K}^{x}_t}{\Delta u}(t) \\
\mathcal{V}^{x}_r=\frac{1}{2}{\Delta \phi}(t) \underbrace{\int_{0}^{l^{\mathcal{A}}}G^{\mathcal{A}}J_{px}^{\mathcal{A}}{{\sigma^\prime}(x){\sigma^\prime}(x) d x}}_{\mathbf{K}^{x}_r}{\Delta \phi}(t) 
\end{cases} 
\end{array} 
\text{, } \\
& \text{with} \quad 
\mathcal{V}=\mathcal{V}^{y}_b+\mathcal{V}^{z}_b+\mathcal{V}^{x}_t+\mathcal{V}^{x}_r=\frac{1}{2} \delta{\mathbf{Q}}_{p}^{\mathrm{T}}(t)\mathbf{K}_{\mathcal{V}}{{\delta{\mathbf{Q}}_{p}}}(t) \quad \text{and} \quad \mathbf{K}_{\mathcal{V}}=\operatorname{diag}\left(\mathbf{0}_{6\times 6}, {\mathbf{K}^{y}_b}, {\mathbf{K}^{z}_b}, {\mathbf{K}^{x}_t}, {\mathbf{K}^{x}_r}\right)
\end{aligned}
\label{potential}
\end{equation}

The Lagrangian $\mathcal{L}=\mathcal{T}-\mathcal{V}$ can ultimately be obtained and the \textsc{Lagrange} derivation (Eqs. \eqref{eq:lagrange1} and \eqref{eq:lagrange2}) leads to the following linear model:
\begin{equation}
\begin{aligned}
& \mathbf{M}_{\mathcal{T}}\delta\dot{{\mathbf{Q}}}_{v}(t)+\left[\frac{1}{2}\left(\mathbf{G}_{\mathcal{T}}^{\mathrm{T}}-\mathbf{G}_{\mathcal{T}}\right)+\left[\begin{array}{c}\mathbf{M}_{\mathcal{L}} \\ \mathbf{0}_{(2N+2) \times (6+2N+2)}\end{array}\right]\right] \delta{\mathbf{Q}}_{v}(t) +\left[\mathbf{K}_{\mathcal{V}}-\mathbf{K}_{\mathcal{T}}+\left[\begin{array}{c}\mathbf{J}_{\mathcal{L}} \\ \mathbf{0}_{(2N+2) \times (6+2N+2)}\end{array}\right]\right]{\delta{\mathbf{Q}}_{p}}(t) \\
& -\mathbf{C}_Q+\left[\begin{array}{c}\mathbf{C}_{\mathcal{L}} \\ \mathbf{0}_{(2N+2) \times 1}\end{array}\right]=\mathbf{W}_{\mathcal{L}}\text{, }  \text{with} \quad  \mathbf{W}_{\mathcal{L}}=\left[\begin{array}{c} \mathbf F_{\mathrm{ext}/\mathcal A,P} \\ \mathbf T_{\mathrm{ext}/\mathcal A,P} \\ {\widehat{\mathbf{U}}}\end{array}\right] 
\end{aligned}
\label{Lagrangianderivation2}
\end{equation}

In Eq. \eqref{Lagrangianderivation2}, the matrices $\mathbf{M}_{\mathcal{L}}$, $\mathbf{J}_{\mathcal{L}}$ and  $\mathbf{C}_{\mathcal{L}}$ arise from the first-order approximation of the following term:

\begin{equation}
\left[\begin{array}{cc}
	({ }^{*} \boldsymbol{\omega}^{\mathcal{A}}_{{P}}) & \mathbf 0_{3} \\  ({ }^{*} \mathbf{v}^{\mathcal{A}}_{{P}}) & ({ }^{*} \boldsymbol{\omega}^{\mathcal{A}}_{{P}}) 
\end{array}\right] \frac{\partial \mathcal{L}}{\partial \left[\begin{array}{c}\delta\mathbf{v}^{\mathcal{A}}_{P} \\ \delta\boldsymbol{\omega}^{\mathcal{A}}_P\end{array}\right]}
\end{equation}


The wrenches applied by the body $\mathcal A$ and to the body $\mathcal A$ are the interaction wrenches $\mathbf{W}_{\mathcal{A}/\bullet,{P}}$ and $\mathbf{W}_{\bullet/\mathcal{A},{C}}$ at the points $P$ and $C$, respectively. Furthermore, the work of the external wrenches due to elastic deformations is equal to:
\begin{equation}
\mathcal W_{\mathrm{ext}}(t)=\left[\begin{array}{cc}\delta \mathbf{PM}^{\mathrm T}(l^{\mathcal{A}},t) & \delta\boldsymbol\Theta_{C/P}^ {\mathrm T}(l^{\mathcal{A}},t)\end{array}\right]^{\mathrm T}\mathbf{W}_{\bullet/\mathcal{A},{C}}(t)
\label{workext}
\end{equation}

For the computation of the work, the distance vector $\mathbf{PM}(x,t)$ is detailed in Eq. \eqref{PMnew}. Additionally, $\delta\boldsymbol\Theta_{C/P}(x,t)$ represents the angular deformation of the point $M(x)$ with respect to the equilibrium and it is given by $\delta\boldsymbol\Theta_{C/P}(x,t)=\left[\Delta\phi(x,t), \; -{\frac{\partial \tilde{w}}{\partial x}^{\mathrm T}(x, t)}, \; {\frac{\partial \tilde{v}}{\partial x}^{\mathrm T}(x, t)} \right]^{\mathrm T}=\mathbf M^{\mathrm T}_2(x)\mathbf q_f(t)$, with:
\begin{equation}
\mathbf{M}^{\mathrm{T}}_2(x)={\left[\begin{array}{cccc}
\mathbf{0}_{1\times N} & \mathbf{0}_{1\times N} & 0 & \sigma(x)\\
\mathbf{0}_{1\times N} & -\boldsymbol{\Phi}_z^{\prime^\mathrm{T}}(x) & 0 & 0 \\
\boldsymbol{\Phi}_y^{\prime^\mathrm{T}}(x) & \mathbf{0}_{1\times N} & 0 & 0 \\
\end{array}\right]}
\label{MT2}
\end{equation}

Consequently, Eq. \eqref{workext} becomes equal to:
\begin{equation}
 	\mathcal W_{\mathrm{ext}}(t)=\left[\begin{array}{cc}\mathbf q_f^{\mathrm{T}}(t)\mathbf M_1(l^{\mathcal{A}})+\mathbf P_2^{\mathrm{T}}(l^{\mathcal{A}},\mathbf q_f) & \mathbf q_f^{\mathrm{T}}(t)\mathbf M_2(l^{\mathcal{A}})\end{array}\right]\mathbf{W}_{\bullet/\mathcal{A},{C}}(t)
\end{equation}


By definition, $\displaystyle{\widehat{\mathbf{U}}}=\frac{\partial 	\mathcal W_{\mathrm{ext}}}{\partial \mathbf q_f}$, meaning that the right-hand side of Eq. \eqref{Lagrangianderivation2} reads:
 \begin{equation}
 	\mathbf{W}_{\mathcal{L}} = \left[\def\arraystretch{1.9}\begin{array}{cc}-\mathbf{I}_6 & \boldsymbol{\tau}^{\mathrm{T}}_{CP}\\ \mathbf{0}_{(2N+2) \times 6} & \left[\begin{array}{cc} \mathbf{M}_1(l^{\mathcal{A}})+\displaystyle\frac{\partial \mathbf{P}^{\mathrm{T}}_2(l^{\mathcal{A}},\mathbf{q}_f)}{\partial {\mathbf{q}_f}} & \mathbf{M}_2(l^{\mathcal{A}})\end{array}\right]\end{array}\right]\left[\begin{array}{c}
 		\mathbf{W}_{\mathcal{A}/\bullet,{P}} \\
 		\mathbf{W}_{\bullet/\mathcal{A},{C}} 
 	\end{array}\right]= \mathbf{N}\left[\begin{array}{c}
 		\mathbf{W}_{\mathcal{A}/\bullet,{P}} \\
 		\mathbf{W}_{\bullet/\mathcal{A},{C}} 
 	\end{array}\right] 
\label{wexteq}
 \end{equation}

\subsection{Equilibrium conditions}

Taking into account the assumptions that have been made, this approach is only valid in the proximity of the equilibrium point that is characterized by $\overline{\mathbf{q}}_f=\mathbf{0}_{(2N+2)\times 1}$. Therefore, it is essential to verify that these conditions are satisfied. This task can be accomplished by employing Eqs. \eqref{Lagrangianderivation2} and \eqref{wexteq} to derive the subsequent equilibrium:

\begin{equation}
\left[\mathbf{K}_{\mathcal{V}}-\mathbf{K}_{\mathcal{T}}+\left[\begin{array}{c}\mathbf{J}_{\mathcal{L}} \\ \mathbf{0}_{(2N+2) \times (6+2N+2)}\end{array}\right]\right]\left[\begin{array}{c} \star \\
\overline{\mathbf{q}}_f \end{array}\right]-\mathbf{C}_Q+\left[\begin{array}{c}\mathbf{C}_{\mathcal{L}} \\ \mathbf{0}_{(2N+2) \times 1}\end{array}\right]=\overline{\mathbf{N}}\left[\begin{array}{c}
\overline{\mathbf{W}}_{\mathcal{A}/\bullet,{P}} \\
\overline{\mathbf{W}}_{\bullet/\mathcal{A},{C}} 
\end{array}\right]
\label{equilibriium}
\end{equation}

Indeed, the geometric equilibrium conditions $\left[\overline{\mathbf{x}}_{P}^{\mathcal{A}{ }^{\mathrm{T}}}, \overline{\mathbf{\Theta}}_{P}^{\mathcal{A} ^\mathrm{T}}\right]_{\mathcal{R}_{a}}^{\mathrm{T}}$ are always multiplied by null terms in the equilibrium computation displayed in Eq. \eqref{equilibriium}. For this reason, they do not have to be taken into account. By means of Eq. \eqref{equilibriium}, $\overline{\mathbf{W}}_{\mathcal{A}/\bullet,{P}}$ and $\overline{\mathbf{q}}_f$ can be calculated from $\overline{\mathbf{W}}_{\bullet/\mathcal{A},{C}}$, making it possible to confirm all the assumptions. After verifying all the hypotheses, $\overline{\mathbf{W}}_{\mathcal{A}/\bullet,{P}}$ can be recomputed with the help of the following equation: 

\begin{equation}
-\mathbf{C}_Q+\left[\begin{array}{c}\mathbf{C}_{\mathcal{L}} \\ \mathbf{0}_{(2N+2) \times 1}\end{array}\right]=\overline{\mathbf{N}}\left[\begin{array}{c}
\overline{\mathbf{W}}_{\mathcal{A}/\bullet,{P}} \\
\overline{\mathbf{W}}_{\bullet/\mathcal{A},{C}} 
\end{array}\right]
\label{equilibriium2}
\end{equation}

\begin{figure}[!ht]
\centering
 \includegraphics[width=1\textwidth]{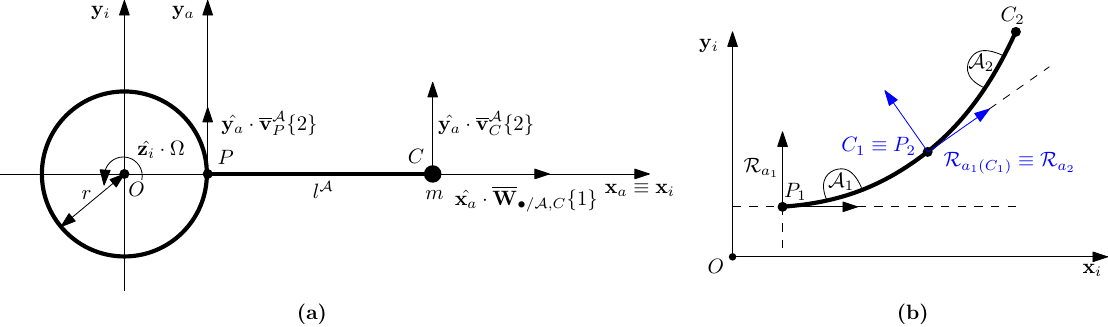}
\caption{(a) Scenario where a beam with length $l^{\mathcal{A}}$ is connected to a spinning rigid hub at the point $P$ and to a tip mass at the point $C$: this figure provides a geometrical interpretation of how the inertial velocity and force components in equilibrium are expressed in $\mathcal{R}_a$ with respect to the configuration of the system. (b) Assembly of two flexible rotating beams in series.}
\label{psiexp} 
\end{figure}

Let us now consider a simple scenario involving a beam with a tip mass. This beam is attached to a spinning rigid hub with radius $r$, as displayed in Fig. \ref{psiexp}a. In this case, the centrifugal force with magnitude $\overline{\mathbf{W}}_{\bullet/\mathcal{A},{C}}\{1\}$ which arises from the spinning motion is equal to $\overline{\mathbf{W}}_{\bullet/\mathcal{A},{C}}\{1\}=m\left(l^{\mathcal{A}}+r\right)\Omega^2$, where $m$ is the tip mass value located at the point $C$ and $l^{\mathcal{A}}$ represents the length of the beam. In this scenario, $\overline{\mathbf{x}}^{\mathcal{A}}_P=\left[ r , 0 , 0\right]^{\mathrm{T}}$, $\overline{\boldsymbol{{\Theta}}}_P^{{\mathcal{A}}}=\mathbf{0}_{3\times 1}$ and $\overline{\boldsymbol{\omega}}_P^{{\mathcal{A}}}=\left[ 0 , 0 ,\Omega\right]^{\mathrm{T}}$. For this reason, it can be concluded that $\overline{\mathbf{W}}_{\bullet/\mathcal{A},{C}}=\left[ m\left(l^{\mathcal{A}}+r\right)\Omega^2 , 0 , 0 , 0 , 0 , 0\right]^{\mathrm{T}}$. Due to the spinning motion, it can also be inferred that $\overline{\mathbf{v}}_{P}^{{\mathcal{A}}}=\left[ 0 , r\Omega , 0\right]^{\mathrm{T}}$. These equilibrium conditions are expressed in the body frame $\mathcal{R}_a$, which is the frame of reference used to define the beam model. A similar procedure can be carried out for the computation of $\overline{\mathbf{v}}_{C}^{{\mathcal{A}}}$, with $\overline{\mathbf{v}}_{C}^{{\mathcal{A}}}=\left[ 0 , \left(l^{\mathcal{A}}+r\right)\Omega , 0\right]^{\mathrm{T}}$. However, $\overline{\mathbf{v}}_{C}^{{\mathcal{A}}}$ does not parameterize the beam model and is only relevant if another body is attached to the beam at the point C. This scenario occurs when a tip mass is present but also when two different beams are connected in series, as displayed in Fig. \ref{psiexp}b.

\subsection{Assembly of beams with other substructures}
\label{sect:assembly}

When the beam $\mathcal A$ is connected to another body $\bullet$ at the point $C$, one has to take into account the projection in the deformed body  frame $\mathcal R_{a(C)}$ of the interaction wrench $\mathbf{W}_{\bullet/\mathcal{A},{C}}$, as illustrated in Fig. \ref{psiexp}b with the connection of two flexible beams in series: the frames $\mathcal R_{a_1(C_1)}$ and $\mathcal R_{a_2}$ are coincident. Therefore, the DCM $\mathbf{P}_{\mathcal{R}_{a(C)}/\mathcal{R}_a}$ between the frames $\mathcal R_{a(C)}$ and  $\mathcal R_{a}$ is introduced, such that:

\begin{equation}
{\left[\mathbf{W}_{\bullet/\mathcal{A},{C}}\right]}_{\mathcal{R}_{a}}=\mathbf{P}_{\mathcal{R}_{a(C)}/\mathcal{R}_a}^{\times2}
{\left[\mathbf{W}_{\bullet/\mathcal{A},{C}}\right]}_{\mathcal{R}_{a(C)} }
\label{frame_deformed_wrenches1}
\end{equation}

This DCM is given by the following first-order approximation:
\begin{equation}
	\mathbf{P}_{\mathcal{R}_{a(C)}/\mathcal{R}_a}=\left[\begin{array}{ccc}
		1 & -\boldsymbol{\Phi}_y^{\prime^\mathrm{T}}(l^{\mathcal{A}})\mathbf{{q}}_{y}(t) & -\boldsymbol{\Phi}_z^{\prime^\mathrm{T}}(l^{\mathcal{A}})\mathbf{{q}}_{z}(t) \\
		\boldsymbol{\Phi}_y^{\prime^\mathrm{T}}(l^{\mathcal{A}})\mathbf{{q}}_{y}(t) & 1 & -\sigma(l^{\mathcal{A}}){\Delta \phi}(t) \\
		\boldsymbol{\Phi}_z^{\prime^\mathrm{T}}(l^{\mathcal{A}})\mathbf{{q}}_{z}(t) & \sigma(l^{\mathcal{A}}){\Delta \phi}(t) & 1 \\
	\end{array}\right]
\end{equation}

Therefore, the linearization of Eq. \eqref{frame_deformed_wrenches1} leads to  ${\left[\overline{\mathbf{W}}_{\bullet/\mathcal{A},{C}}\right]}_{\mathcal{R}_{a}}={\left[\overline{\mathbf{W}}_{\bullet/\mathcal{A},{C}}\right]}_{\mathcal{R}_{a(C)}}$ and also to:

%

\begin{equation}
\begin{aligned}
& {\left[\delta\mathbf{W}_{\bullet/\mathcal{A},{C}}\right]}_{\mathcal{R}_{a}}={\left[\delta\mathbf{W}_{\bullet/\mathcal{A},{C}}\right]}_{\mathcal{R}_{a(C)} }
+\underbrace{\left[\begin{array}{cccc}
-\boldsymbol{\Phi}_y^{\prime^\mathrm{T}}(l^{\mathcal{A}})\overline{\mathbf{W}}_{\bullet/\mathcal{A},{C}}\{2\}  & -\boldsymbol{\Phi}_z^{\prime^\mathrm{T}}(l^{\mathcal{A}})\overline{\mathbf{W}}_{\bullet/\mathcal{A},{C}}\{3\}  & 0 & 0\\
\boldsymbol{\Phi}_y^{\prime^\mathrm{T}}(l^{\mathcal{A}})\overline{\mathbf{W}}_{\bullet/\mathcal{A},{C}}\{1\}  & \mathbf{0}_{1\times N} & 0 & -\sigma(l^{\mathcal{A}})\overline{\mathbf{W}}_{\bullet/\mathcal{A},{C}}\{3\} \\ \mathbf{0}_{1\times N} & \boldsymbol{\Phi}_z^{\prime^\mathrm{T}}(l^{\mathcal{A}})\overline{\mathbf{W}}_{\bullet/\mathcal{A},{C}}\{1\}  & 0 & \sigma(l^{\mathcal{A}})\overline{\mathbf{W}}_{\bullet/\mathcal{A},{C}}\{2\} \\
-\boldsymbol{\Phi}_y^{\prime^\mathrm{T}}(l^{\mathcal{A}})\overline{\mathbf{W}}_{\bullet/\mathcal{A},{C}}\{5\}  & -\boldsymbol{\Phi}_z^{\prime^\mathrm{T}}(l^{\mathcal{A}})\overline{\mathbf{W}}_{\bullet/\mathcal{A},{C}}\{6\}  & 0 & 0\\
\boldsymbol{\Phi}_y^{\prime^\mathrm{T}}(l^{\mathcal{A}})\overline{\mathbf{W}}_{\bullet/\mathcal{A},{C}}\{4\}  & \mathbf{0}_{1\times N} & 0 & -\sigma(l^{\mathcal{A}})\overline{\mathbf{W}}_{\bullet/\mathcal{A},{C}}\{6\} \\ \mathbf{0}_{1\times N} & \boldsymbol{\Phi}_z^{\prime^\mathrm{T}}(l^{\mathcal{A}})\overline{\mathbf{W}}_{\bullet/\mathcal{A},{C}}\{4\}  & 0 & \sigma(l^{\mathcal{A}})\overline{\mathbf{W}}_{\bullet/\mathcal{A},{C}}\{5\} 
\end{array}\right]}_{\mathbf{W}_C}
{\mathbf{q}}_f(t) 
\end{aligned}
\label{frame_deformed_wrenches}
\end{equation}
As previously mentioned, the ultimate goal is to establish a linear model for the rotating beam $\mathcal{A}$ around a specific equilibrium point. For this reason, only the first-order terms originating from the \textsc{Taylor} series expansion of $\mathbf{W}_{\mathcal{L}}$ are taken into consideration. Subsequently, all the various quantities are approximated as the sum of their equilibrium values and slight deviations from these values. 
A first-order linearization is finally applied to Eqs. \eqref{Lagrangianderivation2} and \eqref{wexteq}. Furthermore, the expression of ${\left[\delta\mathbf{W}_{\bullet/\mathcal{A},{C}}\right]}_{\mathcal{R}_{a}}$ displayed in Eq. \eqref{frame_deformed_wrenches} is taken into account in this linearization, leading to the following linearized Lagrangian derivation of the spinning beam:
\begin{equation}
\begin{aligned}
& \mathbf{M}_{\mathcal{T}}\delta\dot{{\mathbf{Q}}}_{v}(t)+{\left[\frac{1}{2}\left(\mathbf{G}_{\mathcal{T}}^{\mathrm{T}}-\mathbf{G}_{\mathcal{T}}\right)+\left[\begin{array}{c}\mathbf{M}_{\mathcal{L}} \\ \mathbf{0}_{(2N+2) \times (6+2N+2)}\end{array}\right]\right]}\delta{\mathbf{Q}}_{v}(t) +{\left[\mathbf{K}_{\mathcal{V}}-\mathbf{K}_{\mathcal{T}}+\left[\begin{array}{c}\mathbf{J}_{\mathcal{L}} \\ \mathbf{0}_{(2N+2) \times (6+2N+2)}\end{array}\right]+\mathbf{F}_c\right]}{\delta{\mathbf{Q}}_{p}}(t)= \\
& \overline{\mathbf{N}}\left[\begin{array}{c}
{\left[\delta{\mathbf{W}_{\mathcal{A}/\bullet,{P}}}\right]}_{\mathcal{R}_{a}} \\
{\left[\delta\mathbf{W}_{\bullet/\mathcal{A},{C}}\right]}_{\mathcal{R}_{a(C)}} 
\end{array}\right]
\end{aligned}
\label{cenas}
\end{equation}

In Eq. \eqref{cenas}, the additional stiffness matrix $\mathbf{F}_c$ depends on $\overline{\mathbf{W}}_{\bullet/\mathcal{A},{C}}$ due to the matrices $\mathbf W_C$ (in Eq. \eqref{frame_deformed_wrenches}) and $\mathbf N$ (in Eq.  \eqref{wexteq}). It can also be noticed that all the constant terms which were present in Eq. \eqref{Lagrangianderivation2} cancelled out. Ultimately, Eq. \eqref{cenas} can be rewritten as:

\begin{equation}
\begin{aligned}
& \left[\begin{array}{c|c}
\mathbf{M}_{rr} & \mathbf{M}_{rf}\\
\hline \mathbf{M}_{fr} & \mathbf{M}_{ff}
\end{array}\right]\left[\begin{array}{c}
\delta\dot{\mathbf{v}}_{P}^{{\mathcal{A}}}\\
\delta\dot{\boldsymbol{{\omega}}}_P^{{\mathcal{A}}}\\
\hline \ddot{\mathbf{q}}_f(t)\\
\end{array}\right]_{{\mathcal{R}_{a}}}+
\left[\begin{array}{c|c}
\mathbf{D}_{rr} & \mathbf{D}_{rf}\\
\hline \mathbf{D}_{fr} & \mathbf{D}_{ff}
\end{array}\right]\left[\begin{array}{c}
\delta \mathbf{v}_{P}^{{\mathcal{A}}}\\
\delta{\boldsymbol{{\omega}}}_P^{{\mathcal{A}}}\\
\hline \dot{\mathbf{q}}_f(t)\\
\end{array}\right]_{{\mathcal{R}_{a}}}+
\left[\begin{array}{c|c}
\mathbf{K}_{rr} & \mathbf{K}_{rf}\\
\hline \mathbf{K}_{fr} & \mathbf{K}_{ff}
\end{array}\right]\left[\begin{array}{c}
\delta \mathbf{x}^{\mathcal{A}}_P\\
\delta \boldsymbol{{\Theta}}_P^{{\mathcal{A}}}\\
\hline {\mathbf{q}}_f(t)\\
\end{array}\right]_{{\mathcal{R}_{a}}}= \\
& \left[\def\arraystretch{1.5}\begin{array}{c|c}
\overline{\mathbf{{\mathbf{N}}}}_{rr} & \overline{\mathbf{{\mathbf{N}}}}_{rf}\\
\hline \overline{\mathbf{{\mathbf{N}}}}_{fr} & \overline{\mathbf{{\mathbf{N}}}}_{ff}
\end{array}\right]\left[\def\arraystretch{1.3}\begin{array}{c}
{\left[\delta{\mathbf{W}_{\mathcal{A}/\bullet,{P}}}\right]}_{\mathcal{R}_{a}} \\ \hline
{\left[\delta\mathbf{W}_{\bullet/\mathcal{A},{C}}\right]}_{\mathcal{R}_{a(C)}} 
\end{array}\right]
\end{aligned}
\label{equivalent}
\end{equation}

\subsection{TITOP model of a flexible spinning beam}

Before deriving the state-space representation of the TITOP beam model, it is essential to compute the output motion vector of the beam at the point $C$. First, it can be deduced that:

\begin{equation}
\mathbf{x}^{\mathcal{A}}_C=\mathbf{O P}+\mathbf{P C}=\mathbf{x}^{\mathcal{A}}_P+\mathbf{P C}
\label{positions2}
\end{equation}

The computation of $\mathbf{P C}$ is given by Eq. \eqref{PMnew}, with $x=l^{\mathcal A}.$ Furthermore, it can be observed from Fig. \ref{beam_v3}, the first spatial derivative of Eq. \eqref{vtilde} and Eq. \eqref{phiprop} that: 

\begin{equation}
{\boldsymbol{\Theta}}^{\mathcal{A}}_C={\boldsymbol{\Theta}}^{\mathcal{A}}_P+ {\delta\boldsymbol{\Theta}}_{C/P}(l^{\mathcal{A}},t)={\boldsymbol{\Theta}}^{\mathcal{A}}_P+\mathbf{M}^{\mathrm{T}}_2(l^{\mathcal{A}}){\mathbf{q}}_f
\label{anglesc}
\end{equation}

For the computation of the inertial velocity vector $\mathbf{v}^{\mathcal{A}}_{C}$, an equivalent approach to the one employed in Eq. \eqref{vm} is used, with:

\begin{equation}
\mathbf{v}^{\mathcal{A}}_{C}=\mathbf{v}^{\mathcal{A}}_{P}+\left.\frac{d \mathbf{P C}}{d t}\right|_{\mathcal{R}_a}+\boldsymbol{{\omega}}_P^{\mathcal{A}} \wedge \mathbf{P C}
\label{vc}
\end{equation}

For the computation of the angular velocity vector at the point C, it is crucial to emphasize that ${\boldsymbol{\omega}}^{\mathcal{A}}_C$ is equivalent to $\boldsymbol{{\omega}}_{\mathcal{R}_{a(C)}/\mathcal{R}_i}$, which represents the angular velocity vector of the deformed body frame $\mathcal{R}_{{a(C)}}$ at the point $C$ with respect to the inertial frame $\mathcal{R}_i$. On that account, it follows that:

\begin{equation}
{\boldsymbol{\omega}}^{\mathcal{A}}_C=\boldsymbol{{\omega}}_{\mathcal{R}_{a(C)}/\mathcal{R}_a}+\boldsymbol{{\omega}}_{\mathcal{R}_a/\mathcal{R}_i}=\mathbf{M}^{\mathrm{T}}_2(l^{\mathcal{A}})\dot{\mathbf{q}}_f+\boldsymbol{{\omega}}^{\mathcal A}_{P}
\label{omegac}
\end{equation}

Ultimately, the computation of the acceleration vectors $\dot{\mathbf{v}}^{\mathcal{A}}_{C}$ and $\dot{{\boldsymbol{\omega}}}^{\mathcal{A}}_C$ can directly be obtained by applying a time derivation to Eqs. \eqref{vc} and \eqref{omegac}, respectively. Eqs. \eqref{positions2}, \eqref{anglesc}, \eqref{vc} and \eqref{omegac} are projected in the body frame of the beam, denoted as $\mathcal{R}_a$. However, the output motion vector computed at the point $C$ also needs to be projected in the beam's deformed body frame ${\mathcal{R}_{a(C)}}$. The need for this projection is driven by the  same reasons that prompted the projection of the input computations ${\mathbf{W}_{\bullet/\mathcal{A},{C}}}$ into ${\mathcal{R}_{a(C)}}$, with:

\begin{equation}
{{\left[{\mathbf{m}}^{\mathcal{A}}_{C}\right]}_{\mathcal{R}_{a(C)}}}=\mathbf{P}_{{\mathcal{R}_{a}}/\mathcal{R}_{a(C)}}^{\times 6}{{\left[{\mathbf{m}}^{\mathcal{A}}_{C}\right]}_{\mathcal{R}_{a}}}
\label{projmotion}
\end{equation}

Subsequently,  Eqs. \eqref{positions2}, \eqref{anglesc}, \eqref{vc} and \eqref{omegac}, along with the time derivations of Eqs. \eqref{vc} and \eqref{omegac}, are incorporated into Eq. \eqref{projmotion}, which is thereafter subjected to a first-order linearization. This process yields the computation of the output motion vector, as follows:

\begin{equation}
{{\left[\delta{\mathbf{m}}^{\mathcal{A}}_{C}\right]}_{\mathcal{R}_{a(C)}}}=
\left[\begin{array}{c|c}
\mathbf{M}_{l} & \mathbf{M}_{r}\\
\end{array}\right]\left[\begin{array}{c}
\delta\dot{\mathbf{v}}_{P}^{{\mathcal{A}}}\\
\delta\dot{\boldsymbol{{\omega}}}_P^{{\mathcal{A}}}\\
\hline \ddot{\mathbf{q}}_f(t)\\
\end{array}\right]_{{\mathcal{R}_{a}}}+
\left[\begin{array}{c|c}
\mathbf{D}_{l} & \mathbf{D}_{r}\\
\end{array}\right]\left[\begin{array}{c}
\delta \mathbf{v}_{P}^{{\mathcal{A}}}\\
\delta{\boldsymbol{{\omega}}}_P^{{\mathcal{A}}}\\
\hline \dot{\mathbf{q}}_f(t)\\
\end{array}\right]_{{\mathcal{R}_{a}}}+
\left[\begin{array}{c|c}
\mathbf{K}_{l} & \mathbf{K}_{r}\\
\end{array}\right]\left[\begin{array}{c}
\delta \mathbf{x}^{\mathcal{A}}_P\\
\delta \boldsymbol{{\Theta}}_P^{{\mathcal{A}}}\\
\hline {\mathbf{q}}_f(t)\\
\end{array}\right]_{{\mathcal{R}_{a}}}
\label{outputcomp}
\end{equation}

Finally, Eqs. \eqref{equivalent} and \eqref{outputcomp} can be transformed into a state-space representation of the linearized two-port beam model ${\left[\mathfrak{T}_{PC}^{\mathcal{A}}(\mathrm{s})\right]}_{\mathcal{R}_{a}}$, as follows:

\begin{equation}
\left[\begin{array}{c}
\left[\begin{array}{c}
\dot{\mathbf{q}}_f(t) \\
\ddot{\mathbf{q}}_f(t) \\
\end{array}\right]_{\mathcal{R}_{a}} \\
\hline
{{\left[\delta{\mathbf{m}}^{\mathcal{A}}_{C}\right]}_{\mathcal{R}_{a(C)}}} \\
{{\left[\delta\mathbf{W}_{\mathcal{A}/\bullet,{P}}\right]}_{\mathcal{R}_{a}}}
\end{array}\right]=\left[\begin{array}{c|c}
\mathbf{A} & \mathbf{B} \\
\hline \mathbf{C} & \mathbf{D}
\end{array}\right]\left[\begin{array}{c}
\left[\begin{array}{c}
{\mathbf{q}}_f(t) \\
\dot{\mathbf{q}}_f(t) \\
\end{array}\right]_{\mathcal{R}_{a}} \\
\hline
{{\left[\delta\mathbf{W}_{\bullet/\mathcal{A},{C}}\right]}_{\mathcal{R}_{a(C)}}} \\
{{\left[\delta{\mathbf{m}}^{\mathcal{A}}_{P}\right]}_{\mathcal{R}_{a}}}
\end{array}\right]
\label{ssinplane}
\end{equation}

In Eq. \eqref{ssinplane}, the matrices $\mathbf{A}$, $\mathbf{B}$, $\mathbf{C}$ and $\mathbf{D}$ are given by:

\begin{equation}
\begin{split}
& \mathbf{A}=\left[\begin{array}{cc}
\mathbf{0}_{(2N+2) \times (2N+2)} & \mathbf{I}_{(2N+2)} \\
-\mathbf{M}_{f f}^{-1}\mathbf{K}_{f f} & -\mathbf{M}_{f f}^{-1}\mathbf{D}_{f f}
\end{array}\right]_{\mathcal{R}_a} \\
& \mathbf{B}=\left[\begin{array}{cc}
\mathbf{0}_{(2N+2) \times 6} & \mathbf{0}_{(2N+2) \times 18} \\
\mathbf{M}_{f f}^{-1}\overline{\mathbf{{\mathbf{N}}}}_{ff} & -\mathbf{M}_{f f}^{-1} \left[\begin{array}{ccc}\mathbf{M}_{fr} & \mathbf{D}_{fr} & \mathbf{K}_{fr}\end{array}\right]
\end{array}\right]_{\mathcal{R}_a} \\
& \mathbf{C}=\left[\begin{array}{cc}
-\mathbf{M}_r\mathbf{M}_{f f}^{-1}\mathbf{K}_{f f}+\mathbf{K}_r & -\mathbf{M}_r\mathbf{M}_{f f}^{-1}\mathbf{D}_{f f}+\mathbf{D}_r \\
-\left({\overline{\mathbf{{\mathbf{N}}}}_{rr}}\right)^{-1}\left(\mathbf{M}_{rf}\mathbf{M}_{f f}^{-1}\mathbf{K}_{ff}-\mathbf{K}_{rf}\right) & -{\left(\overline{\mathbf{{\mathbf{N}}}}_{rr}\right)}^{-1}\left(\mathbf{M}_{rf}\mathbf{M}_{f f}^{-1}\mathbf{D}_{ff}-\mathbf{D}_{rf}\right)
\end{array}\right]_{\mathcal{R}_a} \\
& \mathbf{D}=\left[\begin{array}{cc}\mathbf{M}_{r}\mathbf{M}_{f f}^{-1}\overline{\mathbf{{\mathbf{N}}}}_{ff} & -\mathbf{M}_r\mathbf{M}_{f f}^{-1} \left[\begin{array}{ccc}\mathbf{M}_{fr} & \mathbf{D}_{fr} & \mathbf{K}_{fr}\end{array}\right]+\left[\begin{array}{ccc}\mathbf{M}_{l} & \mathbf{D}_{l} & \mathbf{K}_{l}\end{array}\right] \\
{\left(\overline{\mathbf{{\mathbf{N}}}}_{rr}\right)}^{-1}\left(\mathbf{M}_{rf}\mathbf{M}_{f f}^{-1}\overline{\mathbf{{\mathbf{N}}}}_{ff}-\overline{\mathbf{{\mathbf{N}}}}_{rf}\right) & {\left(\overline{\mathbf{{\mathbf{N}}}}_{rr}\right)}^{-1} \left[\begin{array}{ccc}\mathbf{M}_{rr} & \mathbf{D}_{rr} & \mathbf{K}_{rr}\end{array}\right]-{\left(\overline{\mathbf{{\mathbf{N}}}}_{rr}\right)}^{-1}\mathbf{M}_{rf}\mathbf{M}_{ff}^{-1}\left[\begin{array}{ccc}\mathbf{M}_{fr} & \mathbf{D}_{fr} & \mathbf{K}_{fr}\end{array}\right]
\end{array}\right]_{\mathcal{R}_a}
\end{split}
\label{statespace}
\end{equation}

All the matrices which are employed in this section to build the model of the spinning beam are defined in \ref{append}. In addition, a summary of all the steps that need to be followed to compute the state-space representation of the TITOP beam model is also depicted in Algorithm \ref{alg:cap}.

\begin{algorithm}
\caption{Computation of the TITOP beam model}\label{alg:cap}
{Define the beam's mechanical parameters: $\rho^{\mathcal{A}}$, $S^{\mathcal{A}}$, $l^{\mathcal{A}}$, $E^{\mathcal{A}}$, $\nu^{\mathcal{A}}$, $G^{\mathcal{A}}$, $J^{\mathcal{A}}_{px}$, $J^{\mathcal{A}}_y$ and $J^{\mathcal{A}}_z;$} 

{Define the equilibrium conditions: $\overline{\mathbf{x}}^{\mathcal{A}}_P$, $\overline{\mathbf{v}}_{P}^{{\mathcal{A}}}$, $\overline{\boldsymbol{\omega}}_P^{{\mathcal{A}}}$ and $\overline{\mathbf{W}}_{\bullet/\mathcal{A},{C}};$}  

{Compute $\mathbf{M}_{\mathcal{T}}$, $\mathbf{K}_{\mathcal{T}}$, $\mathbf{G}_{\mathcal{T}}$, $\mathbf{K}_{\mathcal{V}}$, $\mathbf{C}_{Q}$, $\mathbf{M}_{\mathcal{L}}$, $\mathbf{J}_{\mathcal{L}}$, $\mathbf{C}_{\mathcal{L}}$ and $\overline{\mathbf{N}};$} \Comment{see \ref{append}}

{Compute $\overline{\mathbf{W}}_{\mathcal{A}/\bullet,{P}}$ and $\overline{\mathbf{q}}_f$ from the equilibrium equations$;$} \Comment{see Eq. \eqref{equilibriium}}

 \eIf{$\overline{\mathbf{q}}_f$ $\gg \mathbf 0_{
 (2N+2) \times 1}$}{
   The model is not valid\; 
 }{

     {Compute $\mathbf{F}_c$, as well as the Lagrangian derivation$;$} \Comment{see \ref{append} and Eq. \eqref{cenas}}

     {Compute $\mathbf{M}_l$, $\mathbf{M}_r$, $\mathbf{D}_l$, $\mathbf{D}_r$, $\mathbf{K}_l$ and $\mathbf{K}_r;$} \Comment{see \ref{append}}
     
     {Compute the TITOP beam model$;$} \Comment{see Eqs. \eqref{ssinplane} and \eqref{statespace}}

 }

\end{algorithm}

\section{Model Validation}
\label{results}

\subsection{Frequency ratios of a spinning beam with tip mass and root offset}

\begin{figure}[!ht]
\centering
\includegraphics[width=0.65\textwidth]{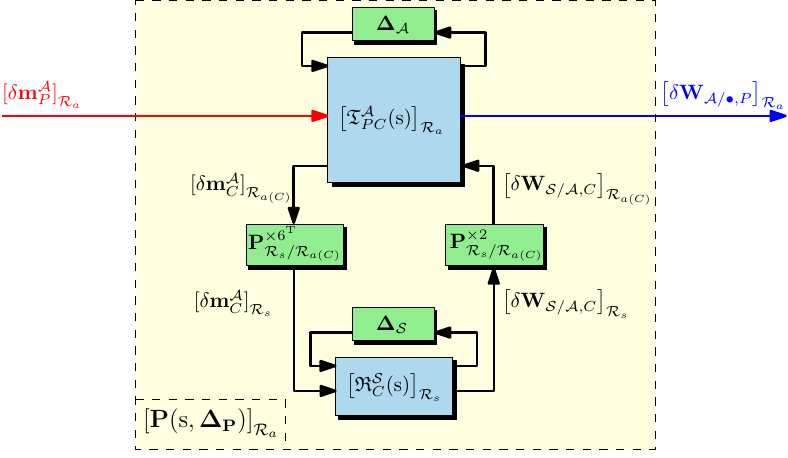}
\caption{Block-diagram representation of a model composed of a flexible beam and a tip mass.}
\label{beamplustip} 
\end{figure}

The simple scenario presented in Fig. \ref{psiexp}a involving a beam with a tip mass is considered in order to validate the TITOP model of a flexible spinning beam. The following parameters are used in this section: the rotation speed ratios $\eta^{\bullet}_{\bullet}$, which are defined differently for the four different types of dynamics, the dimensionless tip mass $\mu=\frac{m}{\rho^{\mathcal{A}} S^{\mathcal{A}} l^{\mathcal{A}}}$ and the dimensionless hub radius $\alpha=\frac{r}{l^{\mathcal{A}}}$. The rotation speed ratios $\eta^{\bullet}_{\bullet}$ are equal to $\eta_{b}^{y}=\Omega \sqrt{\frac{\rho^{\mathcal{A}} S^{\mathcal{A}} l^{\mathcal{A}^4}}{E^{\mathcal{A}} J^{\mathcal{A}}_{z}}}$ for in-plane bending (i.e., in the plane $\left(P, \mathbf{x}_a, \mathbf{y}_a\right)$), $\eta_{b}^{z}=\Omega \sqrt{\frac{\rho^{\mathcal{A}} S^{\mathcal{A}} l^{\mathcal{A}^4}}{E^{\mathcal{A}} J^{\mathcal{A}}_{y}}}$ for out-of-plane bending (i.e., in the plane $\left(P, \mathbf{x}_a, \mathbf{z}_a\right)$), $\eta_{t}^{x}=\Omega \sqrt{\frac{\rho^{\mathcal{A}} l^{\mathcal{A}^2}}{E^{\mathcal{A}}}}$ for traction and $\eta_{r}^{x}=\Omega \sqrt{\frac{\rho^{\mathcal{A}} {l^{\mathcal{A}^2}}}{G^{\mathcal{A}}}}$ for torsion. A variety of results is tabulated for general use in Tables \ref{table1}, \ref{table2} and \ref{table3}, which considers uniform beams with fixed root boundary conditions for different values of tip mass, root offset and angular velocity $\Omega$. To generate these tables, the model displayed in Fig. \ref{beamplustip} is considered. This model involves a beam $\mathcal{A}$ connected to a tip mass $\mathcal{S}$, where the beam is clamped at the point $P$ to a massless rotor. The procedure for determining the equilibrium conditions in the beam model is similar to the one illustrated in Fig. \ref{psiexp}a and explained in section \ref{beam}, since it is once again assumed that $\overline{\mathbf{x}}^{\mathcal{A}}_P=\left[ r , 0 , 0\right]^{\mathrm{T}}$, $\overline{\boldsymbol{{\Theta}}}_P^{{\mathcal{A}}}=\mathbf{0}_{3\times 1}$ and $\overline{\boldsymbol{\omega}}_P^{{\mathcal{A}}}=\left[ 0 , 0 , \Omega\right]^{\mathrm{T}}$. Finally, it is also considered that $\mathbf{P}_{\mathcal{R}_{s}/\mathcal{R}_{a(C)}}=\mathbf{I}_{3}$ and $m=m^{\mathcal{S}}$. 

Given that Tables \ref{table1}, \ref{table2} and \ref{table3} present dimensionless results, it is crucial to emphasize that different combinations of beam parameters, such as length and Young's modulus, must yield identical frequency ratios to those depicted in the tables. However, the choice of beam parameters and rotation speed ratios, denoted as $\eta^{\bullet}_{\bullet}$, can lead to situations where the resulting $\Omega$ may challenge the validity of the beam model. This can be explained by the fact that it could lead to scenarios in which the beam's deformation assumptions no longer hold. For the specific scenario presented here, the geometrical configuration displayed in Fig. \ref{psiexp}a is being considered. For this reason, $\overline{\mathbf{W}}_{\mathcal{S}/\mathcal{A},{C}}=\left[m^{\mathcal{S}}\left(l^{\mathcal{A}}+r\right)\Omega^2 , 0 , 0 , 0 , 0 , 0\right]^{\mathrm{T}}$. Since $\overline{\mathbf{W}}_{\mathcal{S}/\mathcal{A},{C}}\{1\} \neq 0$, very big values of $\Omega$ and $m^{\mathcal{S}}$ might lead to $\overline{\Delta u} \gg 0$, depending on the chosen beam parameters, which would invalidate the beam model.


One of the objectives of this paper is also to demonstrate how to model a system parameterized according to the angular velocity $\Omega$. Therefore, $\Omega$ is considered as an uncertain parameter, as follows:

\begin{equation}
\Omega=\Omega_{0}(1+r_{\Omega}\delta_{\Omega}) 
\label{angunc}
\end{equation}

In Eq. \eqref{angunc}, $\Omega_{0}$ represents the nominal angular velocity, $r_{\Omega}$ is used to set the maximum percent of variation for $\Omega$ and $\delta_{\Omega}  \in [-1, 1]$ is a normalized real uncertainty. The mass of the body $\mathcal{S}$ (tip mass) can also be considered uncertain, with $m^{\mathcal{S}}=m^{\mathcal{S}}_0(1+r_{m^{\mathcal{S}}}\delta_{m^{\mathcal{S}}})$. Therefore, the structured uncertainty blocks (displayed in Fig. \ref{beamplustip}) characterizing the flexible boom and the tip mass are equal to $\boldsymbol{\Delta}_{\mathcal{A}}=\operatorname{diag}\left(\delta_{\Omega}\textbf{I}_{169}, \delta_{m^{\mathcal{S}}}\textbf{I}_{64}\right)$ and $\boldsymbol{\Delta}_{\mathcal{S}}=\operatorname{diag}\left(\delta_{\Omega}\textbf{I}_{4}, \delta_{m^{\mathcal{S}}}\textbf{I}_{5}\right)$, respectively. Lastly, it should also be noted that the number of computed occurrences for the different uncertainties depends on the equilibrium conditions computation of the system.

\begin{table}[!ht]
\setlength{\tabcolsep}{4pt}
\footnotesize
\centering
\caption{Frequency ratios of a uniform cantilever beam with zero root offset and zero tip mass; $\alpha=0$, $\mu=0$.}
\label{table1}
\begin{NiceTabular}{c|cccccccccc|}[cell-space-limits=3pt]
\cline{2-11}
&
\multicolumn{10}{c|}{Frequency ratios of the flexible modes} \\
 \cline{1-11} 
 \multicolumn{1}{|c|}{\multirow{2}{*}{$\eta^{\bullet}_{\bullet}$}} &
 \multicolumn{4}{c|}{\textbf{In-plane bending}} &
 \multicolumn{4}{c|}{\textbf{Out-of-plane bending}} &
 \multicolumn{1}{c|}{\textbf{Traction}} &
 \textbf{Torsion} \\  \cline{2-11} 
\multicolumn{1}{|c|}{} &
 \multicolumn{1}{c|}{$1^{st}$} &
 \multicolumn{1}{c|}{$2^{nd}$} &
 \multicolumn{1}{c|}{$3^{rd}$} &
 \multicolumn{1}{c|}{$4^{th}$} &
 \multicolumn{1}{c|}{$1^{st}$} &
 \multicolumn{1}{c|}{$2^{nd}$} &
 \multicolumn{1}{c|}{$3^{rd}$} &
 \multicolumn{1}{c|}{$4^{th}$} &
 \multicolumn{1}{c|}{$1^{st}$} &
 $1^{st}$ \\ \hline
\multicolumn{1}{|c|}{0} &
\multicolumn{1}{c|}{3.5160} &
 \multicolumn{1}{c|}{22.1578} &
 \multicolumn{1}{c|}{63.3466} &
 \multicolumn{1}{c|}{281.5963} &
 \multicolumn{1}{c|}{3.5160} &
 \multicolumn{1}{c|}{22.1578} &
 \multicolumn{1}{c|}{63.3466} &
 \multicolumn{1}{c|}{281.5963} &
 \multicolumn{1}{c|}{1732} &
  1719 \\ \hline
\multicolumn{1}{|c|}{2} &
 \multicolumn{1}{c|}{3.6218} &
 \multicolumn{1}{c|}{22.6384} &
 \multicolumn{1}{c|}{63.8495} &
 \multicolumn{1}{c|}{281.7849} &
 \multicolumn{1}{c|}{4.1373} &
 \multicolumn{1}{c|}{22.7266} &
 \multicolumn{1}{c|}{63.8809} &
 \multicolumn{1}{c|}{281.7920} &
 \multicolumn{1}{c|}{1732} &
 1719 \\ \hline
\multicolumn{1}{|c|}{4} &
 \multicolumn{1}{c|}{3.8978} &
 \multicolumn{1}{c|}{24.0256} &
 \multicolumn{1}{c|}{65.3351} &
 \multicolumn{1}{c|}{282.3501} &
 \multicolumn{1}{c|}{5.5852} &
 \multicolumn{1}{c|}{24.3564} &
 \multicolumn{1}{c|}{65.4575} &
 \multicolumn{1}{c|}{282.3785} &
 \multicolumn{1}{c|}{1732} &
 1719 \\ \hline
\multicolumn{1}{|c|}{6} &
 \multicolumn{1}{c|}{4.2643} &
 \multicolumn{1}{c|}{26.1815} &
 \multicolumn{1}{c|}{67.7383} &
 \multicolumn{1}{c|}{283.2900} &
 \multicolumn{1}{c|}{7.3614} &
 \multicolumn{1}{c|}{26.8602} &
 \multicolumn{1}{c|}{68.0035} &
 \multicolumn{1}{c|}{283.3535} &
 \multicolumn{1}{c|}{1732} &
 1719 \\ \hline
\multicolumn{1}{|c|}{8} &
 \multicolumn{1}{c|}{4.6626} &
 \multicolumn{1}{c|}{28.9382} &
 \multicolumn{1}{c|}{70.9651} &
 \multicolumn{1}{c|}{284.6012} &
 \multicolumn{1}{c|}{9.2602} &
 \multicolumn{1}{c|}{30.0237} &
 \multicolumn{1}{c|}{71.4147} &
 \multicolumn{1}{c|}{284.7137} &
 \multicolumn{1}{c|}{1732} &
1719  \\ \hline
\multicolumn{1}{|c|}{10} &
 \multicolumn{1}{c|}{5.0653} &
 \multicolumn{1}{c|}{32.1424} &
 \multicolumn{1}{c|}{74.9079} &
 \multicolumn{1}{c|}{286.2793} &
 \multicolumn{1}{c|}{11.2106} &
 \multicolumn{1}{c|}{33.6622} &
 \multicolumn{1}{c|}{75.5725} &
 \multicolumn{1}{c|}{286.4540} &
 \multicolumn{1}{c|}{1732} &
1719  \\ \hline
\end{NiceTabular}
\end{table}

\begin{table}[!ht]
\setlength{\tabcolsep}{4pt}
\footnotesize
\centering
\caption{Frequency ratios of a uniform cantilever beam with zero tip mass; $\alpha=1$, $\mu=0$.}
\label{table2}
\begin{NiceTabular}{c|cccccccccc|}[cell-space-limits=3pt]
\cline{2-11}
&
\multicolumn{10}{c|}{Frequency ratios of the flexible modes} \\
  \cline{1-11} 
 \multicolumn{1}{|c|}{\multirow{2}{*}{$\eta^{\bullet}_{\bullet}$}} &
 \multicolumn{4}{c|}{\textbf{In-plane bending}} &
 \multicolumn{4}{c|}{\textbf{Out-of-plane bending}} &
 \multicolumn{1}{c|}{\textbf{Traction}} &
 \textbf{Torsion} \\  \cline{2-11} 
\multicolumn{1}{|c|}{} &
 \multicolumn{1}{c|}{$1^{st}$} &
 \multicolumn{1}{c|}{$2^{nd}$} &
 \multicolumn{1}{c|}{$3^{rd}$} &
 \multicolumn{1}{c|}{$4^{th}$} &
 \multicolumn{1}{c|}{$1^{st}$} &
 \multicolumn{1}{c|}{$2^{nd}$} &
 \multicolumn{1}{c|}{$3^{rd}$} &
 \multicolumn{1}{c|}{$4^{th}$} &
 \multicolumn{1}{c|}{$1^{st}$} &
 $1^{st}$ \\ \hline
\multicolumn{1}{|c|}{0} &
\multicolumn{1}{c|}{3.5160} &
 \multicolumn{1}{c|}{22.1578} &
 \multicolumn{1}{c|}{63.3466} &
 \multicolumn{1}{c|}{281.5963} &
 \multicolumn{1}{c|}{3.5160} &
 \multicolumn{1}{c|}{22.1578} &
 \multicolumn{1}{c|}{63.3466} &
 \multicolumn{1}{c|}{281.5963} &
 \multicolumn{1}{c|}{1732} &
 1719 \\ \hline
\multicolumn{1}{|c|}{2} &
 \multicolumn{1}{c|}{4.4006} &
 \multicolumn{1}{c|}{23.3799} &
 \multicolumn{1}{c|}{64.5881} &
 \multicolumn{1}{c|}{282.0294} &
 \multicolumn{1}{c|}{4.8339} &
 \multicolumn{1}{c|}{23.4653} &
 \multicolumn{1}{c|}{64.6190} &
 \multicolumn{1}{c|}{282.0365} &
 \multicolumn{1}{c|}{1732} &
 1719 \\ \hline
\multicolumn{1}{|c|}{4} &
 \multicolumn{1}{c|}{6.3165} &
 \multicolumn{1}{c|}{26.7142} &
 \multicolumn{1}{c|}{68.1767} &
 \multicolumn{1}{c|}{283.3255} &
 \multicolumn{1}{c|}{7.4768} &
 \multicolumn{1}{c|}{27.0120} &
 \multicolumn{1}{c|}{68.2939} &
 \multicolumn{1}{c|}{283.3538} &
 \multicolumn{1}{c|}{1732} &
 1719 \\ \hline
\multicolumn{1}{|c|}{6} &
 \multicolumn{1}{c|}{8.5561} &
 \multicolumn{1}{c|}{31.4937} &
 \multicolumn{1}{c|}{73.7678} &
 \multicolumn{1}{c|}{285.4753} &
 \multicolumn{1}{c|}{10.4513} &
 \multicolumn{1}{c|}{32.0602} &
 \multicolumn{1}{c|}{74.0115} &
 \multicolumn{1}{c|}{285.5384} &
 \multicolumn{1}{c|}{1732} &
 1719 \\ \hline
\multicolumn{1}{|c|}{8} &
 \multicolumn{1}{c|}{10.9044} &
 \multicolumn{1}{c|}{37.1564} &
 \multicolumn{1}{c|}{80.9404} &
 \multicolumn{1}{c|}{288.4635} &
 \multicolumn{1}{c|}{13.5265} &
 \multicolumn{1}{c|}{38.0080} &
 \multicolumn{1}{c|}{81.3349} &
 \multicolumn{1}{c|}{288.5745} &
 \multicolumn{1}{c|}{1732} &
 1719 \\ \hline
\multicolumn{1}{|c|}{10} &
 \multicolumn{1}{c|}{13.2997} &
 \multicolumn{1}{c|}{43.3492} &
 \multicolumn{1}{c|}{89.3024} &
 \multicolumn{1}{c|}{292.2694} &
 \multicolumn{1}{c|}{16.6442} &
 \multicolumn{1}{c|}{44.4878} &
 \multicolumn{1}{c|}{89.8606} &
 \multicolumn{1}{c|}{292.4405} &
 \multicolumn{1}{c|}{1732} &
 1719 \\ \hline
\end{NiceTabular}
\end{table}

\begin{table}[!ht]
\setlength{\tabcolsep}{4pt}
\footnotesize
\centering
\caption{Frequency ratios of a uniform cantilever beam with zero root offset; $\alpha=0$, $\mu=1$.}
\label{table3}
\begin{NiceTabular}{c|cccccccccc|}[cell-space-limits=3pt]
\cline{2-11}
&
\multicolumn{10}{c|}{Frequency ratios of the flexible modes} \\
\cline{1-11} 
 \multicolumn{1}{|c|}{\multirow{2}{*}{$\eta^{\bullet}_{\bullet}$}} &
 \multicolumn{4}{c|}{\textbf{In-plane bending}} &
 \multicolumn{4}{c|}{\textbf{Out-of-plane bending}} &
 \multicolumn{1}{c|}{\textbf{Traction}} &
 \textbf{Torsion} \\  \cline{2-11} 
\multicolumn{1}{|c|}{} &
 \multicolumn{1}{c|}{$1^{st}$} &
 \multicolumn{1}{c|}{$2^{nd}$} &
 \multicolumn{1}{c|}{$3^{rd}$} &
 \multicolumn{1}{c|}{$4^{th}$} &
 \multicolumn{1}{c|}{$1^{st}$} &
 \multicolumn{1}{c|}{$2^{nd}$} &
 \multicolumn{1}{c|}{$3^{rd}$} &
 \multicolumn{1}{c|}{$4^{th}$} &
 \multicolumn{1}{c|}{$1^{st}$} &
 $1^{st}$ \\ \hline
\multicolumn{1}{|c|}{0} &
\multicolumn{1}{c|}{1.5573} &
 \multicolumn{1}{c|}{16.2709} &
 \multicolumn{1}{c|}{51.6537} &
 \multicolumn{1}{c|}{184.8364} &
 \multicolumn{1}{c|}{1.5573} &
 \multicolumn{1}{c|}{16.2709} &
 \multicolumn{1}{c|}{51.6537} &
 \multicolumn{1}{c|}{184.8364} &
 \multicolumn{1}{c|}{866.03} &
 1719 \\ \hline
\multicolumn{1}{|c|}{2} &
\multicolumn{1}{c|}{1.7682} &
 \multicolumn{1}{c|}{18.0920} &
 \multicolumn{1}{c|}{53.7584} &
 \multicolumn{1}{c|}{186.4674} &
 \multicolumn{1}{c|}{2.6697} &
 \multicolumn{1}{c|}{18.2022} &
 \multicolumn{1}{c|}{53.7956} &
 \multicolumn{1}{c|}{186.4781} &
 \multicolumn{1}{c|}{866.03} &
 1719 \\ \hline
\multicolumn{1}{|c|}{4} &
 \multicolumn{1}{c|}{2.1544} &
 \multicolumn{1}{c|}{22.6744} &
 \multicolumn{1}{c|}{59.6400} &
 \multicolumn{1}{c|}{191.2771} &
 \multicolumn{1}{c|}{4.5435} &
 \multicolumn{1}{c|}{23.0245} &
 \multicolumn{1}{c|}{59.7740} &
 \multicolumn{1}{c|}{191.3189} &
 \multicolumn{1}{c|}{866.05} &
 1719 \\ \hline
\multicolumn{1}{|c|}{6} &
 \multicolumn{1}{c|}{2.5325} &
 \multicolumn{1}{c|}{28.6878} &
 \multicolumn{1}{c|}{68.3475} &
 \multicolumn{1}{c|}{199.0362} &
 \multicolumn{1}{c|}{6.5132} &
 \multicolumn{1}{c|}{29.3086} &
 \multicolumn{1}{c|}{68.6104} &
 \multicolumn{1}{c|}{199.1266} &
 \multicolumn{1}{c|}{866.09} &
 1719 \\ \hline
\multicolumn{1}{|c|}{8} &
 \multicolumn{1}{c|}{2.8898} &
 \multicolumn{1}{c|}{35.3872} &
 \multicolumn{1}{c|}{78.9597} &
 \multicolumn{1}{c|}{209.4181} &
 \multicolumn{1}{c|}{8.5070} &
 \multicolumn{1}{c|}{36.2802} &
 \multicolumn{1}{c|}{79.3640} &
 \multicolumn{1}{c|}{209.5709} &
 \multicolumn{1}{c|}{866.14} &
 1719 \\ \hline
\multicolumn{1}{|c|}{10} &
 \multicolumn{1}{c|}{3.2380} &
 \multicolumn{1}{c|}{42.4412} &
 \multicolumn{1}{c|}{90.8126} &
 \multicolumn{1}{c|}{222.0561} &
 \multicolumn{1}{c|}{10.5128} &
 \multicolumn{1}{c|}{43.6035} &
 \multicolumn{1}{c|}{91.3616} &
 \multicolumn{1}{c|}{222.2812} &
 \multicolumn{1}{c|}{866.20} &
1719  \\ \hline
\end{NiceTabular}
\end{table}

\subsection{Comparison with other solutions and NASTRAN}

It is also very useful and important to compare the results obtained from the TITOP methodology presented in this paper with other solutions, as well as the results of a conventional finite element code. Table \ref{comp} provides such a comparison for the simple case of a uniform cantilever beam with no root offset or tip mass, where $f_{b\bullet}^{z}$ and $f_{b\bullet}^{y}$ represent the out-of-plane and in-plane frequency ratios, respectively. The method of Frobenius is employed to determine the exact out-of-plane frequencies for rotating beams, where both the flexural rigidity and mass distribution vary linearly \cite{wright}. The exact solutions were not found anywhere in the literature for the in-plane dynamics, but this table also provides a comparison of the results with the outputs of a finite element program, NASTRAN.  As can be seen from Table \ref{comp}, the TITOP approach is extremely accurate even at relatively large rotation rates and does not appear to suffer a significant decrease in accuracy for the higher modes, as long as enough beam elements in series are considered. Indeed, the TITOP method has the versatility of finite elements and can be made to converge to the exact solution by increasing the number of elements, as is the
case with the conventional finite element method. The NASTRAN results, which can provide very reasonable estimates for
many applications, were obtained using 20 beam elements (CBEAM). Furthermore, an Implicit Nonlinear analysis was carried out (SOL 400). The comparison displayed in Table \ref{comp}, which is conducted for a uniform cantilever beam with rotation speed ratios of $\eta_{b}^{\bullet}=0, 3, 6, 12$, also reveals that the calculations presented in this paper closely match the Exact/NASTRAN computations for the first two frequency ratios, even when considering only a single TITOP beam element. It is also evident from the data in Table \ref{comp} that, as the rotation speed ratios increase, the frequency ratios rise, due to the centrifugal stiffening effect. Nevertheless, it is worth noting that the in-plane frequency ratios exhibit a slower growth rate compared to the out-of-plane frequency ratios. This phenomenon occurs because the in-plane dynamics also encompass a centrifugal softening effect, which results in a reduction in the natural frequencies as the speed increases.

\begin{table}[]
\setlength{\tabcolsep}{3.4pt}
\footnotesize
\centering
\caption{Comparison of exact frequency ratios with approximate estimates for the uniform cantilever beam with zero root offset and zero tip mass; $\alpha=0$, $\mu=0$.}
\label{comp}
\begin{NiceTabular}{|c|l|cccc|lclclcllcc}[cell-space-limits=3pt]
\cline{1-6} \cline{8-16}
\multirow{2}{*}{\begin{tabular}[c]{@{}c@{}}\textbf{Freq.} \\ \textbf{ratio}\end{tabular}} &
 \multicolumn{1}{c|}{\multirow{2}{*}{\textbf{Method}}} &
 \multicolumn{4}{c|}{\begin{tabular}[c]{@{}c@{}}\textbf{Rotation} \\ \textbf{speed ratio}\end{tabular}} &
 \multicolumn{1}{l|}{} &
 \multicolumn{1}{l|}{\multirow{2}{*}{\begin{tabular}[c]{@{}l@{}}\textbf{Freq.} \\ \textbf{ratio}\end{tabular}}} &
 \multicolumn{1}{c|}{\multirow{2}{*}{\textbf{Method}}} &
 \multicolumn{7}{c|}{\begin{tabular}[c]{@{}c@{}}\textbf{Rotation} \\ \textbf{speed ratio}\end{tabular}} \\ \cline{3-6} \cline{10-16} 
&
  \multicolumn{1}{c|}{} &
 \multicolumn{1}{c|}{$\eta_{b}^{z}=0$} &
 \multicolumn{1}{c|}{$\eta_{b}^{z}=3$} &
 \multicolumn{1}{c|}{$\eta_{b}^{z}=6$} &
 $\eta_{b}^{z}=12$ &
 \multicolumn{1}{l|}{} &
 \multicolumn{1}{l|}{} &
 \multicolumn{1}{c|}{} &
 \multicolumn{2}{c|}{$\eta_{b}^{y}=0$} &
 \multicolumn{3}{c|}{$\eta_{b}^{y}=3$} &
 \multicolumn{1}{c|}{$\eta_{b}^{y}=6$} &
 \multicolumn{1}{c|}{$\eta_{b}^{y}=12$} \\ \cline{1-6} \cline{8-16} 
\multirow{4}{*}{$f_{b1}^{z}$} &
 Exact \cite{wright} &
 \multicolumn{1}{c|}{3.5160} &
 \multicolumn{1}{c|}{4.7973} &
 \multicolumn{1}{c|}{7.3604} &
 13.1702  &
 \multicolumn{1}{l|}{} &
 \multicolumn{1}{c|}{\multirow{3}{*}{$f_{b1}^{y}$}} &
 \multicolumn{1}{l|}{TITOP beam (1 el.)} &
 \multicolumn{2}{c|}{3.5160} &
 \multicolumn{3}{c|}{3.7435} &
 \multicolumn{1}{c|}{4.2643} &
 \multicolumn{1}{c|}{5.4640} \\ \cline{2-6} \cline{9-16} 
&
 TITOP beam (1 el.) &
 \multicolumn{1}{c|}{3.5160} &
 \multicolumn{1}{c|}{4.7974} &
 \multicolumn{1}{c|}{7.3614} &
  13.1868 &
 \multicolumn{1}{l|}{} &
 \multicolumn{1}{c|}{} &
 \multicolumn{1}{l|}{TITOP beam (5 el.)} &
 \multicolumn{2}{c|}{3.5160} &
 \multicolumn{3}{c|}{3.7434} &
 \multicolumn{1}{c|}{4.2625} &
 \multicolumn{1}{c|}{5.4233} \\ \cline{2-6} \cline{9-16} 
&
 TITOP beam (5 el.) &
 \multicolumn{1}{c|}{3.5160} &
 \multicolumn{1}{c|}{4.7973} &
 \multicolumn{1}{c|}{7.3604} &
  13.1702 &
 \multicolumn{1}{l|}{} &
 \multicolumn{1}{c|}{} &
 \multicolumn{1}{l|}{NASTRAN (20 el.)} &
 \multicolumn{2}{c|}{3.5118} &
 \multicolumn{3}{c|}{3.7395} &
 \multicolumn{1}{c|}{4.2595} &
 \multicolumn{1}{c|}{5.4264} \\ \cline{2-6} \cline{8-16} 
&
 NASTRAN (20 el.) &
 \multicolumn{1}{c|}{3.5118} &
 \multicolumn{1}{c|}{4.7941} &
 \multicolumn{1}{c|}{7.3582} &
 13.1699 &
 \multicolumn{1}{l|}{} &
 \multicolumn{1}{c|}{\multirow{3}{*}{$f_{b2}^{y}$}} &
 \multicolumn{1}{l|}{TITOP beam (1 el.)} &
 \multicolumn{2}{c|}{22.1578} &
 \multicolumn{3}{c|}{23.2260} &
 \multicolumn{1}{c|}{26.1815} &
 \multicolumn{1}{c|}{35.6729} \\ \cline{1-6} \cline{9-16} 
\multirow{4}{*}{$f_{b2}^{z}$} &
 Exact \cite{wright} &
 \multicolumn{1}{c|}{22.0345} &
 \multicolumn{1}{c|}{23.3203 } &
 \multicolumn{1}{c|}{26.8091} &
 37.6031 &
 \multicolumn{1}{l|}{} &
 \multicolumn{1}{c|}{} &
 \multicolumn{1}{l|}{TITOP beam (5 el.)} &
 \multicolumn{2}{c|}{22.0345} &
 \multicolumn{3}{c|}{23.1263} &
 \multicolumn{1}{c|}{26.1284} &
 \multicolumn{1}{c|}{35.6338} \\ \cline{2-6} \cline{9-16} 
&
 TITOP beam (1 el.) &
 \multicolumn{1}{c|}{22.1578} &
 \multicolumn{1}{c|}{23.4189} &
 \multicolumn{1}{c|}{26.8602} &
 37.6372 &
 \multicolumn{1}{l|}{} &
 \multicolumn{1}{c|}{} &
 \multicolumn{1}{l|}{NASTRAN (20 el.)} &
 \multicolumn{2}{c|}{21.9402} &
 \multicolumn{3}{c|}{23.0296} &
 \multicolumn{1}{c|}{26.0246} &
 \multicolumn{1}{c|}{35.5066} \\ \cline{2-6} \cline{8-16} 
&
 TITOP beam (5 el.) &
 \multicolumn{1}{c|}{22.0345} &
 \multicolumn{1}{c|}{23.3203} &
 \multicolumn{1}{c|}{26.8091} &
 37.6031 &
 \multicolumn{1}{l|}{} &
 \multicolumn{1}{c|}{\multirow{3}{*}{$f_{b3}^{y}$}} &
 \multicolumn{1}{l|}{TITOP beam (1 el.)} &
 \multicolumn{2}{c|}{63.3466} &
 \multicolumn{3}{c|}{64.4727} &
 \multicolumn{1}{c|}{67.7383} &
 \multicolumn{1}{c|}{79.4582} \\ \cline{2-6} \cline{9-16} 
&
 NASTRAN (20 el.) &
 \multicolumn{1}{c|}{21.9402} &
 \multicolumn{1}{c|}{23.2242} &
 \multicolumn{1}{c|}{26.7073} &
 37.4796 &
 \multicolumn{1}{l|}{} &
 \multicolumn{1}{c|}{} &
 \multicolumn{1}{l|}{TITOP beam (5 el.)} &
 \multicolumn{2}{c|}{61.6973} &
 \multicolumn{3}{c|}{62.9134} &
 \multicolumn{1}{c|}{66.4130} &
 \multicolumn{1}{c|}{78.7026} \\ \cline{1-6} \cline{9-16} 
\multirow{4}{*}{$f_{b3}^{z}$} &
 Exact \cite{wright} &
 \multicolumn{1}{c|}{61.6972} &
 \multicolumn{1}{c|}{62.9850} &
 \multicolumn{1}{c|}{66.6840} &
  79.6145&
 \multicolumn{1}{l|}{} &
 \multicolumn{1}{c|}{} &
 \multicolumn{1}{l|}{NASTRAN (20 el.)} &
 \multicolumn{2}{c|}{61.2502} &
 \multicolumn{3}{c|}{62.4602} &
 \multicolumn{1}{c|}{65.9411} &
 \multicolumn{1}{c|}{78.1579} \\ \cline{2-6} \cline{8-16} 
&
 TITOP beam (1 el.) &
 \multicolumn{1}{c|}{63.3466} &
 \multicolumn{1}{c|}{64.5425} &
 \multicolumn{1}{c|}{68.0035} &
  80.3593 &
  &
 \multicolumn{1}{l}{} &
  &
 \multicolumn{1}{l}{} &
  &
 \multicolumn{1}{l}{} &
  &
  &
 \multicolumn{1}{l}{} &
 \multicolumn{1}{l}{} \\ \cline{2-6}
&
 TITOP beam (5 el.) &
 \multicolumn{1}{c|}{61.6973} &
 \multicolumn{1}{c|}{62.9850} &
 \multicolumn{1}{c|}{66.6840} &
  79.6146 &
  &
 \multicolumn{1}{l}{} &
  &
 \multicolumn{1}{l}{} &
  &
 \multicolumn{1}{l}{} &
  &
  &
 \multicolumn{1}{l}{} &
 \multicolumn{1}{l}{} \\ \cline{2-6}
&
 NASTRAN (20 el.) &
 \multicolumn{1}{c|}{61.2502} &
 \multicolumn{1}{c|}{62.5322} &
 \multicolumn{1}{c|}{66.2135} &
 79.0738 &
  &
 \multicolumn{1}{l}{} &
  &
 \multicolumn{1}{l}{} &
  &
 \multicolumn{1}{l}{} &
  &
  &
 \multicolumn{1}{l}{} &
 \multicolumn{1}{l}{} \\ \cline{1-6}
\end{NiceTabular}
\end{table}

\begin{figure}[!ht]
\centering
 \includegraphics[width=1\textwidth]{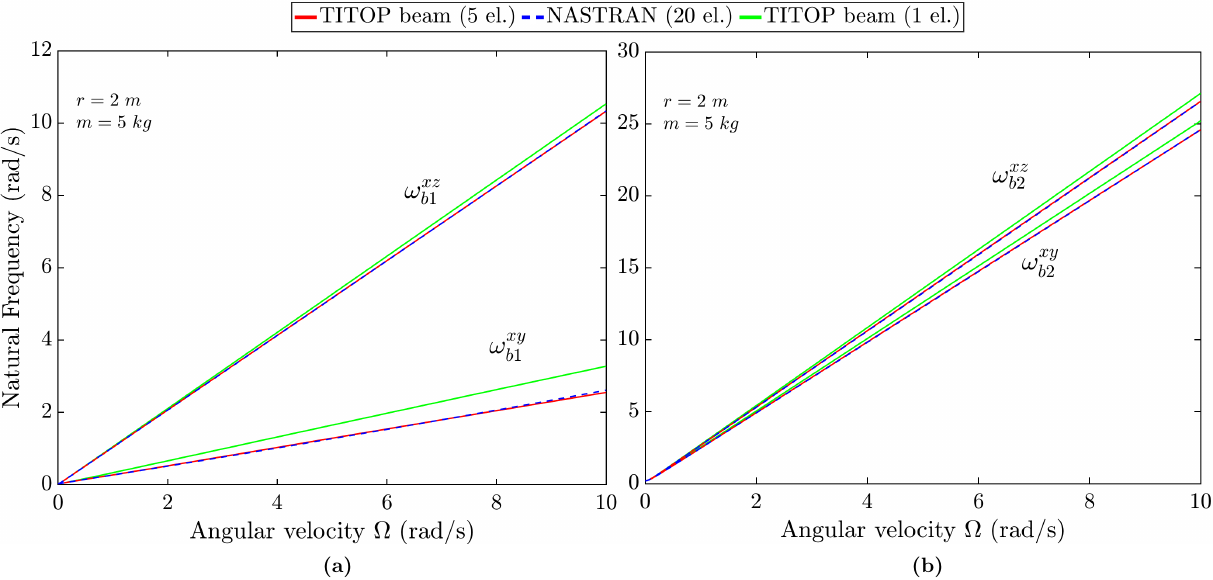}
\caption{Campbell diagrams for a system composed of a flexible beam and a tip mass: (a) evolution of the first in-plane and out-of-plane natural frequencies $\omega_{b1}^{\bullet}$ with respect to $\Omega$. (b) evolution of the second in-plane and out-of-plane natural frequencies $\omega_{b2}^{\bullet}$ with respect to $\Omega$.}
\label{campbell} 
\end{figure}

Let us now consider the system displayed in Fig. \ref{beamplustip}, composed of a beam and a tip mass, where the beam and tip mass parameters displayed in Table \ref{tab:Sat_prop} are used to define the system. Furthermore, $\overline{\mathbf{x}}^{\mathcal{A}}_P=\left[ r, 0 , 0\right]^{\mathrm{T}}$, with $r=2$ \si{\meter}, $\overline{\boldsymbol{\Theta}}_P^{{\mathcal{A}}}=\mathbf{0}_{3\times 1}$, $\overline{{\boldsymbol{\omega}}}_P^{{\mathcal{A}}}=\left[0 , 0 , \Omega\right]^{\mathrm{T}}$ and $m=m^{\mathcal{S}}=5$ \si{\kilogram}. Fig. \ref{campbell} displays two Campbell diagrams regarding this model, where $\omega_{b\bullet}^{z}$ and $\omega_{b\bullet}^{y}$ represent the out-of-plane and in-plane natural frequencies, respectively. These plots depict how the first and second natural frequencies for both the in-plane and out-of-plane dynamics change as the angular velocity $\Omega$ increases. Once more, the influence of the centrifugal stiffening effect is clearly visible. Additionally, it is also noteworthy that, as the number of TITOP beam elements used to model the system is increased, the natural frequencies converge to the results obtained from NASTRAN, since the red and dashed blue lines overlap.

\section{Case study: Spinning and flexible spacecraft}
\label{casestudy}

In-situ plasma measurements are used as a crucial tool for examining the characteristics of plasma in space. In order to conduct such measurements, spinning spacecraft platforms are commonly employed due to their numerous advantages. A primary benefit of utilizing a spinning spacecraft is the ability to achieve more consistent sampling of the plasma surroundings. By spinning, the spacecraft allows plasma instruments to capture plasma properties in all directions, thereby obtaining a more comprehensive understanding of the plasma environment \cite{thor}. Another advantage of employing a spinning spacecraft platform is its capacity to mitigate spacecraft charging effects. When a spacecraft remains stationary, it can accumulate a negative charge caused by electron buildup on its surface. This charge can disrupt in-situ plasma measurements by causing the spacecraft to repel the plasma. However, when a spacecraft spins, the charge is distributed more evenly across its surface, ultimately reducing the overall charging effect. Furthermore, spinning spacecraft platforms aid in diminishing instrument noise, which can stem from various sources, including electromagnetic interference and thermal fluctuations. By spinning the spacecraft, the noise can be averaged out, resulting in more precise and reliable measurements.

\subsection{Complete model of the system}

\begin{figure}[!ht]
\centering
 \includegraphics[width=1\textwidth]{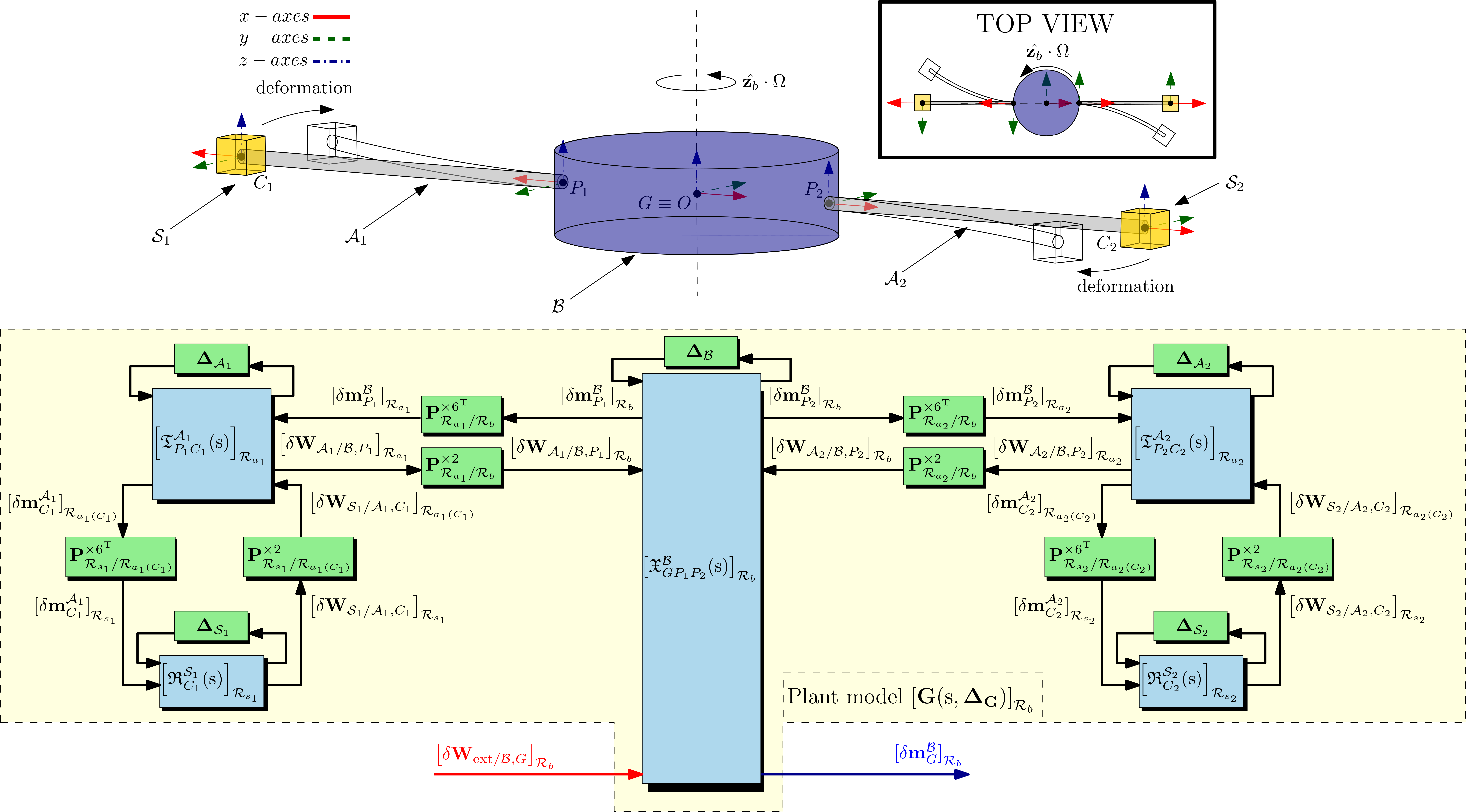}
\caption{3D representation and detailed block-diagram representation of a spinning satellite mission scenario composed of one main rigid body, two flexible booms and two tip masses representing two measurement instruments attached to the tip of each boom (Note: for the sake of simplicity, the
x-axes are displayed in solid red lines, the y-axes in dashed green lines and the z-axes in dash-dotted blue lines).}
\label{studycasefullmodel} 
\end{figure}

For the mission scenario being studied in this paper, a spinning spacecraft is considered, as depicted in Fig. \ref{studycasefullmodel}. The shape of this spacecraft is inspired by the THOR and Cluster missions. The central cylinder $\mathcal{B}$ represents the main equipment platform. Two long rod-shaped and flexible booms $\mathcal{A}_{\bullet}$ are attached to the rigid hub of the spacecraft $\mathcal{B}$ and operate when the spacecraft begins to spin. Their function can involve the measurement of fluctuating electrical and magnetic fields in the vicinity of the spacecraft, as demonstrated in the Cluster mission. Alternatively, they may serve as carriers for plasma measurement instruments located at their tips, as exemplified in the THOR project. Within the scope of this paper, the primary focus is directed towards the second scenario, where the measurement instruments are regarded as tip masses.

A detailed block-diagram representation of the complete system is shown in Fig. \ref{studycasefullmodel}, where $\overline{\boldsymbol{\omega}}_{\bullet}^{\bullet}=\left[ 0 , 0 , \Omega \right]^{\mathrm{T}}$. Furthermore, the uncertainty block regarding the central rigid hub $\mathcal{B}$ can be written as $\boldsymbol{\Delta}_{\mathcal{B}}=\delta_{\Omega}\textbf{I}_{8}$. Following the same logic, the structured uncertainty blocks characterizing the flexible booms and the tip masses are equal to $\boldsymbol{\Delta}_{\mathcal{A}_{\bullet}}=\operatorname{diag}\left(\delta_{\Omega}\textbf{I}_{169}, \delta_{m^{\mathcal{S}_{\bullet}}}\textbf{I}_{64}\right)$ and $\boldsymbol{\Delta}_{\mathcal{S}_{\bullet}}=\operatorname{diag}\left(\delta_{\Omega}\textbf{I}_{4}, \delta_{m^{\mathcal{S}_{\bullet}}}\textbf{I}_{5}\right)$, respectively, as explained in section \ref{results}. In this case, the number of occurrences of $\delta_{\Omega}$ and $\delta_{m^{\mathcal{S}_{\bullet}}}$ corresponds to the spinning satellite mission scenario displayed in Fig. \ref{studycasefullmodel}. For the flexible boom models, the equilibrium conditions are calculated by following the instructions depicted in section \ref{beam}.

When the full system is assembled, a global LFR representation is obtained, which fully captures the dynamics and interactions between all the subsystems of the scenario being studied. Furthermore, this model also takes into account the various uncertainty effects in a very compact representation. Fig. \ref{studycasefullmodel} illustrates the internal structure of the overall LFR model, as well as the interconnections between the several subsystems. In this representation, all the block uncertainties are isolated at the component level. 

Since the complete system is spinning, its rotations about the principal axes that correspond to the largest and the smallest moments of inertia are stable. However, the rotation about the axis corresponding to the
intermediate principal moment of inertia becomes unstable, leading to unexpected rotations that are the basis of the phenomenon commonly known as the \textsc{Dzhanibekov} effect \cite{Murakami2016}. For this reason, it must be ensured that the rotation about the $\mathbf{z}_b$-axis does not correspond to the
intermediate principal moment of inertia. Lastly, the presence of symmetry within the system is essential for an equilibrium to be established. All the numerical values and range of variations of the numerous system parameters which are employed in this section are described in Table \ref{tab:Sat_prop}.

\subsection{Analysis of the spinning and flexible spacecraft dynamics}

The Rayleigh damping model is employed to introduce a damping matrix $\mathbf{D}_{\xi}$ for the flexible booms $\mathcal{A}_{\bullet}$ of the spinning spacecraft. This model characterizes damping as a linear combination of the mass and stiffness matrices, facilitating the inclusion of an adaptable damping ratio for the associated modes. Therefore, the method presented in \cite{hall} is used to generate the damping matrix $\mathbf{D}_{\xi}=\alpha \mathbf{M}_{ff}+\beta \mathbf{K}_{ff}$, with $\alpha=0.0001$ \si{\per\second} and $\beta=0.0012$ \si{\second}. Ultimately, the matrix $\mathbf{D}_{ff}$ is replaced by $\widehat{\mathbf{D}}_{ff}=\mathbf{D}_{ff}+\mathbf{D}_{\xi}$ in the state-space representations of the flexible spinning booms (displayed in Eq. \eqref{statespace}), so that the selected damping is taken into account. 

\begin{figure}[!ht]
\centering
 \includegraphics[width=1\textwidth]{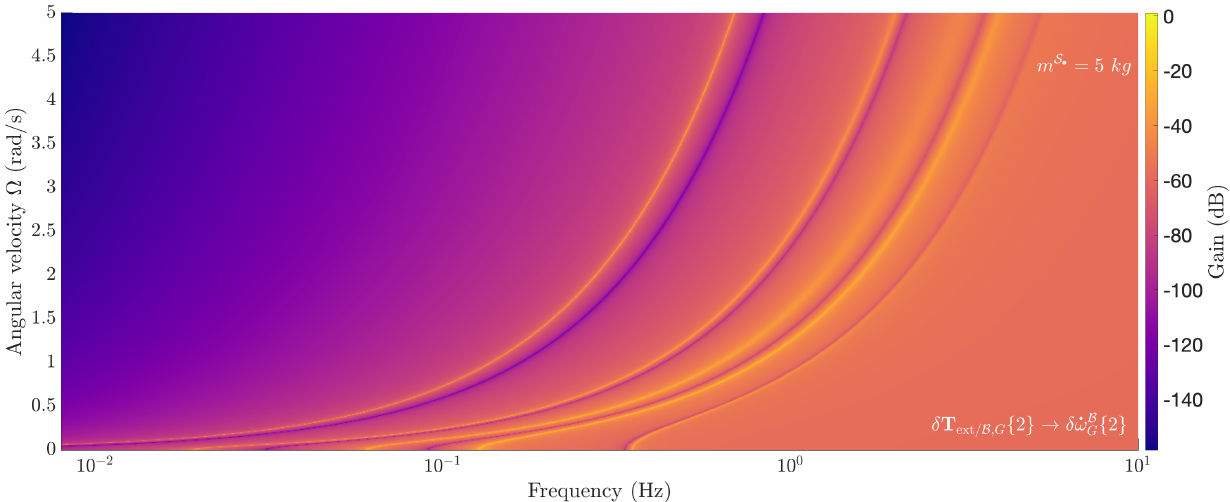}
\caption{Singular values of the model ${\left[\mathbf{G}({\mathrm{s}, \boldsymbol{\Delta}_{\mathbf{G}}})\right]_{\mathcal{R}_{b}}}$  with respect to the angular velocity $\Omega$ and along a dense grid of frequencies ($\delta{\mathbf{T}_{\mathrm{ext} / \mathcal{B}, G}}\{2\} \rightarrow \delta\boldsymbol{\dot{{\omega}}}_{G}^{\mathcal{B}}\{2\}$ channel).}
\label{compomegam} 
\end{figure}

\begin{figure}[!ht]
\centering
 \includegraphics[width=1\textwidth]{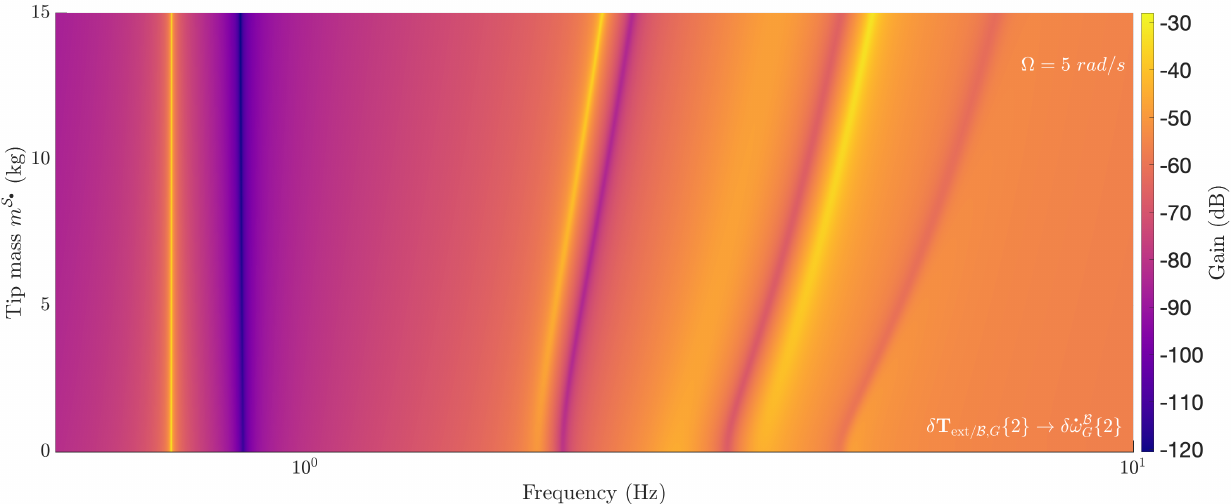}
\caption{Singular values of the model ${\left[\mathbf{G}({\mathrm{s}, \boldsymbol{\Delta}_{\mathbf{G}}})\right]_{\mathcal{R}_{b}}}$  for different tip mass values $m^{\mathcal{S}_{\bullet}}$ and along a dense grid of frequencies ($\delta{\mathbf{T}_{\mathrm{ext} / \mathcal{B}, G}}\{2\} \rightarrow \delta\boldsymbol{\dot{{\omega}}}_{G}^{\mathcal{B}}\{2\}$ channel).}
\label{surftipmass} 
\end{figure}

Let us now analyze the singular values of the spinning and flexible spacecraft model ${\left[\mathbf{G}({\mathrm{s}, \boldsymbol{\Delta}_{\mathbf{G}}})\right]_{\mathcal{R}_{b}}}$ for different values of $\Omega$ and $m^{\mathcal{S}_{\bullet}}$, where $\boldsymbol{\Delta}_{\mathbf{G}}=\operatorname{diag}\left(\boldsymbol{\Delta}_{\mathcal{B}}, \boldsymbol{\Delta}_{\mathcal{A}_1}, \boldsymbol{\Delta}_{\mathcal{A}_2}, \boldsymbol{\Delta}_{\mathcal{S}_1}, \boldsymbol{\Delta}_{\mathcal{S}_2}\right)$. For instance, if the transfer function from the second component of the external torque $\delta{\mathbf{T}_{\mathrm{ext} / \mathcal{B}, G}}$ to the second component of the angular acceleration $\delta\boldsymbol{\dot{{\omega}}}_{G}^{\mathcal{B}}$ is considered, Figs. \ref{compomegam} and \ref{surftipmass} show how the natural frequencies of the system evolve with respect to $\Omega$ and $m^{\mathcal{S}_{\bullet}}$. The complete system has 52 poles, where 12 poles are associated with the rigid hub model ${\left[\mathfrak{X}_{GP_1P_2}^{\mathcal{B}}(\mathrm{s})\right]}_{\mathcal{R}_{b}}$ and the other 40 poles are related to the two spinning beam models ${\left[\mathfrak{T}_{P_{\bullet}C_{\bullet}}^{\mathcal{A}_{\bullet}}(\mathrm{s})\right]}_{\mathcal{R}_{a_{\bullet}}}$. In Figs. \ref{compomegam} and \ref{surftipmass}, the number of flexible modes which are visible correspond to the number of observable poles linked with the minimal realization of the system ${\left[\mathbf{G}({\mathrm{s}, \boldsymbol{\Delta}_{\mathbf{G}}})\right]_{\mathcal{R}_{b}}}$ when taking into account the selected channel $\delta{\mathbf{T}_{\mathrm{ext} / \mathcal{B}, G}}\{2\} \rightarrow \delta\boldsymbol{\dot{{\omega}}}_{G}^{\mathcal{B}}\{2\}$. Since ${\left[\mathbf{G}({\mathrm{s}, \boldsymbol{\Delta}_{\mathbf{G}}})\right]_{\mathcal{R}_{b}}}$ represents the inverse linearized dynamic model of the whole system projected in $\mathcal{R}_{b}$ and the channel  $\delta{\mathbf{T}_{\mathrm{ext} / \mathcal{B}, G}}\{2\} \rightarrow \delta\boldsymbol{\dot{{\omega}}}_{G}^{\mathcal{B}}\{2\}$ is being analyzed, the static gain of the plot in Fig. \ref{compomegam} where $\Omega=0$ \si{\radian\per\second} represents the inverse of the second moment of inertia of the complete spacecraft, measured at the point $G$ and written in $\mathcal{R}_{b}$.

In Fig. \ref{studycasefullmodel}, the centrifugal forces applied by the tip masses $\mathcal{S}_{\bullet}$ to the spinning beams $\mathcal{A}_{\bullet}$ are determined by the equation $\overline{\mathbf{W}}_{\mathcal{S}_{\bullet}/\mathcal{A}_{\bullet},{C}_{\bullet}}\{1\}=m^{\mathcal{S}_{\bullet}}\left(l^{\mathcal{A}_{\bullet}}+r\right)\Omega^2$. 
These forces are projected onto the beams' body frames, labeled as $\mathcal{R}_{a_{\bullet}}$. They are also accountable for the centrifugal stiffening effect, which causes the natural frequencies of the system to shift to the right in the singular values plots as the values of $\Omega$ and $m^{\mathcal{S}_{\bullet}}$ increase, as displayed in Figs. \ref{compomegam} and \ref{surftipmass}. Notably, this effect is more pronounced for $\Omega$, since it appears squared in the expression for the centrifugal forces.

\section{Conclusion}

This paper outlined a comprehensive study on the analytical modeling and analysis of a spinning, flexible and uniform Euler-Bernoulli beam, with a specific focus on its application to a spinning spacecraft mission scenario. First and foremost, the presented framework shows how to build a very compact representation of a six degrees of freedom analytical model that fully captures the complex dynamics of a spinning beam, including the bending, traction and torsion effects. This model is represented as a linear fractional representation and is parameterized based on the angular velocity of the beam. Such a model is instrumental for robust control system design in spacecraft applications, where precise control of the flexible structures is critical. Moreover, the TITOP beam model was validated by comparing the natural frequencies with results from a traditional finite element code, ensuring the accuracy and reliability of this analytical approach. The paper also provides a valuable insight into the dynamics of a spinning spacecraft mission, which includes a central body, two flexible booms and two tip masses. This multibody modeling and analysis offers a practical understanding of the behavior of such systems under spinning conditions, which can be beneficial for the design and operation of similar missions, like THOR and Cluster.

\begin{table} [!ht]
	\caption{Spinning spacecraft mechanical data and equilibrium conditions.}
	\label{tab:Sat_prop}	
	\centering
	\resizebox{\textwidth}{!}{
	\begin{tabular}{p{1.5cm} l l r }
		\toprule
		& \textbf{Parameter} & \textbf{Description}  & \textbf{Value} \\
		\midrule
		\multirow{12}{1.5cm}{\centering Central rigid body $\mathcal{\mathcal{B}}$}
		& $\overline{\mathbf{G P}}_{\bullet}$ & distance vector between $G$ and $P_{\bullet}$ in equilibrium and written in $\mathcal{R}_{b}$ & $\left[\pm 2, 0, 0\right]$ \si{\meter} \\
		&  $m^{{\mathcal{B}}}$ & mass of $\mathcal{B}$ & $500\ \si{\kilogram}$ \\

		& ${\left[\begin{array}{ccc}
{J}^{{\mathcal{B}}}_{xx} & {J}^{{\mathcal{B}}}_{xy} & {J}^{{\mathcal{B}}}_{xz} \\
 & {J}^{{\mathcal{B}}}_{yy} & {J}^{{\mathcal{B}}}_{yz} \\
 & & {J}^{{\mathcal{B}}}_{zz} 
\end{array}\right]}$ & inertia of $\mathcal{\mathcal{B}}$ computed at the point $G$ and written in $\mathcal{R}_{b}$& $\begin{bmatrix}
		570.42 & 0 & 0 \\
		 &  570.42 & 0 \\
		 &  &  1000 \\
		\end{bmatrix}\, \si{\kilogram\square\meter} $ \\
&  $\mathcal{T}_{\mathcal{R}_{a_1}/\mathcal{R}_{b} }$ & change of frame DCM between $\mathcal{R}_{a_1}$ and $\mathcal{R}_{b}$ & $\begin{bmatrix}
		-1 & 0 & 0 \\
		 0 &  -1 & 0 \\
		0 & 0 &  1 \\
		\end{bmatrix} $ \\

&  $\mathcal{T}_{\mathcal{R}_{a_2}/\mathcal{R}_{b} }$ & change of frame DCM between $\mathcal{R}_{a_2}$ and $\mathcal{R}_{b}$ & $\mathbf{I}_{3}$ \\ 

  & $\overline{\mathbf{v}}_{G}^{{\mathcal{B}}}$ & inertial velocity vector of the body $\mathcal{B}$ computed at the point $G$ in equilibrium & $\mathbf{0}_{3\times 1}$ \si{\meter\per\second} \\

    & $\overline{\boldsymbol{\omega}}_{G}^{{\mathcal{B}}}$ & angular velocity vector of the body $\mathcal{B}$ computed at the point $G$ in equilibrium & $\left[ 0 , 0 , \Omega\right]^{\mathrm{T}}$ \si{\radian\per\second} \\

   \midrule
		\multirow{16}{1.5cm}{\centering Flexible booms ${\mathcal{A}_{\bullet}}$}
		& $l^{\mathcal{A}_\bullet}$ & length of ${\mathcal{A}_{\bullet}}$  & $50$ \si{\meter} \\
		& $\rho^{\mathcal{A}_\bullet}$ & mass density of  $\mathcal{A}_{\bullet}$ & $2700$ \si{\kilogram\per\cubic\meter} \\
  & $S^{\mathcal{A}_\bullet}$ & cross-sectional area of $\mathcal{A}_{\bullet}$ & $3.14\times 10^{-4}$ \si{\square\meter} \\
    & $E^{\mathcal{A}_\bullet}$ & Young's modulus of $\mathcal{A}_{\bullet}$ & $7\times 10^{10}$ \si{\newton\per\square\meter} \\
    & $\nu^{\mathcal{A}_\bullet}$ & Poisson's ratio of $\mathcal{A}_{\bullet}$ & $0.33$ \\

      & $J^{\mathcal{A}_\bullet}_{px}$ & second polar moment of area of ${\mathcal{A}_\bullet}$ with respect to the $\mathbf{x}_a$-axis & $1.57\times 10^{-8}$ \si{\meter}$^4$ \\
       & $J^{\mathcal{A}_\bullet}_y$ & second moment of area of ${\mathcal{A}_\bullet}$ with respect to the $\mathbf{y}_a$-axis & $7.85\times 10^{-9}$ \si{\meter}$^4$ \\
        &  $J^{\mathcal{A}_\bullet}_z$ & second moment of area of ${\mathcal{A}_\bullet}$ with respect to the $\mathbf{z}_a$-axis & $7.85\times 10^{-9}$ \si{\meter}$^4$ \\
        
     &  $\mathcal{T}_{\mathcal{R}_{s_\bullet}/{\mathcal{R}_{a_\bullet(C_{\bullet})}}}$ & change of frame DCM between $\mathcal{R}_{s_\bullet}$ and ${\mathcal{R}_{a_\bullet(C_\bullet)}}$ & $\mathbf{I}_{3}$ \\

    & $\overline{\mathbf{x}}^{\mathcal{A}_{\bullet}}_{P_\bullet}$ & distance vector between $O$ and ${P_\bullet}$ in equilibrium and written in $\mathcal{R}_{a_{\bullet}}$ & $\left[ r = 2 , 0 , 0\right]^{\mathrm{T}}$ \si{\meter} \\

& $\overline{\boldsymbol{{\Theta}}}_{P_\bullet}^{{\mathcal{A}_{\bullet}}}$ & Euler angles of the beam $\mathcal{A}_{\bullet}$ computed at the point ${P_\bullet}$ in equilibrium & $\mathbf{0}_{3\times 1}$ \si{\radian}\\

& $\overline{\mathbf{v}}_{{P_\bullet}}^{{\mathcal{A}_{\bullet}}}$ & inertial velocity vector of the beam $\mathcal{A}_{\bullet}$ computed at the point ${P_\bullet}$ in equilibrium & $\left[ 0 , r\Omega , 0\right]^{\mathrm{T}}$ \si{\meter\per\second} \\

& $\overline{{\boldsymbol{{\omega}}}}_{P_\bullet}^{{\mathcal{A}_{\bullet}}}$ & angular velocity vector of the beam $\mathcal{A}_{\bullet}$ computed at the point ${P_\bullet}$ in equilibrium & $\left[ 0 , 0 , \Omega\right]^{\mathrm{T}}$ \si{\radian\per\second}  \\
 
& $\overline{\mathbf{W}}_{\mathcal{S}_{\bullet}/\mathcal{A}_{\bullet},{C}_{\bullet}}$ & wrench vector applied by the body $\mathcal{S}_{\bullet}$ to the body $\mathcal{A}_{\bullet}$ at the point $C_{\bullet}$ in equilibrium &  $\left[ m^{\mathcal{S}_{\bullet}}\left(l^{\mathcal{A}_{\bullet}}+r\right)\Omega^2 , 0 , 0 , 0 , 0 , 0\right]^{\mathrm{T}}\ \left[\si{\newton}\;\; \si{\newton\meter}\right]$ \\
 
		 \midrule

  		\multirow{4}{1.5cm}{\centering Tip masses ${\mathcal{S}_{\bullet}}$}
		& ${\left[\begin{array}{ccc}
{J}^{{\mathcal{S}_{\bullet}}}_{xx} & {J}^{{\mathcal{S}_{\bullet}}}_{xy} & {J}^{{\mathcal{S}_{\bullet}}}_{xz} \\
 & {J}^{{\mathcal{S}_{\bullet}}}_{yy} & {J}^{{\mathcal{S}_{\bullet}}}_{yz} \\
 & & {J}^{{\mathcal{S}_{\bullet}}}_{zz} 
\end{array}\right]}$ & inertia of $\mathcal{S}_{\bullet}$ computed at the point $C_{\bullet}$ and written in $\mathcal{R}_{s_{\bullet}}$& $\begin{bmatrix}
		0 & 0 & 0 \\
		 &  0 & 0 \\
		 &  &  0 \\
		\end{bmatrix}\, \si{\kilogram\square\meter} $ \\

  & $\overline{\mathbf{v}}_{{C_\bullet}}^{{\mathcal{S}_{\bullet}}}$ & inertial velocity vector of the beam $\mathcal{S}_{\bullet}$ computed at the point ${C_\bullet}$ in equilibrium & $\left[ 0 , \left(l^{\mathcal{A}_{\bullet}}+r\right)\Omega , 0\right]^{\mathrm{T}}$ \si{\meter\per\second}\\

    & $\overline{\boldsymbol{\omega}}_{C_{\bullet}}^{{\mathcal{S}_{\bullet}}}$ & angular velocity vector of the body $\mathcal{S}_{\bullet}$ computed at the point $C_{\bullet}$ in equilibrium & $\left[ 0 , 0 , \Omega\right]^{\mathrm{T}}$ \si{\radian\per\second} \\

		 \midrule
	\end{tabular}
}
\end{table}

\appendix
\section{Choice of the modal decomposition}
\label{modal}

A decomposition based on fifth-order polynomial shape functions \cite{MURALI, CHEBBI} is selected to build the TITOP model ${\left[\mathfrak{T}_{PC}^{\mathcal{A}}(\mathrm{s})\right]}_{\mathcal{R}_{a}}$. In this approach, the deflection $\tilde{v}(x, t)$ in the moving frame ${\mathcal{R}}_a$ is expressed as:

\begin{equation}
\tilde{v}(x, t)=a_2^{y}(t) x^2+a_3^{y}(t) x^3+a_4^{y}(t) x^4+a_5^{y}(t) x^5 
\end{equation}

This fifth-order polynomial ensures that $\tilde{v}(0, t)=\frac{\partial \tilde{v}}{\partial x}(0, t)=0, \forall t$. The coefficients $a_i^{y}(t)$ are then expressed as a function of four time-domain functions $q_i^{y}(t)$ defined by:

\begin{equation}
\mathbf{q}_{y}(t)=\left[\begin{array}{llll}
\frac{\partial^{2} \tilde{v}}{\partial x^{2}}(0, t) & \tilde{v}(l^{\mathcal{A}}, t) & \frac{\partial \tilde{v}}{\partial x}(l^{\mathcal{A}}, t) & \frac{\partial^{2} \tilde{v}}{\partial x^{2}}(l^{\mathcal{A}}, t)\end{array}\right]^{\mathrm{T}}
\end{equation}

The shape functions vector reads:

\begin{equation}
\boldsymbol{\Phi}_y(x)=\left[\def\arraystretch{1.2}\begin{array}{c}
\frac{1}{2} x^2-\frac{3}{2 l^{\mathcal{A}}} x^3+\frac{3}{2 l^{\mathcal{A}^2}} x^4-\frac{1}{2 l^{\mathcal{A}^3}} x^5 \\
\frac{10}{l^{\mathcal{A}^3}} x^3-\frac{15}{l^{\mathcal{A}^4}} x^4+\frac{6}{l^{\mathcal{A}^5}} x^5 \\
\frac{-4}{l^{\mathcal{A}^2}} x^3+\frac{7}{l^{\mathcal{A}^3}} x^4-\frac{3}{l^{\mathcal{A}^4}} x^5 \\
\frac{1}{2 l^{\mathcal{A}}} x^3-\frac{1}{l^{\mathcal{A}^2}} x^4+\frac{1}{2 l^{\mathcal{A}^3}} x^5
\end{array}\right]
\end{equation}

A similar process is achieved for the deflection $\tilde{w}(x, t)$ in the moving frame ${\mathcal{R}}_a$, with:

\begin{equation}
\mathbf{q}_{z}(t)=\left[\begin{array}{llll}
\frac{\partial^{2} \tilde{w}}{\partial x^{2}}(0, t) & \tilde{w}(l^{\mathcal{A}}, t) & \frac{\partial \tilde{w}}{\partial x}(l^{\mathcal{A}}, t) & \frac{\partial^{2} \tilde{w}}{\partial x^{2}}(l^{\mathcal{A}}, t)\end{array}\right]^{\mathrm{T}} \quad \text{and} \quad \boldsymbol{\Phi}_z(x)=\boldsymbol{\Phi}_y(x)
\end{equation}

\section{Matrices used to model the state-space representation of the spinning beam}
\label{append}

\begin{flalign}
& \def\arraystretch{1.45}\begin{array}{ll}
\mathbf{M}_{p}=\rho^{\mathcal{A}}S^{\mathcal{A}} \int_{0}^{l^{\mathcal{A}}}{{({ }^{*} \mathbf{P}_0(x)) d x}}  & \mathbf{M}_{\sigma\sigma}=\rho^{\mathcal{A}}S^{\mathcal{A}} \int_{0}^{l^{\mathcal{A}}}{{\sigma(x)\sigma(x) d x}} \\
\mathbf{M}_{m}=\rho^{\mathcal{A}}S^{\mathcal{A}} \int_{0}^{l^{\mathcal{A}}}{{ \mathbf{M}_1(x)  d x}} 
& \mathbf{M}_{\sigma}=\rho^{\mathcal{A}}S^{\mathcal{A}} \int_{0}^{l^{\mathcal{A}}}{{\sigma(x) d x}} \\
\mathbf{K}_{soft}=\rho^{\mathcal{A}}S^{\mathcal{A}} \int_{0}^{l^{\mathcal{A}}}{{ -\mathbf{M}_1(x)({ }^{*} \overline{\boldsymbol{\omega}}^{\mathcal{A}}_{{P}})^2\mathbf{M}^{\mathrm{T}}_1(x) d x}} & \mathbf{D}_{soft}=\rho^{\mathcal{A}}S^{\mathcal{A}} \int_{0}^{l^{\mathcal{A}}}{{ \mathbf{M}_1(x)({ }^{*} \overline{\boldsymbol{\omega}}^{\mathcal{A}}_{{P}})\mathbf{M}^{\mathrm{T}}_1(x) d x}} \\
\mathbf{M}_{mp\omega}=\rho^{\mathcal{A}}S^{\mathcal{A}} \int_{0}^{l^{\mathcal{A}}}{{ \mathbf{M}_1(x)({ }^{*} \overline{\boldsymbol{\omega}}^{\mathcal{A}}_{{P}})({ }^{*} \mathbf{P}_0(x)) d x}}  &
\mathbf{E}_{i}=\rho^{\mathcal{A}}S^{\mathcal{A}} \int_{0}^{l^{\mathcal{A}}}{{\left[\begin{array}{cc}\mathbf{E}^{yz}(x) & \mathbf{0}_{2N\times 2} \\ \mathbf{0}_{2\times 2N} & \mathbf{0}_{2\times 2}\end{array}\right] d x}} \\
\mathbf{E}_{ix}=\rho^{\mathcal{A}}S^{\mathcal{A}} \int_{0}^{l^{\mathcal{A}}}{{\left[\begin{array}{cc}x\mathbf{E}^{yz}(x) & \mathbf{0}_{2N\times 2} \\ \mathbf{0}_{2\times 2N} & \mathbf{0}_{2\times 2}\end{array}\right] d x}}  &
\mathbf{E}_{l}=\left[\begin{array}{cc}\mathbf{E}^{yz}(l^{\mathcal{A}}) & \mathbf{0}_{2N\times 2} \\ \mathbf{0}_{2\times 2N} & \mathbf{0}_{2\times 2}\end{array}\right] 
\end{array} &
\end{flalign}

\begin{equation}
\def\arraystretch{1.45}\begin{array}{lll}
\mathbf{P}_{\sigma1}=\left[\begin{array}{ccc}m^{\mathcal{A}}\frac{J^{\mathcal{A}}_{px}}{S^{\mathcal{A}}} & 0 & 0 \\
0 & 0 & 0 \\
0 & 0 & 0 \end{array}\right] &
\mathbf{P}_{\sigma2}=\left[\begin{array}{ccc} \mathbf{0}_{N\times 1} & \mathbf{0}_{N\times 1} & \mathbf{0}_{N\times 1} \\
\mathbf{0}_{N\times 1} & \mathbf{0}_{N\times 1} & \mathbf{0}_{N\times 1} \\
0 & 0 & 0 \\
\frac{J^{\mathcal{A}}_{px}}{S^{\mathcal{A}}}\mathbf{M}_{\sigma} & 0 & 0
\end{array}\right] &
\mathbf{P}_{\sigma3}=\left[\begin{array}{cccc}\mathbf{0}_{N\times N} & \mathbf{0}_{N\times N} & \mathbf{0}_{N\times 1} & \mathbf{0}_{N\times 1} \\
\mathbf{0}_{N\times N} & \mathbf{0}_{N\times N} & \mathbf{0}_{N\times 1} & \mathbf{0}_{N\times 1} \\
\mathbf{0}_{1\times N} & \mathbf{0}_{1\times N} & 0 & 0 \\
\mathbf{0}_{1\times N} & \mathbf{0}_{1\times N} & 0 & \frac{J^{\mathcal{A}}_{px}}{S^{\mathcal{A}}}\mathbf{M}_{\sigma\sigma}\end{array}\right]
\end{array}
\end{equation}

\begin{flalign}
& \def\arraystretch{1.45}\begin{array}{ll}
\mathbf{J}_{P} =\rho^{\mathcal{A}}S^{\mathcal{A}} \int_{0}^{l^{\mathcal{A}}}-{{({ }^{*} \mathbf{P}_0(x))^2 d x}}+\mathbf{P}_{\sigma1} &
\mathbf{M}_{mp}=\rho^{\mathcal{A}}S^{\mathcal{A}} \int_{0}^{l^{\mathcal{A}}}{{ \mathbf{M}_1(x) ({ }^{*} \mathbf{P}_0(x))^\mathrm{T} d x}} + \mathbf{P}_{\sigma2} \\
\mathbf{M}_{mm}=\rho^{\mathcal{A}}S^{\mathcal{A}} \int_{0}^{l^{\mathcal{A}}}{{ \mathbf{M}_1(x)\mathbf{M}^{\mathrm{T}}_1(x)  d x}}+ \mathbf{P}_{\sigma3}
\end{array} &
\end{flalign}

\begin{equation}
\mathbf{M}_{\mathcal{T}}=\left[\def\arraystretch{1.2}\begin{array}{ccc}
m^{\mathcal{A}}\mathbf{I}_3 & \mathbf{M}^{\mathrm{T}}_{p} & \mathbf{M}^{\mathrm{T}}_{m} \\
\mathbf{M}_{p} & \mathbf{J}_{P} & \mathbf{M}_{mp}^{\mathrm{T}} \\
\mathbf{M}_{m} & \mathbf{M}_{mp} & \mathbf{M}_{mm}
\end{array}\right]
\end{equation}

\begin{equation}
\mathbf{K}_{\mathcal{T}}=\left[\def\arraystretch{1.2}\begin{array}{ccc}
\mathbf{0}_{3\times 3} & \mathbf{0}_{3\times 3} & \mathbf{0}_{3\times (2N+2)} \\
\mathbf{0}_{3\times 3} & \mathbf{0}_{3\times 3} & \mathbf{0}_{3\times (2N+2)} \\
\mathbf{0}_{(2N+2) \times 3} & \mathbf{0}_{(2N+2) \times 3} & \mathbf{K}_{soft}+\left(\overline{\boldsymbol{\omega}}_{P}^{\mathcal{A}}\{2\}\overline{\mathbf{v}}_{P}^{{\mathcal{A}}}\{3\}-\overline{\boldsymbol{\omega}}_{P}^{{\mathcal{A}}}\{3\}\overline{\mathbf{v}}_{P}^{{\mathcal{A}}}\{2\}\right)\mathbf{E}_i-\left((\overline{\boldsymbol{\omega}}_{P}^{{\mathcal{A}}}\{3\})^2+(\overline{\boldsymbol{\omega}}_{P}^{{\mathcal{A}}}\{2\})^2\right)\mathbf{E}_{ix}
\end{array}\right]
\end{equation}

\begin{equation}
\mathbf{G}_{\mathcal{T}}=\left[\def\arraystretch{1.2}\begin{array}{ccc}
\mathbf{0}_{3\times 3} & \mathbf{0}_{3\times 3} & \mathbf{0}_{3\times (2N+2)} \\
\mathbf{0}_{3\times 3} & \mathbf{0}_{3\times 3} & \mathbf{0}_{3\times (2N+2)} \\
-2\mathbf{M}_{m}({ }^{*} \overline{\boldsymbol{\omega}}^{\mathcal{A}}_{{P}}) & 2\mathbf{M}_{m}({ }^{*} \overline{\mathbf{v}}^{\mathcal{A}}_{{P}})+2\mathbf{M}_{mp}({ }^{*} \overline{\boldsymbol{\omega}}^{\mathcal{A}}_{{P}})+4\mathbf{M}_{mp\omega} & -2\overline{\mathbf{v}}_{P}^{{\mathcal{A}}}\{1\}\mathbf{E}_i -2\mathbf{D}_{soft}
\end{array}\right]
\end{equation}

\begin{equation}
\def\arraystretch{1.2}\begin{array}{ll}
\mathbf{C}_{Q}^{\mathrm{T}}=\left[\begin{array}{ccc}
\mathbf{0}_{1\times 3} & \mathbf{0}_{1\times 3} &  \overline{\mathbf{v}}^{\mathcal{A}^{\mathrm{T}}}_{{P}}({ }^{*} \overline{\boldsymbol{\omega}}^{\mathcal{A}}_{{P}})\mathbf{M}^{\mathrm{T}}_{m}+({ }^{*} \overline{\boldsymbol{\omega}}^{\mathcal{A}}_{{P}})^{\mathrm{T}}\mathbf{M}^{\mathrm{T}}_{mp\omega}
\end{array}\right]  &
\mathbf{C}_{\dot{Q}}^{\mathrm{T}}=\left[\begin{array}{c} \overline{\mathbf{v}}^{\mathcal{A}}_{{P}} \\ \overline{\boldsymbol{\omega}}^{\mathcal{A}}_{{P}}\end{array}\right]^{\mathrm{T}}\mathbf{M}_{\mathcal{T}{(1:6,:)}}
\end{array}
\end{equation}

\begin{equation}
\left[\begin{array}{cc} ({ }^{*} \boldsymbol{\omega}^{\mathcal{A}}_{{P}}) & \mathbf{0}_{3 \times 3}\\
({ }^{*} \mathbf{v}^{\mathcal{A}}_{{P}}) & ({ }^{*} \boldsymbol{\omega}^{\mathcal{A}}_{{P}})
\end{array}\right]\frac{\partial \mathcal{L}}{\partial\left[\begin{array}{c}\delta\mathbf{v}^{\mathcal{A}}_{{P}}\\\delta\boldsymbol{\omega}^{\mathcal{A}}_{{P}}\end{array}\right]}
=\mathbf{M}_{\mathcal{L}}\delta{{\mathbf{Q}}}_{v}(t)+\mathbf{J}_{\mathcal{L}}\delta{{\mathbf{Q}}}_{p}(t)+\mathbf{C}_{\mathcal{L}} 
\end{equation}

\begin{equation}
\mathbf{M}_{\mathcal{L}}=\left[\begin{array}{cc} ({ }^{*} \overline{\boldsymbol{\omega}}^{\mathcal{A}}_{{P}}) & \mathbf{0}_{3 \times 3}\\
({ }^{*} \overline{\mathbf{v}}^{\mathcal{A}}_{{P}}) & ({ }^{*} \overline{\boldsymbol{\omega}}^{\mathcal{A}}_{{P}})
\end{array}\right]\mathbf{M}_{\mathcal{T}{(1:6,:)}} - \left[\begin{array}{ccc}\mathbf{0}_{3 \times 3} & m^{\mathcal{A}}({ }^{*} \overline{\mathbf{v}}^{\mathcal{A}}_{{P}})+ ({ }^{*}( \mathbf{M}^{\mathrm{T}}_{p}\overline{\boldsymbol{\omega}}^{\mathcal{A}}_{{P}})) & \mathbf{0}_{3 \times (2N+2)}\\
m^{\mathcal{A}}({ }^{*} \overline{\mathbf{v}}^{\mathcal{A}}_{{P}})+ ({ }^{*} (\mathbf{M}^{\mathrm{T}}_{p}\overline{\boldsymbol{\omega}}^{\mathcal{A}}_{{P}})) & ({ }^{*} (\mathbf{M}_{p}\overline{\mathbf{v}}^{\mathcal{A}}_{{P}}))+({ }^{*} (\mathbf{J}_P\overline{\boldsymbol{\omega}}^{\mathcal{A}}_{{P}})) & \mathbf{0}_{3 \times (2N+2)}
\end{array}\right]
\end{equation}

\begin{equation}
\def\arraystretch{1.2}\begin{array}{ll}
\mathbf{J}_{\mathcal{L}}=\left[\begin{array}{cc} ({ }^{*} \overline{\boldsymbol{\omega}}^{\mathcal{A}}_{{P}}) & \mathbf{0}_{3 \times 3}\\
({ }^{*} \overline{\mathbf{v}}^{\mathcal{A}}_{{P}}) & ({ }^{*} \overline{\boldsymbol{\omega}}^{\mathcal{A}}_{{P}})
\end{array}\right]\frac{1}{2}\left(\mathbf{G}_{\mathcal{T}{(:,1:6)}}\right)^{\mathrm{T}}   &
\mathbf{C}_{\mathcal{L}}=\left[\begin{array}{cc} ({ }^{*} \overline{\boldsymbol{\omega}}^{\mathcal{A}}_{{P}}) & \mathbf{0}_{3 \times 3}\\
({ }^{*} \overline{\mathbf{v}}^{\mathcal{A}}_{{P}}) & ({ }^{*} \overline{\boldsymbol{\omega}}^{\mathcal{A}}_{{P}})
\end{array}\right]\mathbf{C}_{\dot{Q}{(1:6,1)}}
\end{array}
\end{equation}

\begin{equation}
\def\arraystretch{1.2}\begin{array}{ll}
\overline{\boldsymbol{\tau}}_{CP}=\left[\begin{array}{c|c}
\mathbf{I}_{3} & \left[\begin{array}{ccc} 0 & 0 & 0 \\  0 & 0 & l^{\mathcal{A}} \\
0 & -l^{\mathcal{A}} & 0
\end{array}\right]\\ \hline
\mathbf{0}_{3\times 3} & \mathbf{I}_{3}
\end{array}\right]  &
\mathbf{G}_{vv}=\left[\begin{array}{c}
\mathbf{M}^{\mathrm{T}}_1(l^{\mathcal{A}}) \\
\mathbf{M}^{\mathrm{T}}_2(l^{\mathcal{A}})
\end{array}\right]
\end{array}
\end{equation}

\begin{equation}
\mathbf{M}^{\mathrm{T}}_3=\left[\begin{array}{cccc}
\boldsymbol{\Phi}_y^{\prime^\mathrm{T}}(l^{\mathcal{A}})\left(\overline{\mathbf{v}}_{P}^{{\mathcal{A}}}\{2\}+l^{\mathcal{A}}\overline{\boldsymbol{\omega}}_{P}^{{\mathcal{A}}}\{3\}\right) & \boldsymbol{\Phi}_z^{\prime^\mathrm{T}}(l^{\mathcal{A}})\left(\overline{\mathbf{v}}_{P}^{{\mathcal{A}}}\{3\}-l^{\mathcal{A}}\overline{\boldsymbol{\omega}}_{P}^{{\mathcal{A}}}\{2\}\right) & 0 & 0 \\
-\boldsymbol{\Phi}_y^{\prime^\mathrm{T}}(l^{\mathcal{A}})\overline{\mathbf{v}}_{P}^{{\mathcal{A}}}\{1\} & \mathbf{0}_{1\times N} & 0 & \sigma(l^{\mathcal{A}})\left(\overline{\mathbf{v}}_{P}^{{\mathcal{A}}}\{3\}-l^{\mathcal{A}}\overline{\boldsymbol{\omega}}_{P}^{{\mathcal{A}}}\{2\}\right) \\
\mathbf{0}_{1\times N} & -\boldsymbol{\Phi}_z^{\prime^\mathrm{T}}(l^{\mathcal{A}})\overline{\mathbf{v}}_{P}^{{\mathcal{A}}}\{1\} & 0 & -\sigma(l^{\mathcal{A}})\left(\overline{\mathbf{v}}_{P}^{{\mathcal{A}}}\{2\}+l^{\mathcal{A}}\overline{\boldsymbol{\omega}}_{P}^{{\mathcal{A}}}\{3\}\right)
\end{array}\right]
\end{equation}

\begin{equation}
\mathbf{G}_{vp}=\left[\def\arraystretch{1.5}\begin{array}{c}
({ }^{*} \overline{\boldsymbol{\omega}}^{\mathcal{A}}_{{P}})\mathbf{M}^{\mathrm{T}}_1(l^{\mathcal{A}})+\mathbf{M}^{\mathrm{T}}_3 \\
\left[\def\arraystretch{1.2}\begin{array}{cccc}
\boldsymbol{\Phi}_y^{\prime^\mathrm{T}}(l^{\mathcal{A}})\overline{\boldsymbol{\omega}}_{P}^{{\mathcal{A}}}\{2\} & \boldsymbol{\Phi}_z^{\prime^\mathrm{T}}(l^{\mathcal{A}})\overline{\boldsymbol{\omega}}_{P}^{{\mathcal{A}}}\{3\} & 0 & 0 \\
-\boldsymbol{\Phi}_y^{\prime^\mathrm{T}}(l^{\mathcal{A}})\overline{\boldsymbol{\omega}}_{P}^{{\mathcal{A}}}\{1\} & \mathbf{0}_{1\times N} & 0 & \sigma(l^{\mathcal{A}})\overline{\boldsymbol{\omega}}_{P}^{{\mathcal{A}}}\{3\} \\
\mathbf{0}_{1\times N} & -\boldsymbol{\Phi}_z^{\prime^\mathrm{T}}(l^{\mathcal{A}})\overline{\boldsymbol{\omega}}_{P}^{{\mathcal{A}}}\{1\} & 0 & -\sigma(l^{\mathcal{A}})\overline{\boldsymbol{\omega}}_{P}^{{\mathcal{A}}}\{2\}
\end{array}\right]
\end{array}\right]
\end{equation}

\begin{equation}
\mathbf{G}_{p}=\left[\def\arraystretch{1.5}\begin{array}{c}
\left[\def\arraystretch{1.2}\begin{array}{cccc}
\boldsymbol{\Phi}_y^{\prime^\mathrm{T}}(l^{\mathcal{A}})\overline{\mathbf{x}}^{\mathcal{A}}_P\{2\} & \boldsymbol{\Phi}_z^{\prime^\mathrm{T}}(l^{\mathcal{A}})\overline{\mathbf{x}}^{\mathcal{A}}_P\{3\} & \tau(l^{\mathcal{A}}) & 0 \\
-\boldsymbol{\Phi}_y^{\prime^\mathrm{T}}(l^{\mathcal{A}})\left(\overline{\mathbf{x}}^{\mathcal{A}}_P\{1\}+l^{\mathcal{A}}\right)+\boldsymbol{\Phi}_y^{\mathrm{T}}(l^{\mathcal{A}}) & \mathbf{0}_{1\times N} & 0 & \sigma(l^{\mathcal{A}})\overline{\mathbf{x}}^{\mathcal{A}}_P\{3\} \\
\mathbf{0}_{1\times N} & -\boldsymbol{\Phi}_z^{\prime^\mathrm{T}}(l^{\mathcal{A}})\left(\overline{\mathbf{x}}^{\mathcal{A}}_P\{1\}+l^{\mathcal{A}}\right)+\boldsymbol{\Phi}_z^{\mathrm{T}}(l^{\mathcal{A}}) & 0 & -\sigma(l^{\mathcal{A}})\overline{\mathbf{x}}^{\mathcal{A}}_P\{2\}\end{array}\right] \\
\mathbf{M}^{\mathrm{T}}_2(l)
\end{array}\right]
\end{equation}

\begin{equation}
\def\arraystretch{1.2}\begin{array}{llllll}
\mathbf{M}_{l}=\left[\begin{array}{c}
\overline{{\boldsymbol{\tau}}}_{CP} \\
\mathbf{0}_{6\times 6} \\
\mathbf{0}_{6\times 6} 
\end{array}\right] &
\mathbf{D}_{l}=\left[\begin{array}{c}
\mathbf{0}_{6\times 6} \\
\overline{{\boldsymbol{\tau}}}_{CP} \\
\mathbf{0}_{6\times 6} 
\end{array}\right] & 
\mathbf{K}_{l}=\left[\begin{array}{c}
\mathbf{0}_{6\times 6} \\
\mathbf{0}_{6\times 6} \\
\mathbf{I}_{6}
\end{array}\right] &
\mathbf{M}_{r}=\left[\begin{array}{c}
\mathbf{G}_{vv}\\
\mathbf{0}_{6\times 10} \\
\mathbf{0}_{6\times 10} 
\end{array}\right] &
\mathbf{D}_{r}=\left[\begin{array}{c}
\mathbf{G}_{vp}\\
\mathbf{G}_{vv}\\
\mathbf{0}_{6\times 10} 
\end{array}\right] &
\mathbf{K}_{r}=\left[\begin{array}{c}
\mathbf{0}_{6\times 10}\\
\mathbf{G}_{vp}\\
\mathbf{G}_{p}
\end{array}\right]
\end{array}
\end{equation}

\begin{equation}
\overline{\mathbf{{\mathbf{N}}}}=\left[\begin{array}{cc}-\mathbf{I}_6 & \overline{\boldsymbol{\tau}}_{CP}^{\mathrm{T}}\\ \mathbf{0}_{(2N+2) \times 6} & \left[\begin{array}{cc} \mathbf{M}_1(l^{\mathcal{A}}) & \mathbf{M}_2(l^{\mathcal{A}})\end{array}\right]\end{array}\right]
\end{equation}

\begin{equation}
\mathbf{K}_{c}=-\left[\begin{array}{cccc}
\boldsymbol{\Phi}_y^{\mathrm{T}}(l^{\mathcal{A}})\overline{\mathbf{W}}_{{\bullet}/\mathcal{A},{C}}\{3\} & -\boldsymbol{\Phi}_z^{\mathrm{T}}(l^{\mathcal{A}})\overline{\mathbf{W}}_{{\bullet}/\mathcal{A},{C}}\{2\} & 0 & 0 \\ 
\mathbf{0}_{1\times N} & \boldsymbol{\Phi}_z^{\mathrm{T}}(l^{\mathcal{A}})\overline{\mathbf{W}}_{{\bullet}/\mathcal{A},{C}}\{1\} & -{\tau}(l^{\mathcal{A}})\overline{\mathbf{W}}_{{\bullet}/\mathcal{A},{C}}\{3\} & 0 \\
-\boldsymbol{\Phi}_y^{\mathrm{T}}(l^{\mathcal{A}})\overline{\mathbf{W}}_{{\bullet}/\mathcal{A},{C}}\{1\} & \mathbf{0}_{1\times N} & {\tau}(l^{\mathcal{A}})\overline{\mathbf{W}}_{{\bullet}/\mathcal{A},{C}}\{2\} & 0
\end{array}\right]
\end{equation}

\begin{equation}
\mathbf{F}_{c}=\left[\begin{array}{ccc}\mathbf{0}_{3 \times 3} & \mathbf{0}_{3 \times 3} & \mathbf{0}_{3 \times (2N+2)}\\
\mathbf{0}_{3 \times 3} & \mathbf{0}_{3 \times 3} & \mathbf{K}_{c}\\
\mathbf{0}_{(2N+2) \times 3} & \mathbf{0}_{(2N+2) \times 3} & \mathbf{E}_{l}\overline{\mathbf{W}}_{{\bullet}/\mathcal{A},{C}}\{1\}\\
\end{array}\right]+\left[\begin{array}{cc}\mathbf{0}_{(6+2N+2) \times 6} & \overline{\mathbf{{\mathbf{N}}}}_{(:,7:12)}\mathbf{W}_C\end{array}\right]
\end{equation}

\section*{Funding Sources}

This work was funded by ISAE-SUPAERO.




\bibliographystyle{model1-num-names}
\bibliography{library.bib}







\end{document}